%% file: review_draft.tex
\documentclass[chreay]{achemso}
\setkeys{acs}{articletitle = true, doi = true, maxauthors = 10}
\SectionNumbersOn
\usepackage[T1]{fontenc}
\usepackage[utf8]{inputenc}
\usepackage[activate={true,nocompatibility},final, tracking=true,factor=1100,stretch=10,shrink=10]{microtype}
\SetProtrusion{encoding={*},family={bch},series={*},size={6,7}}
              {1={ ,750},2={ ,500},3={ ,500},4={ ,500},5={ ,500},
               6={ ,500},7={ ,600},8={ ,500},9={ ,500},0={ ,500}}
\usepackage[sc, osf]{mathpazo}
\usepackage{csquotes}
\usepackage[hidelinks]{hyperref}

\usepackage{listings}
\usepackage{longtable}
\usepackage{xcolor}
\usepackage{wrapfig}

 
\definecolor{codegreen}{rgb}{0,0.6,0}
\definecolor{codegray}{rgb}{0.5,0.5,0.5}
\definecolor{codepurple}{rgb}{0.58,0,0.82}
\definecolor{backcolour}{rgb}{0.95,0.95,0.92}
 
\lstdefinestyle{mystyle}{
    commentstyle=\color{codegreen},
    keywordstyle=\color{magenta},
    numberstyle=\tiny\color{codegray},
    stringstyle=\color{codepurple},
    basicstyle=\ttfamily\footnotesize,
    breakatwhitespace=false,         
    breaklines=true,                 
    keepspaces=true,                 
    numbers=left,                    
    numbersep=5pt,                  
    showspaces=false,                
    showstringspaces=false,
    showtabs=false,                  
    tabsize=2
}
\usepackage{float}
\usepackage{booktabs}
\usepackage{tabularx}

\usepackage[version=3]{mhchem} 
\usepackage{amssymb}
\usepackage{amsmath}
\usepackage{nicefrac}
\usepackage{siunitx}
\usepackage{braket}

\usepackage{todonotes}
\usepackage[acronym]{glossaries}
\makeglossaries
\include{acronyms}

\usepackage{tikz}
\usepackage{pgf}
\usetikzlibrary{mindmap,trees,shadows}

\newcommand{\br}{\mathbf{r}}

\title[ML for Porous Materials]{Big-Data Science in Porous Materials: Materials Genomics and Machine Learning}

\author{Kevin~Maik~Jablonka}
\affiliation{Laboratory of Molecular Simulation (LSMO), Institut des Sciences et Ingénierie Chimiques (ISIC), École Polytechnique Fédérale de Lausanne (EPFL), Sion, Switzerland}

\author{Daniele~Ongari}
\affiliation{Laboratory of Molecular Simulation (LSMO), Institut des Sciences et Ingénierie Chimiques (ISIC), École Polytechnique Fédérale de Lausanne (EPFL), Sion, Switzerland}

\author{Seyed~Mohamad~Moosavi}
\affiliation{Laboratory of Molecular Simulation (LSMO), Institut des Sciences et Ingénierie Chimiques (ISIC), École Polytechnique Fédérale de Lausanne (EPFL), Sion, Switzerland}

\author{Berend~Smit}
\email{berend.smit@epfl.ch}
\affiliation{Laboratory of Molecular Simulation (LSMO), Institut des Sciences et Ingénierie Chimiques (ISIC), École Polytechnique Fédérale de Lausanne (EPFL), Sion, Switzerland}

\keywords{Metal organic frameworks; covalent organic frameworks, Machine learning, molecular simulation}

\begin{document}
\setstretch{1.2}
\maketitle
\begin{abstract}
    By combining metal nodes with organic linkers we can potentially synthesize millions of possible metal-organic frameworks (MOFs).
    The fact that we have so many materials opens many exciting avenues, but also create new challenges. We simply have too many material to be processed using conventional, brute force, methods.
    In this review, we show that having so many materials allows us to use big-data methods as a powerful technique to study these materials and to discover complex correlations.
    The first part of the review gives an introduction to the principles of big-data science. We show how to select appropriate training sets, survey approaches that are used to represent these materials in feature space, review different learning architectures,  as well as evaluation and interpretation strategies.
    In the second part, we review how the different approaches of machine learning have been applied to porous materials. In particular, we discuss applications in the field of gas storage and separation, the stability of these materials, their electronic properties, and their synthesis.
    Given the increasing interest of the scientific community in machine learning, we expect this list to rapidly expand in the coming years.
\end{abstract}

\begin{tocentry}
    \begin{center}
        \includegraphics[width=5cm]{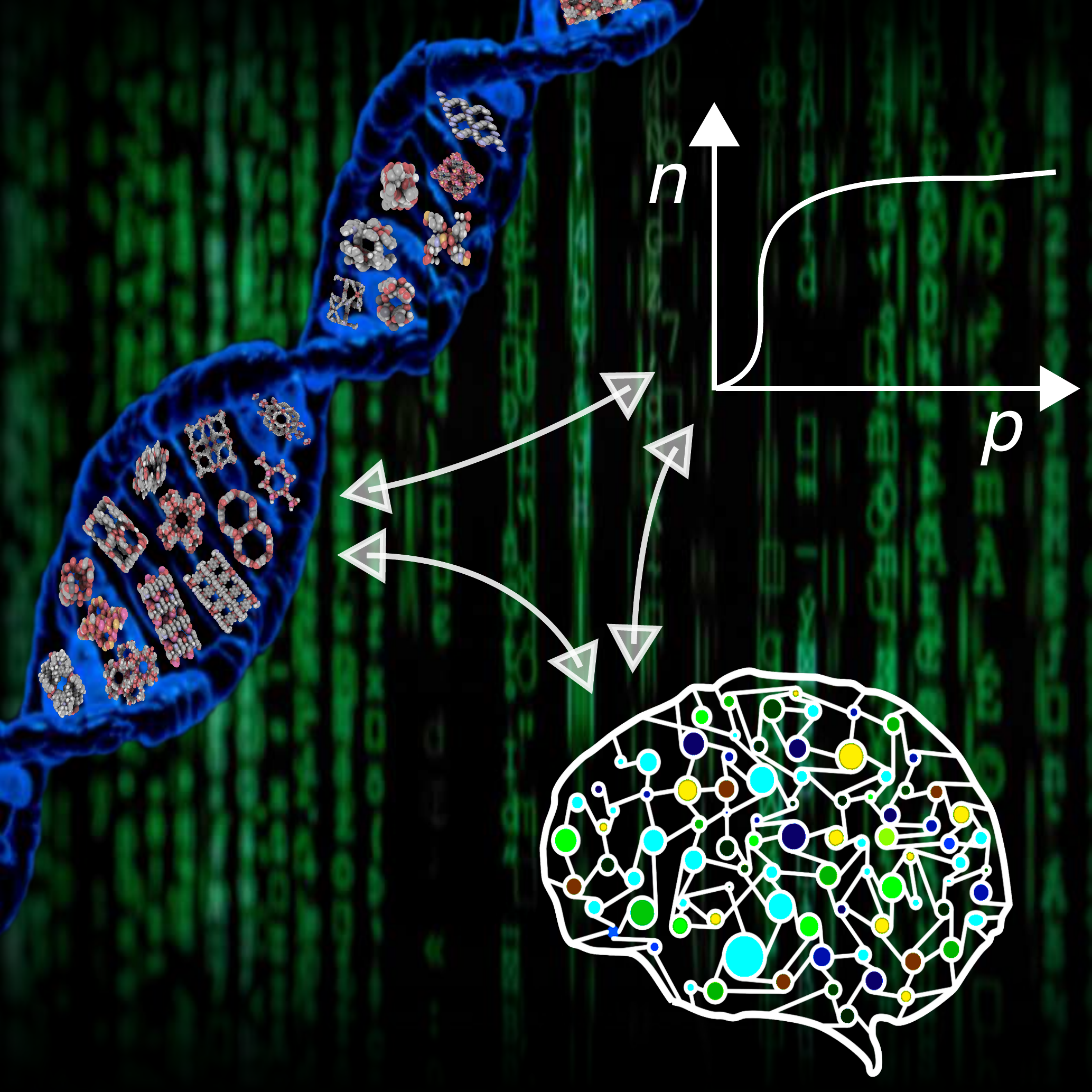}
    \end{center}
\end{tocentry}

\clearpage
\tableofcontents
\clearpage

\input{main/0_introduction.tex}
\input{main/1_data_selection.tex}
\input{main/2_featurization.tex}
\input{main/3_learning.tex}
\input{main/4_learn_well.tex}
\input{main/5_metrics.tex}
\input{main/6_interpretation.tex}
\input{main/9_applications_to_mof.tex}
\input{main/7_outlook.tex}
\input{main/8_summary_conclusion.tex}
\clearpage
\printglossaries
\clearpage
\bibliography{cleaned_references}

\clearpage
\input{main/biographies.tex}

\end{document}

%% file: acronyms.tex
\newacronym{aa}{AA}{archetypal analysis}
\newacronym{adasyn}{ADASYN}{adaptive synthetic oversampling}
\newacronym{ap}{AP}{atomic property}
\newacronym{auc}{AUC}{area under the curve}
\newacronym{baml}{BAML}{bond-angles machine learning}
\newacronym{bet}{BET}{Brunauer–Emmett–Teller}
\newacronym{bob}{BoB}{bag of bonds}
\newacronym{cadd}{CADD}{computer aided drug design}
\newacronym{cart}{CART}{classification and regression tree}
\newacronym{cas}{CAS}{chemical abstract services}
\newacronym{ccsdt}{CCSD(T)}{Coupled Cluster Single-double with pertubative triple excitations}
\newacronym{ccs}{CCS}{carbon capture and storage}
\newacronym{cnn}{CNN}{convolutional neural network}
\newacronym{cof}{COF}{covalent organic framework}
\newacronym{core}{CoRE}{computationally ready experimental}
\newacronym{crisp-dm}{CRISP-DM}{cross-industry standard process for data mining}
\newacronym{csd}{CSD}{Cambridge Structure Database}
\newacronym{dfa}{DFA}{density-functional approximation}
\newacronym{dftb}{DFTB}{density functional based tight-binding}
\newacronym{dft}{DFT}{density-functional theory}
\newacronym{dl}{DL}{deep learning}
\newacronym{dnn}{DNN}{deep neural networks}
\newacronym{doe}{DoE}{design of experiment}
\newacronym{dos}{DOS}{density of states}
\newacronym{dtnn}{DTNN}{deep tensor neural network}
\newacronym{dt}{DT}{decision tree}
\newacronym{eda}{EDA}{exploratory data analysis}
\newacronym{fair}{FAIR}{findable, accessible, interoperable, reusable}
\newacronym{ff}{FF}{force field}
\newacronym{fps}{FPS}{farthest point sampling}
\newacronym{gam}{GAM}{generalized additive model}
\newacronym{gan}{GAN}{generative adverserial network}
\newacronym{gap}{GAP}{Gaussian approximation potential}
\newacronym{ga}{GA}{genetic algorithm}
\newacronym{gcnn}{GCNN}{graph-convolutional \gls{nn}}
\newacronym{gbdt}{GBDT}{gradient boosted decision trees}
\newacronym{gcmc}{GCMC}{grand-canonical Monte Carlo}
\newacronym{gga}{GGA}{general-gradient approximation}
\newacronym{gpr}{GPR}{Gaussian process regression}
\newacronym{gp}{GP}{Gaussian process}
\newacronym{hdnpp}{HDNNP}{high-dimensional neural network potential}
\newacronym{hf}{HF}{Hartree-Fock}
\newacronym{hip}{HIP}{hierarchically interacting particle}
\newacronym{iid}{i.i.d}{independently and identically distributed}
\newacronym{kmm}{KMM}{kernel-mean matching}
\newacronym{knn}{\ensuremath{k}NN}{\(k\) nearest neighbor}
\newacronym{krr}{KRR}{kernel ridge regression}
\newacronym{lammps}{LAMMPS}{large-scale atomic/molecular massively parallel simulator}
\newacronym{lasso}{Lasso}{least absolute shrinkage and selection operator}
\newacronym{lhs}{LHS}{latin hypercube sampling}
\newacronym{lococv}{lococv}{leave-on-cluster-out cross-validation}
\newacronym{loob}{LOOB}{leave-one-out bootstrap}
\newacronym{loocv}{LOOCV}{leave-one-out cross validation}
\newacronym{mae}{MAE}{mean absolute error}
\newacronym{mbtr}{MBTR}{many-body tensor representation}
\newacronym{mc}{MC}{Monte-Carlo}
\newacronym{mdp}{MDP}{maximum diversity problem}
\newacronym{md}{MD}{molecular dynamics}
\newacronym{mlp}{MLP}{multilayer perceptron}
\newacronym{ml}{ML}{machine learning}
\newacronym{mof}{MOF}{metal-organic framework}
\newacronym{monc}{MONC}{metal-organic nanocapsules}
\newacronym{mse}{MSE}{mean squared error}
\newacronym{nlp}{NLP}{natural language processing}
\newacronym{nn}{NN}{neural network}
\newacronym{np}{NP}{non-deterministic polynomial-time}
\newacronym{obd}{OBD}{optimal brain damage}
\newacronym{oms}{OMS}{open metal site}
\newacronym{oob}{obb}{out-of-bag}
\newacronym{oqmd}{OQMD}{open quantum materials database}
\newacronym{osda}{OSDA}{organic structure directing agent}
\newacronym{pca}{PCA}{principal component analysis}
\newacronym{pes}{PES}{potential energy surface}
\newacronym{plmf}{PLMF}{property labeled materials fragments}
\newacronym{ppn}{PPN}{porous polymer network}
\newacronym{psd}{PSD}{pore size distribution}
\newacronym{qsar}{QSAR}{quantitative structure activity relationship}
\newacronym{qspr}{QSPR}{quantitative structure property relationship}
\newacronym{rac}{RAC}{revised autocorrelation}
\newacronym{rdf}{RDF}{radial distribution function}
\newacronym{reach}{REACH}{registration evaluation and authorization of chemicals}
\newacronym{relu}{ReLU}{rectified linear unit}
\newacronym{rfa}{RFA}{recursive feature addition}
\newacronym{rfe}{RFE}{recursive feature elimination}
\newacronym{rf}{RF}{random forest}
\newacronym{rmse}{RMSE}{root \gls{mse}}
\newacronym{rnn}{RNN}{recurrent neural network}
\newacronym{roc}{ROC}{receiver-operating characteristic}
\newacronym{rpa}{RPA}{random phase approximation}
\newacronym{scan}{SCAN}{strongly constrained and appropriately normed}
\newacronym{sgd}{SGD}{stochastic gradient descent}
\newacronym{shap}{SHAP}{SHapley Additive exPlanations}
\newacronym{sisso}{SISSO}{sure independence screening and sparsifying operator}
\newacronym{si}{si}{sure independence}
\newacronym{smbo}{SMBO}{sequential model-based optimization}
\newacronym{smiles}{SMILES}{simplified molecular input line entry system}
\newacronym{smote}{SMOTE}{synthetic minority oversampling technique}
\newacronym{soap}{SOAP}{smooth overlap of atomic positions}
\newacronym{svc}{SVC}{support vector classifier}
\newacronym{svm}{SVM}{support vector machine}
\newacronym{tda}{TDA}{topological data analysis}
\newacronym{tpe}{TPE}{tree-Parzen estimator}
\newacronym{tsne}{t-SNE}{t-distributed stochastic neighbor embedding}
\newacronym{vae}{VAE}{variational autoencoders}
\newacronym{vif}{VIF}{variance inflation factor}
\newacronym{xrd}{XRD}{x-ray diffraction}
\newacronym{xrpd}{XRPD}{x-ray powder diffraction}
\newacronym{zif}{ZIF}{zeolitic imidazolate framework}

%% file: main/0_introduction.tex
\section{Introduction}\label{sec:introduction}
One of the fascinating aspects of \glspl{mof} is that by combining linkers and metal nodes we can synthesize millions of different materials.\cite{furukawa_chemistry_2013}
Over the last decade, over 10,000 porous\cite{chung_computation-ready_2014, chung_advances_2019} and 80,000 non-porous \glspl{mof} have been synthesized.\cite{moghadam_development_2017}
In addition, one also has \glspl{cof}, \glspl{ppn}, zeolites, and related porous materials.
Because of their potential in many applications, ranging from gas separation and storage, sensing, catalysis, etc.\ these materials have attracted a lot of attention.
From a scientific point of view, these materials are interesting as their chemical tunability allows us to tailor-make materials with exactly the right properties.
As one can only synthesize a tiny fraction of all possible materials, these experimental efforts are often combined with computational approaches, often referred to as materials genomics,\cite{boyd_computational_2017} to generate libraries of predicted or hypothetical \glspl{mof}, \glspl{cof}, and other related porous materials.
These libraries are subsequently computationally screened to identify the most promising material for a given application.

That we now have of the order of ten thousand synthesized porous crystals and over a hundred thousand predicted materials does create new challenges; we simply have too many structures and too much data.
Issues related to having so many structures can be simple questions on how to manage so much data, but also more profound on how to use the data to discover new science. Therefore, a logical next step in materials genomics is to apply the tools of big-data science and to exploit \enquote{the unreasonable effectiveness of data}.\cite{halevy_unreasonable_2009}
In this review, we discuss how \gls{ml} has been applied to porous materials and review some aspects of the underlying techniques in each step.
Before discussing the specific applications of \gls{ml} to porous materials, we give an overview over the \gls{ml} landscape to introduce some terminologies, and also give a short overview over the technical terms we will use throughout this review in Table~\ref{tab:terminology}.

In this review, we focus on applications of \gls{ml} in materials science and chemistry with a particular focus on porous materials. For more general discussion on \gls{ml}, we refer the reader to some excellent reviews.\cite{mehta_high-bias_2019, butler_machine_2018}

\input{main/terminology.tex}

\section{The Machine Learning Landscape}\label{sec:introductionML}
Nowadays it is difficult, if not impossible, to avoid \gls{ml} in science. Because of recent developments in technology, we now routinely store and analyze large amounts of data.
The underlying idea of big-data science is that if one has large amounts of data, one might be able to discover statistically significant patterns that are correlated to some specific properties or events.
Arthur Samuel was among the first to use the term \enquote{machine learning} for the algorithms he developed in 1959 to teach a computer to play the game of checkers. His \gls{ml} algorithm let the computer look ahead a few moves.\cite{samuel_studies_2000}
Initially, each possible move had the same weight, and hence probability of being executed. By collecting more and more data from actual games, the computer could learn which move for a given board configuration would develop a winning strategy. One of the reasons why Arthur Samuel looked at checkers was that in the practical sense the game of checkers is not deterministic; there is no known algorithm that leads to winning the game and the complete evaluation of all \(10^{40}\) possible moves is beyond the capacity of any computer.

\begin{figure}
	\centering
	\includegraphics[width=\textwidth]{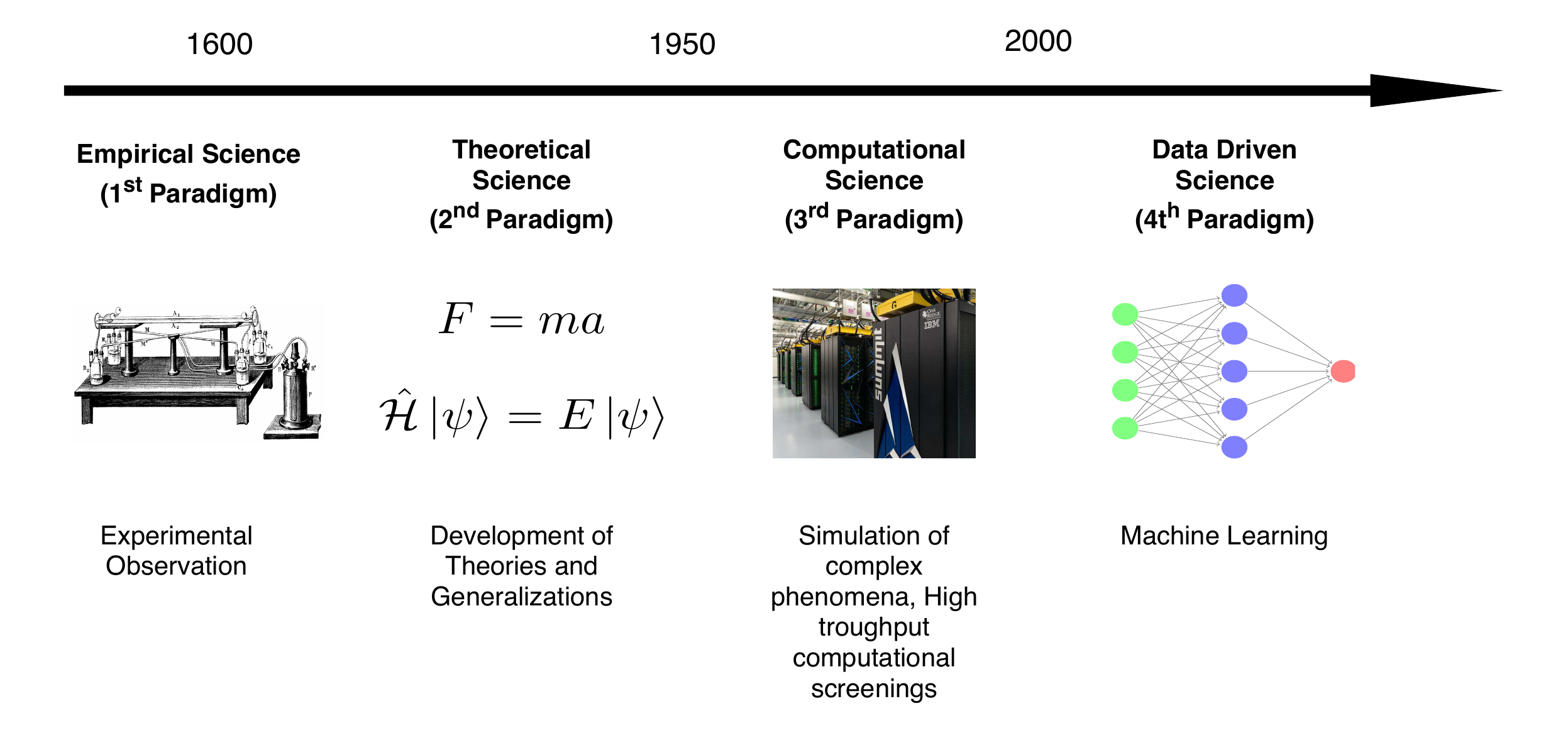}
	\caption{Different approaches to science that evolved over time, starting from empirical observation, generalizations to theories and simulation of different, complex, phenomena. The latest addition is the data-driven discovery (\enquote{fourth paradigm of science}). The supercomputer image was taken from the Oak Ridge National Laboratory.}\label{fig:forth_paradigm}
\end{figure}

There are some similarities between the game checkers and the science of discovering new materials.
Making a new material is in practice equally non-deterministic. The number of possible ways we can combine atoms is simply too large to evaluate all possible materials.
For a long time, materials discovery has been based on empirical knowledge.
Significant advances were made, once some of this empirical knowledge was generalized in the form of theoretical frameworks.
Combined with supercomputers these theoretical frameworks resulted in accurate predictions of the properties of materials.
Yet, the number of atoms and possible materials is simply too large to predict all properties of all possible materials.
Hence, there will be large parts of our material space that are, in practical terms, out of reach of the conventional paradigms of science.
Some phenomena are simply too complex to be explicitly described with theory.
Teaching the computer the concepts using big data might be an interesting route to study some of these problems. The emergence of off-the-shelf machine learning methods that can be used by domain experts\cite{hutson_bringing_2019}---not only specialized data scientists---in combination with big data is thought to spark the \enquote{fourth industrial revolution} and the \enquote{fourth paradigm of science} (cf.\ Figure~\ref{fig:forth_paradigm}).\cite{gray_escience-transformed_2007, hey_fourth_2009}
In this context, big data can add a new dimension to material discovery. One needs to realize that even though \gls{ml} might appear as \enquote{black box} engineering in some instances, good predictions from a black box are indefinitely better than no prediction at all.
This is to some extent similar to an engineer that can make things work without understanding all the underlying physics. And, as we will discuss below, there are many techniques to investigate the reliability and domain of applicability of a \gls{ml} model as well as techniques that can help in understanding the predictions made by the model.

Material science and chemistry may not be the most obvious topics for big-data science.
Experiments are labor-intensive and the amount of data about materials that have been collected in the last centuries is minute compared to what Google and the likes collect every single second. 
However, recently the field of materials genomics has changed the landscape.\cite{boyd_data_2019}
High-throughput \gls{dft} calculations\cite{curtarolo_high-throughput_2013} and molecular simulations\cite{lin_silico_2012} have become routine tools to study the properties of real and even hypothetical materials.
In these studies, \gls{ml} is becoming more popular and widely used as a filter in the computational funnel of high-throughput screenings.\cite{pyzer-knapp_learning_2015}
But also to assist and guide simulations\cite{curtarolo_predicting_2003, collins_materials_2016, duan_learning_2019, heinen_machine_2019} or experiments,\cite{moosavi_capturing_2019} or to even replace them,\cite{aspuru-guzik_matter_2018, de_luna_use_2017} and to design new high-performing materials.\cite{sanchez-lengeling_inverse_2018}

Another important factor is the prominent role patterns play in chemistry.
The most famous example is Mendeleev's periodic table, but also Pauling's rules,\cite{pauling_principles_1929} Pettifor's maps,\cite{pettifor_bonding_1995} and many other structure-property relationships were guided by a combination of empirical knowledge and chemical intuition.
What we hope to show in this review is that \gls{ml} holds the promise to discover much more complex relationships from (big) data.

We continue this section with a broad overview of the main principles of \gls{ml}.
This section will be followed with a more detailed and technical discussion on the different subtopics introduced in this section.

\subsection{The Machine Learning Pipeline}
\subsubsection{Machine Learning Workflow}
\Gls{ml} is no different from any other method in science.
There are questions for which \gls{ml} is an extremely powerful method to find an answer, but if one sees \gls{ml} as the modern solution to any ill-posed problem one is bound to be disappointed. In section~\ref{sec:applications}, we will discuss the type of questions that have been successfully addressed using \gls{ml} in the contexts of synthesis and applications of porous materials.

Independent of the learning algorithm or goal, the \gls{ml} workflow from materials' data to prediction and interpretation can be divided into the following blueprint of a workflow, which also this review follows:
\begin{enumerate}
	\item \textbf{Understanding the problem:} An understanding of the phenomena that need to be described is important. For example, if we are interested in methane storage in porous media, the key performance parameter is the deliverable capacity, which can be obtained directly for the experimental adsorption isotherms at a given temperature.
	      In more general terms, an understanding of the phenomena helps us to guide the generation and transformation of the data (discussed in more detail in the next step).

	      In case of the deliverable capacity we have a continuous variable and hence a regression problem, which can be more difficult to learn compared to classification problems (e.g., whether the channels in our porous material form a 1, 2 or 3-dimensional network or classifying the deliverable capacity as \enquote{high} or \enquote{low}).

	      Importantly, the problem definition guides the choice of the strategies for model evaluation, selection, and interpretation (cf.\ section~\ref{sec:metrics}): In some classification cases, such as in a part of the high-throughput funnel, in which we are interested in finding the top-performing materials by down selecting materials, missing the highest-performing material is worse than doing an additional simulation for a mediocre material---this is something one should realize before building the model.

	\item \textbf{Generating and exploring data:} Machine learning needs data to learn from.
	      In particular, one needs to ensure that we have suitable training data. Suitable, in the sense that the data are reliable and provide sufficient coverage of the design space we would like to explore.
	      Sometimes, suitable training data must be generated or augmented. The process of exploring a suitable data set (known as \gls{eda}\cite{tukey_exploratory_1977})  and its subsequent featurization can help to understand the problem better and inform the modeling process.

	      Once we have collected a data set, the next steps involve:
	      \begin{enumerate}
		      \item \emph{Data selection:} If the goal is to predict materials properties, which is the focus of this review, it is crucial to ensure that the available labels \(\mathbf{y}\), i.e., the targets we want to predict, are consistent, and special care has to be taken when data from different sources are used. We discuss this step in more detail in section~\ref{sec:datasource} and the outlook.

		      \item \emph{Featurization} is the process in which the structures or raw data are mapped into feature (or design) matrices \(\mathbf{X}\), where one row in this matrix characterizes one material.
		            Domain knowledge in the context of the problem we are addressing can be particularly useful in this step. For example, to select the relevant length scales (atomistic, coarse-grained, or global) or properties (electronic, geometric, or involved experimental properties). We give an overview of this process in section~\ref{sec:featurization}.

		      \item \emph{Sampling:} Often, training data are randomly selected from a large database of training points. But this is not necessarily the best choice as most likely the materials are not uniformly distributed for all possible labels we are potentially interested in. For example, one class (often the low-performing structures) might constitute the majority of the training set and the algorithm will have problems in making predictions for the minority class (which are often the most interesting cases). Special methods, e.g., \gls{fps}, have been developed to sample the design space more uniformly. In section~\ref{sec:sampling} we discuss ways to mitigate this problem and approaches to deal with little data.
	      \end{enumerate}

	\item \textbf{Learning and Prediction:} In section~\ref{sec:learning_algorithms} we examine several ways in which one can learn from data, and what one should consider when choosing a particular algorithm.
	      We then describe different methods with which one can improve predictive performance and avoid overfitting (cf.\ section~\ref{sec:learning_well}). \\
	      To guide the modeling and model selection, methods for performance evaluation are needed.  In section~\ref{sec:metrics}  we describe best practices for model evaluation and comparison.

	\item \textbf{Interpretation:} Often it is interesting to understand what and how the model learned---e.g., to better grasp structure-property relationships or to debug \gls{ml} models.
	      \Gls{ml} is often seen as a black-box approach to predict numerical values with zero understanding---defeating the goal of science to understand and explain phenomena.
	      Therefore, the need for causal models is seen as a step towards machines \enquote{that learn and think like people} (learning as model building instead of mere pattern recognition).\cite{lake_building_2016}
	      In section~\ref{sec:interpretatbility} we present different approaches to look into black-box models, or how to avoid them in the first place.
\end{enumerate}
It is important to remember that model development is an iterative process;  the understanding gained from the first model evaluations can help to understand the model better and help in refining the data, the featurization and the model architecture. For this, interpretable models can be particularly valuable.\cite{rudin_stop_2019}

The scope of this review is to provide guidance along this path and to highlight the caveats, but also to point to more detailed resources and useful Python packages that can be used to implement a specific step.

An excellent general overview that digs deeper into the mathematical background than this review is the \enquote{high-bias, low variance introduction to Machine Learning} by Mehta et al.\cite{mehta_high-bias_2019}, recent applications of \gls{ml} to materials science are covered by Schmidt et al.\cite{schmidt_recent_2019}
But also many textbooks cover the fundamentals of machine learning; e.g., Tibshirani and Friedman\cite{tibshirani_elements_2017}, Shalev-Shwartz and Ben-David\cite{shalev-shwartz_understanding_2014} as well as Bishop (from a more Bayesian point of view)\cite{bishop_pattern_2006} focus more on the theoretical background of statistical learning, whereas Géron provides a \enquote{how-to} for the actual implementation, also of \gls{nn} architectures, using popular Python frameworks,\cite{geron_hands-machine_2019} which were recently reviewed by Rascka et al.\cite{raschka_machine_2020}

\subsubsection{Machine Learning Algorithms}
Step three of the workflow described in the previous section, learning and predictions,  usually receives the most attention.
Broadly, there are three classes, though with fuzzy boundaries, for this step, namely supervised, unsupervised and reinforcement learning.
We will focus only on supervised learning in this review, and only briefly describe possible applications of the other categories and highlight good starting points to help the reader orient in the field.

\paragraph{Supervised Learning: Feature Matrix and Labels are Given}
The most widely used flavor, which is also the focus of this review, is supervised learning.
Here, one has access to features that describe a material and the corresponding labels (the property one wants to predict).

A common use case is to completely replace expensive calculations with the calculation of features that can be then fed into a model to make a prediction.
A different use case can be to still perform molecular simulations---but to use \gls{ml} to generate better \gls{pes}, e.g., using \enquote{machine learned} force fields.
Another promising avenue is \(\Delta\)-ML in which a model is trained to predict a correction to a coarser level of theory:\cite{ramakrishnan_big_2015} One example would be to predict the correction to \gls{dft} energies to predict coupled-cluster energies.

Supervised learning can also be used as part of an active learning loop for self-driving laboratories and to efficiently optimize reaction conditions.
In this review, we do not focus on this aspect---good starting points are reports from the groups around Alán Aspuru-Guzik\cite{hase_next-generation_2019, macleod_self-driving_2019, hase_chimera_2018, tabor_accelerating_2018} and Lee Cronin.\cite{gromski_how_2019, gromski_universal_2019, keenan_nanomaterials_2019, dragone_autonomous_2017}

\paragraph{Unsupervised Learning: Using Only the Feature Matrix}
Unsupervised learning differs from supervised learning in the sense that it only uses the feature matrix and not the labels (which are often unknown when unsupervised learning is used).
Unsupervised learning can help to find patterns in the data, which in turn might provide chemical insight.

\begin{figure}
	\centering
	\includegraphics[width=.4\textwidth]{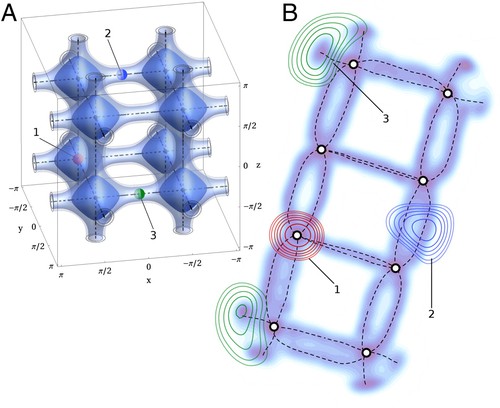}
	\caption{A, Three-dimensional energy landscape and B, its two-dimensional projection using sketchmap, which is a dimensionality reduction technique. The biasing potentials can now be represented in terms of sketchmap coordinates. Figure reproduced from \protect{\cite{tribello_using_2012}}, Copyright (2012) National Academy of Sciences.}\label{fig:dimensional}
\end{figure}

\subparagraph{Dimensionality Reduction and Clustering}
The importance of unsupervised methods becomes clear when dealing with high-dimensional data which are notoriously difficult to visualize and understand (cf.\ section~\ref{sec:curse_dimensionality}).
And in fact some of the earliest applications of these techniques were to analyze\cite{gasparotto_recognizing_2018, das_low-dimensional_2006, xie_graph_2019} and then speed up molecular simulations.\cite{tribello_using_2012, hashemian_modeling_2013}
The challenge with molecular simulations is that we explore a \(3N\) dimensional space, where \(N\) is the number of particles. For large \(N\), as it is, for example, the case for the simulation of protein dynamics, it can be hard to identify low energy states.\cite{tribello_using_2012}
To accelerate the sampling, one can apply biasing potentials that help the simulation to move over barriers between metastable states.
Typically, such potentials are constructed in terms of a small number of variables, known as collective variables---but it can be a challenge to identify what a good choice of the collective variables is when the dimensionality of the system is high.
In this context, \gls{ml} has been employed to lower the dimensionality of the system (cf.\ Figure~\ref{fig:dimensional} for an example of such a dimensionality reduction) and to express the collective variables in this low-dimensional space.

Dimensionality reduction techniques, like \gls{pca}, ISOMAP, \gls{tsne}, self-organizing maps,\cite{kohonen_exploration_1997, beckonert_visualizing_2003} growing cell structures,\cite{fritzke_growing_1994} or sketchmap,\cite{ceriotti_demonstrating_2013, de_comparing_2016} can be used to do so.\cite{tribello_using_2012}
But they can also be used for \enquote{materials cartography},\cite{isayev_materials_2015} i.e., to present the high-dimensional space of material properties in two dimensions to help identify patterns in big and high-dimensional data.\cite{kunkel_knowledge_2019}
A book chapter Samudrala et al.\cite{samudrala_data_2013} and a perspective by Ceriotti\cite{ceriotti_unsupervised_2019} give an overview of applications in materials science.

Recently, unsupervised learning---in the form of word-embeddings, which are vectors in the multidimensional \enquote{vocabulary space} that are usually used for \gls{nlp}---has also been used to discover chemistry in form of structure-property relationships in chemical literature. This technique could also be used to make recommendations based on the distance of a word-embedding of a compound, to the vector of a concept such as thermoelectricity in the word-embedding space.\cite{tshitoyan_unsupervised_2019}

\subparagraph{Generative Models}\label{sec:gen_intro}
One ultimate goal of \gls{ml} is to design new materials (which recently has been also popularized as \enquote{inverse design}). Generative models, like \glspl{gan} or \glspl{vae} hold the promise to do this.\cite{sanchez-lengeling_machine_2019}
\Glspl{gan} and \glspl{vae} can create new molecules,\cite{gomez-bombarelli_automatic_2018} or probability distributions,\cite{noe_boltzmann_2019} with the desired properties on the computer.\cite{collins_materials_2016}
One example for the success of generative techniques (in combination with reinforcement learning) is the discovery of inhibitors for a kinase target implicated in fibrosis, that were discovered in 21 days on the computer and also showed promising results in experiments.\cite{zhavoronkov_deep_2019}
An excellent outline of the promises of generative models and their use for the design of new compounds is given by Sanchez\cite{sanchez-lengeling_inverse_2018} and Elton.\cite{elton_deep_2019}

The interface between unsupervised and supervised learning is known as \textbf{semi-supervised learning}. In this setting, only some labels are known, which is often the case when labeling is expensive.
This was also the case in a recent study of the group around Ceder,\cite{huo_semi-supervised_2019} where they attempted to classify synthesis descriptions in papers according to different categories like hydrothermal or solid-state synthesis.
The initial labeling for a small subset was performed manually, but they could then use semi-supervised techniques to leverage the full datasets, i.e., also the unlabeled parts.

\paragraph{Reinforcement Learning: Agents Maximizing Rewards}
\begin{figure}
	\centering
	\includegraphics[width=.7\textwidth]{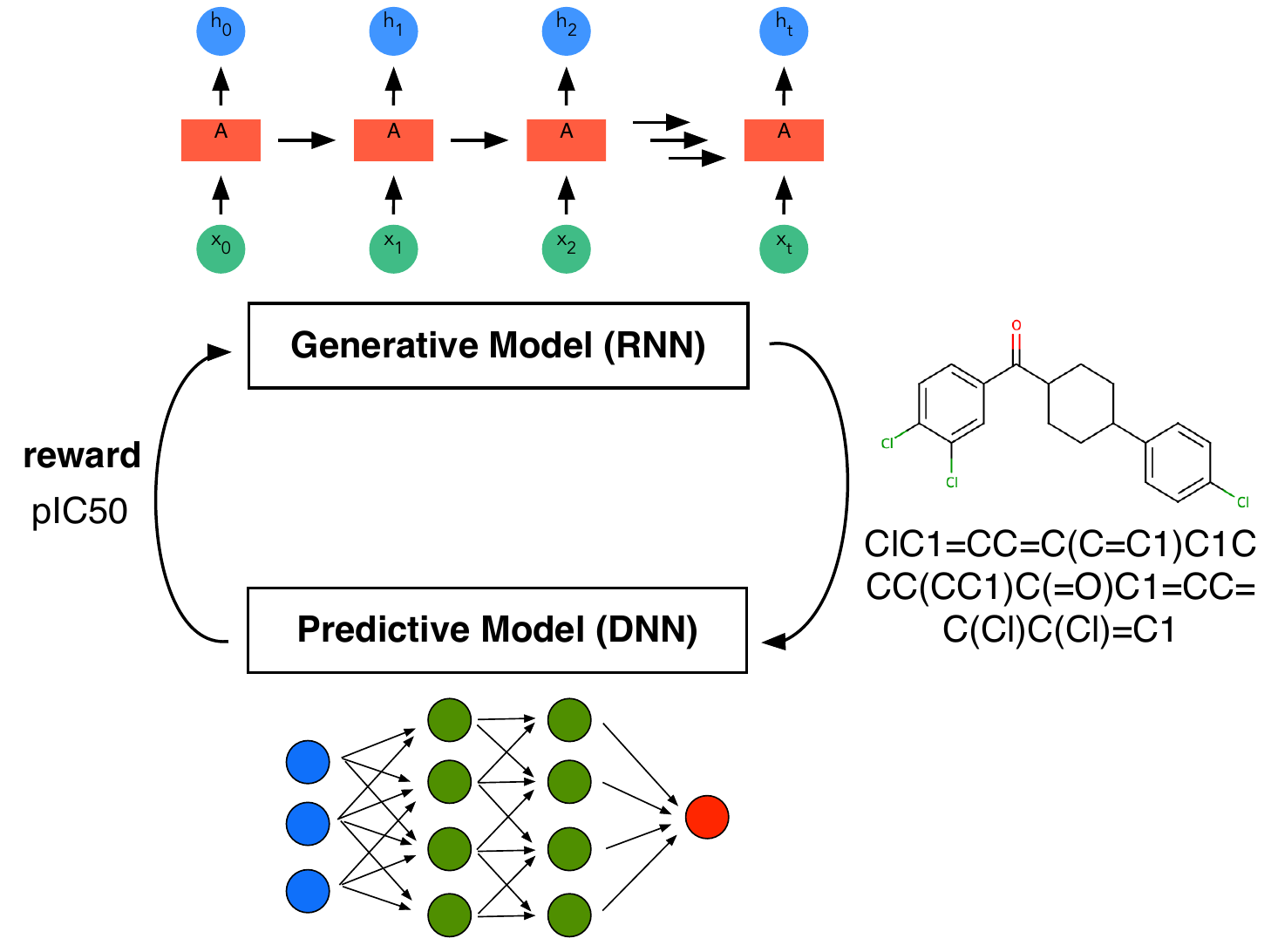}
	\caption{Reinforcement learning scheme illustrated based on the approach chosen by Popova et al.\protect{\cite{popova_deep_2018}} for drug design. They use a \gls{rnn} (cf.\ section~\ref{sec:rnn}) for the generation of \gls{smiles} strings and a deep \gls{nn} for property prediction.
		In a first stage, both models are trained separately, and then they are used jointly to bias, using the target properties as the reward, the generation of new molecules. This example also nicely shows that the boundary between the different \enquote{flavors} of \gls{ml} is fuzzy and that they are often used together.}\label{fig:reinforcement_scheme}
\end{figure}
In reinforcement learning\cite{sutton_reinforcement_2018} agents try to figure out the optimal sequence of actions (which is known as policy) in some environment to maximize a reward.
An interesting application of this sub-field of \gls{ml} in chemistry is to find the optimal reaction conditions to maximize the yield or to create structures with desired properties (cf.\ Figure~\ref{fig:reinforcement_scheme}).\cite{zhou_optimizing_2017, popova_deep_2018}
Reinforcement learning has also been in the news for the superhuman performance achieved on some video games.\cite{mnih_playing_2013, silver_mastering_2016} Still, it tends to require a lot of training. AlphaGo Zero, for example, needed nearly 5 million matches, requiring millions of dollars of investment in hardware and computational time.\cite{carey_interpreting_2018}

\subsection{Theory-Guided Data Science}
We are at an age in which some argue that \enquote{the end of theory} is near,\cite{anderson_end_2008} but throughout this review we will find that many successful \gls{ml} models are guided by physics and physical insights.\cite{andreoni_machine_2018, childs_embedding_2019, maier_learning_2019}
We will see that the symmetry of the systems guides the design of the descriptors and can guide the design of the models (e.g., by decomposing the problems into sub-problems) or the choice of constraints.
Sometimes, we will also encounter hybrid approaches where one component of the problem (often the local part, as locality is often an assumption for the \gls{ml} models, cf.\ section~\ref{sec:locality}) is solved using \gls{ml} and that for example the electrostatic, long-range interaction, is added using well-known theory.

Generally, the decomposition of the problem can help to debug the model, and make the model more interpretable and physical.\cite{lake_human-level_2015}
For example, physics-guided breakdown of the target proved to be useful in the creation of a model for the equation of state of fluid methane.\cite{veit_equation_2019}

Physical insight can also be introduced using sparsity,\cite{constantine_many_2016} or physics-based functional forms.\cite{bereau_non-covalent_2018}
Constraints, introduced for example via Euler-Lagrange constrained minimization or coordinate scaling (stretching the coordinates should also stretch the density), have also proven to be successful in the development of \gls{ml} learned density functionals.\cite{li_understanding_2016, hollingsworth_can_2018}

That physical insight can guide model development has been shown by Chmiele et al., who built a model of potential energy surfaces using forces instead of energies to respect energy conservation (also, the force is a quantity that is well-defined for atoms, whereas the energy is only defined for the full system).\cite{chmiela_machine_2017, huan_universal_2017}

This paradigm of incorporating domain knowledge into the \gls{ml} workflow is also known as theory-guided data science.\cite{karpatne_theory-guided_2017, wagner_theory-guided_2016}
Theory-guided data science can help to get the right answers for the right reasons, and we will revisit it in every chapter of this review.

\subsection{The Scientific Method in Machine Learning: Strong Inference and Multiple Models}
Throughout this review we will encounter the method of strong inference,\cite{platt_strong_1964, chamberlin_method_1965} i.e., the need for alternative hypotheses, or more generally the integral role of critical thinking, at different places---mostly in the later stages of the \gls{ml} pipeline when one analyzes a model.
The idea here is to always pursue multiple alternative hypotheses that could explain the performance of a model: Is the improved performance really because of a more complex architecture or rather due to better hyperparameter optimization (cf.\ ablation testing in section~\ref{sec:ablation}) or does the model really learn sensible chemical relationships or could we achieve similar performance with random labels (cf.\ randomization tests as discussed in section~\ref{sec:randomization_tests}\cite{van_gunsteren_seven_2013, chuang_adversarial_2018})?

\Gls{ml} comes with many opportunities, but also many pitfalls.
In the following, we review the details of the supervised \gls{ml} workflow to aid the use of \gls{ml} for the progress of our field.

%% file: main/terminology.tex
\begin{longtable}{lp{22em}}
    \caption{Common technical terms used in \gls{ml} and their meaning.}\label{tab:terminology}                                                                                                                                              \\
    \toprule
    technical term                     & explanation                                                                                                                                                                                         \\
    \midrule
    \endhead
    \midrule
    \multicolumn{2}{r}{{Continued on next page}}                                                                                                                                                                                             \\
    \midrule
    \endfoot
    \bottomrule
    \endlastfoot
    bagging                            & acronym for bootstrap aggregating, ensemble technique in which models are fitted on bootstrapped samples from the data and then averaged                                                            \\
    bias                               & error that remains for infinite number of training examples, e.g., due to limited expressivity                                                                                                      \\
    boosting                           & ensemble technique in which weak learners are iteratively combined to build a stronger learner                                                                                                      \\
    bootstrapping                      & calculate statistics by randomly drawing samples with replacement                                                                                                                                   \\
    classification                     & process of assigning examples to a particular class                                                                                                                                                 \\
    confidence interval                & interval of confidence around  predicted mean response                                                                                                                                              \\
    feature                            & vector with numeric encoding of a description of a material that the \gls{ml} uses for learning                                                                                                     \\
    fidelity                           & measure of how close a model represents the real case                                                                                                                                               \\
    fitting                            & estimating parameters of some models with high accuracy                                                                                                                                             \\
    gradient descent                   & optimization by following the gradient, stochastic gradient descent approximates the gradient using a mini-batch of the available data                                                              \\
    hyperparameters                    & are tuning parameters of the learner (like learning rate, regularization strength) which, in contrast to model parameters, are not learned during training and have to be specified before training \\
    instance based learning            & learning by heart, query data are compared to training examples to make a prediction                                                                                                                \\
    irreducible error                  & error that cannot be reduced (e.g., due to noise in the data), i.e., that is also there for a perfect model. Also known as Bayes error rate                                                         \\
    label (target)                     & the property one wants to predict                                                                                                                                                                   \\
    objective function (cost function) & the function that a \gls{ml} algorithm tries to minimize                                                                                                                                            \\
    one-hot encoding                   & method to represent categorical variables by creating a feature column for each category and using value of one to encode the presence and zero to encode the absence.                              \\
    overfitting                        & the gap between training and test error is large, i.e., the model solely \enquote{remembers} the training data but fails to predict on unseen examples                                              \\
    predicting                         & making predictions for future samples with high accuracy                                                                                                                                            \\
    prediction interval                & interval of confidence around predicted sample response, always wider than confidence interval                                                                                                      \\
    regression                         & process of estimating the continuous relationship between a dependent variable and one or more independent variables                                                                                \\
    regularization                     & describes techniques that add terms or information to the model to avoid overfitting                                                                                                                \\
    stratification                     & data is divided in homogeneous subgroups (strata) such that sampling will not disturb the class distributions                                                                                       \\
    structured data                    & data that is organized in tables with rows and columns, i.e., data that resides in relational databases                                                                                             \\
    test set                           & collection of labels and feature vectors that is used for model evaluation and which must not overlap with the training set                                                                         \\
    training set                       & collection of labels and feature vectors that is used for training                                                                                                                                  \\
    transfer                           & use knowledge gained on one distribution to perform inference on another distribution                                                                                                               \\
    unstructured data                  & e.g., image, video, audio, text. I.e., data that is not organized in a tabular form                                                                                                                 \\
    validation set                     & also known as development set, collection of labels and feature vectors that is used for hyperparameter tuning and which must not overlap with the test and training sets                           \\
    variance                           & part of the error that is due to finite-size effects (e.g., fluctuations due to random split in training and test set)                                                                              \\
\end{longtable}

%% file: main/1_data_selection.tex
\section{Selecting the Data: Dealing with Little, Imbalanced and Non-Representative Data}\label{sec:datasource}

The first, but most important step in \gls{ml} is to generate good training data.\cite{domingos_few_2012}
This is also captured in the \enquote{garbage in garbage out} saying among \gls{ml} practitioners.
Data matters more than algorithms.\cite{halevy_unreasonable_2009, banko_scaling_2001}
In this section, we will mostly focus on the rows of the feature matrix, \(\mathbf{X}\), and discuss the columns of it, the descriptors, in the next section.

\begin{figure}
    \centering
    \includegraphics[]{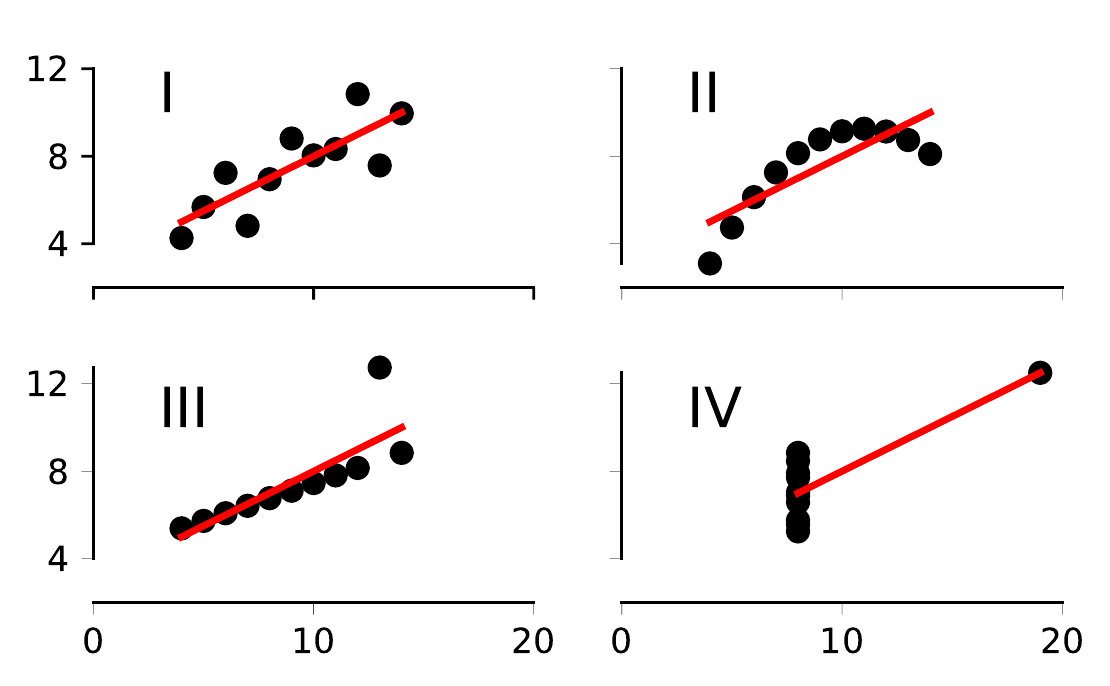}
    \caption{Anscombe's quartet shows the importance of visualization.\protect{\cite{anscombe_graphs_1973}} The four datasets have the same mean (7.50), standard deviation (1.94), and regression line, but still look completely different.}\label{fig:anscombe}
\end{figure}

That the selection of suitable data can be far from trivial is illustrated with Anscombe's quartet (cf.\ Figure~\ref{fig:anscombe}).\cite{anscombe_graphs_1973}
In this archetypal example four different distributions, with distinct graphs, have the same statistics, e.g., due to single high-leverage points.
This example emphasizes the notion in \gls{ml} that statistics can be deceiving, and why in \gls{ml} so much emphasis is placed on the visualization of the data sets.

\subsection{Limitations of Hypothetical Databases}\label{sec:hypothetical_databases}
Hypothetical databases of \glspl{cof}, \glspl{mof} and zeolites have become popular and are frequently used as a training set for \gls{ml} models---mostly because they are the largest self-consistent data sources that are available in this field.
But due to the way in which the databases are constructed they can only cover a limited part of the design space (as one uses a finite, small, numbers of linkers and nodes)---which is also not necessarily representative of the \enquote{real world}.

The problem of idealized models and hypothetical structures is even more pronounced for materials with unconventional electronic properties.
Many features that favor topological materials, which are materials with special shape of their electronic bands due to the symmetries of the atom positions, work against stability.
For example, creating a topological insulator (which is insulating in the bulk, but conductive on the surface) involves moving electrons into antibonding orbitals, which weakens the lattice.\cite{zunger_beware_2019}
Also, in the real world one often has to deal with defects and kinetic phenomena---real materials are often non-equilibrium structures\cite{zunger_beware_2019, olson_computational_1997}---while most databases assume ideal crystal structures.

\subsection{Sampling to Improve Predictive Performance}\label{sec:sampling}
A widespread technique in \gls{ml} is to randomly split the all available data into a training and a test set.
But this is not necessarily the best approach as random sampling might not sample some sparsely populated regions of the chemical space.
A more reasonable sampling approach would cover as much of the chemical space as feasible to construct a maximally informed training set.
This is especially important when one wants to minimize the number of training points. Limiting the number of training points can be reasonable or even essential when the featurization or labeling is expensive e.g.\ when it involves experiment or ab initio calculations.
But it can also be necessary for computational reasons as in the case of kernel methods (cf.\ section~\ref{sec:kernel_methods}), for which the data needs to be kept in memory and for which the computational cost scales cubically with the number of training points.

\subsubsection{Diverse Set Selection}\label{sec:diverse_set_seletion}
\paragraph{(Greedy) Farthest Point Sampling}\label{sec:fps}
Instead of randomly selecting training points, one can try to create a maximally diverse dataset to ensure a more uniform sampling of the design space and to avoid redundancy.
Creating such as dataset, in which the distances between the chosen data points are maximized, is known as the \gls{mdp}.\cite{ghosh_computational_1996}
Unfortunately, the \gls{mdp}, is of factorial computational cost, and hence becomes computationally prohibitive for large datasets.\cite{kennard_computer_1969, bartok_machine_2017,  dral_structure-based_2017}
Therefore, in practice, one usually uses a greedy algorithm to perform \gls{fps}.
Those algorithms add points for which the minimum distance to the already chosen points is maximal (i.e., using the max-min criterion, this sampling approach is also known as Kennard-Stone sampling, cf.\ pseudocode in Listing~\ref{lst:kennard_stone}).

This \gls{fps} is also a key to the work by Moosavi et al.\cite{moosavi_capturing_2019}, in which they use a diverse set of initial reaction conditions, most of which will yield to failed reactions, to build their model for reaction condition prediction.

\begin{lstlisting}[language=python, caption={Pseudocode for the greedy implementation of a \gls{fps} scheme. The initialization could also be to choose a point that is maximally distant from the center or using the two most separated points, as in the original Kennard-Stone framework.}, label=lst:kennard_stone]
# initialize by choosing a random point from the dataset p
Q =[]
Q.append(random.choice(p))

# perform the greedy search
while len(Q) < k: 
    # select the point with the maximal minimal distance
    new_point_index = argmax(min(d(p, Q))) 
    # add point to the selected subset
    Q.append(p[new_point_index])
    # remove point from the old set 
    p.remove(new_point_index) 
\end{lstlisting}

\paragraph{Design of Experiments}\label{sec:doe}
The efficient exploration is also the main goal of most \gls{doe} methods,\cite{montgomery_design_2020, fisher_arrangement_1992} which in chemistry have been widely used for reaction condition or process optimization,\cite{tye_use_2004, m.murray_application_2016, weissman_design_2015, delmonte_kilogram_2011} where the task is to understand the relationship between input variables (temperature, reaction time, \ldots) and the reaction outcome in the least time and effort possible.
But they also have been used in computer science to generate good initial guesses for computer codes.\cite{mckay_comparison_1979, park_optimal_1994}

If our goal is to perform reaction condition prediction, the use of \gls{doe} techniques can be a good starting point to get an initial training set that covers the design space.
Similarly, they can also be a good starting point if we want to build a model that correlates polymer building blocks with the properties of the polymer: since also in this case, we want to make sure that we sample all relevant combinations of building blocks efficiently.
The most trivial approach in \gls{doe} is to use a full-factorial design in which the combination of all factors in all possible levels (e.g., all relevant temperatures and reaction times) are tested. But this can easily lead to a combinatorial problem.
As we discussed in section~\ref{sec:fps}, one could cover the design space using \gls{fps}.
But the greedy \gls{fps} also has some properties that might not be desirable in all cases.\cite{pardalos_sampling_2016}
For instance, it tends to preferentially select points that lie at the boundaries of design space.
Also, one might prefer that the samples are equally spaced along the different dimensions.

Different classical \gls{doe} techniques can help to overcome these issues.\cite{pardalos_sampling_2016}
In \gls{lhs} the range of each variable is binned in equally spaced intervals and the data is randomly sampled from each of these intervals---but in this way, some regions of space might remain unexplored.
For this reason, the max-min-\gls{lhs} has been developed in which evenly spread samples are selected from \gls{lhs} samples using the max-min criterion.

\paragraph{Alternative Techniques}
An alternative for the selection of a good set of training points can be the use of special matrix decompositions.
CUR is a low-rank matrix decomposition into matrices of actual columns (C) and rows (R) of the original matrix, which main advantage over other matrix decompositions, such as \gls{pca}, is that the decomposition is much more interpretable due to use of actual columns and rows of the original matrix.\cite{mahoney_cur_2009}
In the case of \gls{pca}, which builds linear combinations of features, one would have to analyze the loadings of the principal components to get an understanding.
In contrast, the CUR algorithm selects the columns (features) and rows (structures) which have the highest influence on the low-rank fit of the matrix.
And selecting structures with high statistical leverage is what we aim for in diverse set selection.
Bernstein et al.\ found that the use of CUR to select the most relevant structures was the key for their self-guided learning of \gls{pes}, in which a \gls{ml} force-field is built in an automated fashion.\cite{bernstein_novo_2019}

Further, also D-optimal design algorithms have been put to use, in which samples are selected that maximize the \(\|\mathbf{X}^T\mathbf{X}\|\) matrix, where \(\mathbf{X}\) is the information matrix (in some references it is also called dispersion matrix) which contains the model coefficients in the columns and the different examples in the rows.\cite{de_aguiar_d-optimal_1995, podryabinkin_active_2017, podryabinkin_accelerating_2019}
Since it requires the model coefficients, it was mostly used with multivariate linear regression models in cheminformatics.

Moreover, other unsupervised learning approaches such as self-organizing maps,\cite{kohonen_exploration_1997} \gls{knn},\cite{zheng_novel_2000} sphere exclusion\cite{golbraikh_predictive_2002} or hierarchical clustering\cite{rannar_novel_2010, yu_making_2005} have been used, though mostly for cheminformatics applications.\cite{wu_artificial_1996}

\paragraph{Sampling Configurations}
For fitting of models for potential energy surfaces, non-equilibrium configurations are needed.
Here, it can be practical to avoid arbitrarily sampling from trajectories of molecular simulations as consecutive frames are usually highly correlated.
To avoid this, normal mode sampling, where the atomic positions are displaced along randomly scaled normal modes, has been suggested to generate out-of-equilibrium chemical environments and has been successfully applied in the training of the ANI-1 potential.\cite{s.smith_ani-1_2017}
Similarly, binning procedures, where e.g., the amplitude of the force in images of a trajectory is binned have been proposed. When generating the training data, one can then sample from all bins (like in \gls{lhs}).\cite{huan_universal_2017}

Still, one needs to remember that the usage of rational sampling techniques does not necessarily improve the predictive performance on a brand-new dataset which might have a different underlying distribution.\cite{martin_does_2012}
For example, hypothetical databases of \glspl{cof} contain mainly large pore structures, which are not as frequent in experimental structures.
Training a model on a diverse set of hypothetical \glspl{cof} will hence not guarantee that our model can predict properties of experimental structures, which might be largely non-porous.

An alternative to rationally chosen (e.g., using \gls{doe} techniques or \gls{fps}), and hence static, datasets is to let the model (actively) decide which data to use. We discuss this active learning technique next.

\subsection{Active Learning}\label{sec:active_learning}
An alternative to using static training sets, which are assembled before training, is to let the machine decide which data are most effective to improve the model at its current state.\cite{warmuth_active_2003}
This is known as active learning.\cite{settles_active_2012}
And it is especially valuable in cases where the generation of training data is expensive, such as for experimental data or high-accuracy quantum chemical calculations where a simple \enquote{Edisonian} approach, in which we create a large library of reference data by brute force, might not be feasible.

Similar ideas, like adding quantum-mechanical data to a force field when needed, have already been used in molecular dynamics simulations before they became widespread among the \gls{ml} practitioners in materials science and chemistry.\cite{de_vita_novel_1997, csanyi_learn_2004-1}

One of the ways to determine where the current model is ambiguous, i.e., to decide when new data is useful, is to use an ensemble of models (which is also known as \enquote{query by committee}).\cite{behler_constructing_2015, gastegger_machine_2017}
The idea here is to train an ensemble of models, which are slightly different and hence will likely give different, wrong, answers if the model is used outside its domain of applicability (cf.\ section~\ref{sec:domain_applicability}); but the answers will tend to agree mostly when the model is used within the domain of applicability.

Another form of uncertainty sampling is to use a model that can directly output a probability estimate---like the width of the posterior (target) distribution of a Gaussian process (cf.\ section~\ref{sec:bayesian} for more details).
One can then add training points to the space where the distribution is wide and the model is uncertain.\cite{proppe_gaussian_2019}

Botu and Ramprasad reported a simpler strategy, which is related to the concept of the domain of applicability, which we will discuss below (cf.\ section~\ref{sec:domain_applicability}).
The decision if a configuration needs new training data is not made based on an uncertainty measure but merely by using the distance of the fingerprints to the already observed ones.\cite{botu_adaptive_2015}
Active learning is closely linked to Bayesian hyperparameter optimization (cf.\ section~\ref{sec:hyperpar_tuning}) and self-driving laboratories, as they have the goal to choose experiments in the most efficient way, where active learning tries to choose data in the most efficient way.\cite{hernandez-lobato_parallel_2017, lookman_active_2019-2}

\subsection{Dealing With Little Data}\label{sec:augmentation}
Often, one can use tricks to artificially enlarge the dataset to improve model performance.
But these tricks generally require some domain knowledge to decide which transformations are applicable to the problem, i.e.\ which invariances exist. For example, if we train a force field for a porous crystal, one can use the symmetry of the crystal to generate configurations with equivalent energies (which would be a redundant operation when one uses descriptors that already respect this symmetry).
For image data, like steel microstructures\cite{azimi_advanced_2018} or 2D diffraction patterns,\cite{ziletti_insightful_2018} several techniques have been developed, which include to randomly rotate, flip or mirror the image which is, for example, implemented in the \texttt{ImageDataGenerator} module of the \texttt{keras} Python package.
Notably, there is also effort to automate the augmentation process and promising results have been reported for images.\cite{cubuk_autoaugment_2019}
However, data augmentation always relies on assumptions about the equivariances and invariances of the data, wherefore it is difficult to develop general rules for any type of dataset.

Still, the addition of Gaussian noise is a method that can be applied on most datasets.\cite{cortes-ciriano_improved_2015}
This works effectively as data augmentation if the data is presented multiple times to the model (e.g., in \glspl{nn} where one has multiple forward and backward passes of the data through the network). By the addition of random noise, the model will then see a slightly different example upon each pass of the data.
The addition of noise also acts as \enquote{smoother}, which we will explore in more detail when we discuss regularization in section~\ref{sec:regularization}.

Oviedo et al.\ reported the impact data augmentation can have in materials science.
Thin-film \gls{xrd} patterns are often distorted and shifted due to strain or lattice contraction or expansion.
Also, the orientations of the grains are not randomized, as they are in a powder, and some reflexes will have an increased intensity depending on the orientation of the film.
For this reason, conventional simulations cannot be used to form a training set for a \gls{ml} model to predict the space group based on the diffraction pattern.
To combat the data scarcity problem, the authors expanded the training set, generated by simulating diffraction patterns from a crystal structure database, by taking data from the training set and by scaling, deleting or shifting of reflexes in the patterns. In this way, the authors generated new training data that correspond to the typically experimental distortions.\cite{oviedo_fast_2019}
A similar approach was also chosen by Wang et al.\ who built a \gls{cnn} to identify \glspl{mof} based on their \gls{xrpd} patterns.
Wang et al.\ predicted the patterns for \glspl{mof} in the \gls{csd} and then augmented their dataset by creating new patterns by merging the main peaks of the predicted patterns with (shuffled) noise from pattern they measured in their own lab.\cite{wang_rapid_2019}

Sometimes, data augmentation techniques have also been used to solve non-uniqueness or invariance problems.
The Chemception model is a \gls{cnn}, inspired by models for image recognition, that is trained to predict chemical properties based on images of molecular drawings.\cite{goh_chemception_2017}
The prediction should, of course, not depend on the relative orientation of the molecule in the drawing.
For this reason, the authors applied augmentation methods such as rotations. Interestingly, many image augmentation techniques also use cropping. However, the local information density in drawings of molecules is higher than in usual images and hence losing a part of the image would be a more significant problem.

Another issue is that not all datasets are unique. For example, if one uses (non-canonical) \gls{smiles} strings to describe molecules, one has to realize that they are not unique.
Therefore, Bjerrum trained this model on all possible \gls{smiles} string for a molecule and obtained a dataset that was 130 times bigger than the original dataset.\cite{bjerrum_smiles_2017}
This idea was also used for the Coulomb matrix, a popular descriptor that encodes the structure by capturing all pairwise Coulomb terms, based on the nuclear charges, in a matrix (cf.\ section~\ref{sec:distance_matrix_based}).
Without additional steps, this representation is not permutation invariant (swapping rows or columns does not change the molecule but would change the representation).
Montavon used an augmented dataset and in which they mapped each molecule to a set of randomly sorted Coulomb matrices and could improve upon other techniques of enforcing permutation symmetry---likely due to the increased dataset size.\cite{montavon_learning_2012}

But also simple physical heuristics can help if there is only little data to learn from.
Rhone et al.\ used \gls{ml} to predict the outcome of reactions in heterogeneous catalysis, where only little curated data is available.\cite{rhone_predicting_2019}
Hence, they aided their model with a reaction tree and chose the prediction of the model that is closest to a point in the reaction tree (and hence a chemically meaningful reaction).
Moreover, they also added heuristics like conservation rules and penalties for some transformations (e.g., based on the difference of heavy atoms in educts and products) to support the model.

Another promising avenue are multitask learning approaches where a model, like a \gls{dnn}, is trained to predict several properties.
The intuition here is to capture the implicit information in the relationship between the multimodal variables.\cite{ramsundar_massively_2015, zubatyuk_accurate_2019}
Closely related to this are transfer learning approaches (cf.\ section~\ref{sec:transfer}), where on trains a model on a large dataset and then \enquote{refines} the weights of the model using a smaller dataset.\cite{hutchinson_overcoming_2017} Again, this approach is a well-established practice in the \enquote{mainstream} \gls{ml} community.

Given the importance of the data scarcity problem, there is a lot of ongoing effort in developing alternative solutions to combat this challenge, many of which build on encoding-decoding architectures.
Generative models like \glspl{gan} or \gls{vae} can be used to create new examples by learning how to generate underlying distribution of the data.\cite{antoniou_data_2017}

Some problems may also be suitable for so-called one-shot learning approaches.\cite{vinyals_matching_2016, lake_human-level_2015, li_fei-fei_one-shot_2006}
In the field of image recognition, the problem of correctly classifying an image after seeing only one training example for this class (e.g.\ correctly assigning names to images of persons after having seen only one image for each person) has received a lot of interest.
Supposedly, because this is what humans are able to do---but machines are not, at least not in the \enquote{usual} classification setting.\cite{lake_building_2016}

One- or few-shot learning is based on learning a so-called attention mechanism.\cite{olah_attention_2016}
Upon inference, the attention mechanism, which is distance measure to the memory, can be exploited to compare the new example to all training points and express the prediction as a linear combination of all labels in the support set.\cite{koch_siamese_2015}
One approach to do this is Siamese learning, using a \gls{nn} that takes two inputs and then learns an attention mechanism.
This has also been used, in a refined formulation, by Pande and co-workers to classify the activity of small molecules on different assays for pharmaceutical activity.\cite{altae-tran_low_2017} Such techniques are especially appealing for problems where only little data is available.

Still, one always should remember that there is no absolute number that defines what \enquote{little data} is.
This number depends on the problem, the model, and the featurization.
But it can be estimated using learning curves, in which one plots the error of the model against the number of training points (cf.\ section~\ref{sec:metrics}).

\subsection{Dealing With Imbalanced Data Labels}
Often, data is imbalanced, meaning that different classes which we attempt to predict (e.g.\ \enquote{stable} and \enquote{unstable}, or \enquote{low performing} and \enquote{high performing}) do not have the same number of examples in our training set.
Balachandran et al.\ faced this challenge when they tried to predict compounds that break spatial inversion symmetry, and hence could be interesting for e.g.\ their piezoelectric properties.\cite{balachandran_learning_2017}
They found that one symmetry group was misclassified to \SI{100}{\percent} due to imbalanced data.
To remedy this problem, they used an oversampling technique, which we will briefly discuss next.

Oversampling, which means adding points to the underrepresented class, is one of the most widely used approaches to deal with imbalanced data.
The opposite approach is undersampling, in which instances of the majority class are removed.
Since random oversampling can cause overfitting (due to replication of training points) and undersampling can lead to poorer predictive performance (as training points are eliminated) both strategies have been refined by means of interpolative procedures.\cite{he_learning_2009}

The \gls{smote} for example, creates new (synthetic) data for the minority class by randomly selecting a point on the vector connecting a data point from the minority class with one of its nearest neighbors.
In \gls{smote}, each point in the minority class is treated equally---which might not be ideal since one would expect that examples close to class boundaries are more likely to be misclassified.
Borderline-\gls{smote} and \gls{adasyn} try to improve on this point.
In a similar vein, it can also be easier to learn clear classification rules when so-called Tomek links\cite{tomek_two_1976} are deleted.
Tomek links are pairs of two points from different classes for which the distance to the example from the alternative class is smaller than to any other example from their class.

Still, care needs to be taken in the case of very imbalanced data in which algorithms can have difficulties to recognize class structures.
In this case over- or undersampling can even deteriorate the performance.\cite{krawczyk_learning_2016}

A useful Python package to address data imbalance problems is \texttt{imbalanced-learn}, which implements all the methods we mentioned and which are analyzed in more detail in a review by He and Garcia.\cite{he_learning_2009}
There they also discuss cost-sensitive techniques.
In these approaches, a cost matrix is used to describe a higher penalty for misclassifying examples from a certain class---which can be an alternative strategy to deal with imbalanced data.\cite{he_learning_2009}
Importantly, oversampling techniques should only be applied---as all data transformations---after the split into training and test sets.

\begin{figure}
    \centering
    \includegraphics[width=.5\textwidth]{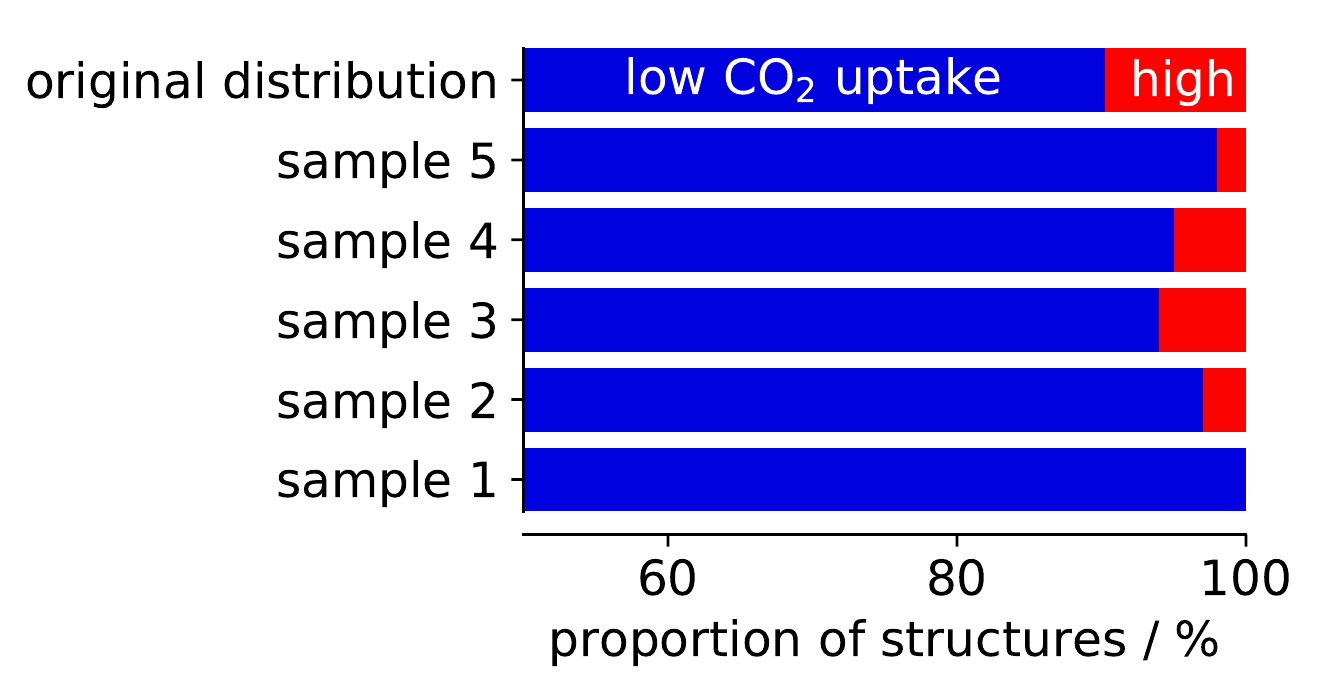}
    \caption{Example for the importance of stratification.
        For this example, we use a threshold of \SI{2.5}{\milli\mole\, \ce{CO2} \per \gram} to group structures in low and high performing materials, which is slightly higher than the threshold chosen by  Boyd et al.\protect{\cite{boyd_data_2019}}
        Then, we randomly draw 100 structures and can observe that the class distribution gets distorted---sometimes we do not have any high performing materials in our sample. Stratification can be used to remedy this effect.}\label{fig:pmof_stratification}
\end{figure}

In any case, it is also advisable to use stratified sampling which ensures that the class proportions in the training set are equal to the ones in the test set.
An example of the influence of stratified sampling is shown in Figure~\ref{fig:pmof_stratification} where we contrast the random with the stratified splitting of structures from the database of Boyd et al.\cite{boyd_data_2019}

%% file: main/2_featurization.tex
\section{What To Learn From: Translating Structures Into Feature Vectors}\label{sec:featurization}

After having reviewed the rows of the feature matrix, we now focus on the columns and discuss ways to generate those columns (descriptors), and how to select the best ones (as more is not always better in the case of feature columns).
The possibilities for structural descriptors are so vast that it is impossible to give a comprehensive overview.
Especially since there is no silver bullet and the performance of descriptors depends on the problem and the learning setting.
In some cases, local fingerprints based on symmetry functions might be more appropriate, e.g., for potential energy surfaces, whereas in other cases, where structure-property insights are needed, higher-level features such as pore shapes and sizes can be more instructive.

An important distinction of \glspl{nn} compared to classical \gls{ml} models, like kernel methods (cf.\ section~\ref{sec:kernel_methods}) is that \glspl{nn} can perform representation learning, i.e., the need for highly engineered structural descriptors is less pronounced than for \enquote{classical} learners as \gls{nn} can learn their own features from unstructured data.
Therefore, one will find \gls{nn} models that directly use the positions and the atomic charges whereas such an approach is deemed to fail with classical \gls{ml} models, like \gls{krr}, that rely on structured data.
The representation learning of \glspl{nn} can potentially leverage regularities in the data that cannot be described with classical descriptors---but it only works with large amounts of data.
We will discuss this in more detail when we revisit special \gls{nn} architectures in section~\ref{sec:messagepassing}.

The quest for good structural descriptors is not new.
Cheminformatics researchers tried to devise strategies to describe structures, e.g., to determine whether a compound has already been deposited on the \gls{cas} database, which led to the development of Morgan fingerprints.\cite{morgan_generation_1965}
Also the demand for \gls{qsar} in drug development led to the development of a range of descriptors that are often highly optimized for a specific application (also because simple linear models have been used) as well as heuristics (e.g., Lipinkski's rule of five\cite{lipinski_experimental_2001}).
But also fingerprints (e.g., Daylight fingerprints)---i.e., representations of the molecular graphs have been developed.
We will not discuss them in detail in this review as most of them are not directly applicable to solid-state systems.\cite{tropsha_predictive_2007, danishuddin_descriptors_2016}
Still, one needs to note that for the description of \glspl{mof} one needs to combine information about organic molecules (linkers), metal centers, and the framework topologies wherefore not all standard featurization approaches are ideally suited for \glspl{mof}.
Therefore, molecular fingerprints can still be interesting to encode the chemistry of the linkers in \glspl{mof}, which can be important for electronic properties or more complex gas adsorption phenomena (e.g., involving \ce{CO2}, \ce{H2O}).

\begin{figure}
    \centering
    \includegraphics[width=\textwidth]{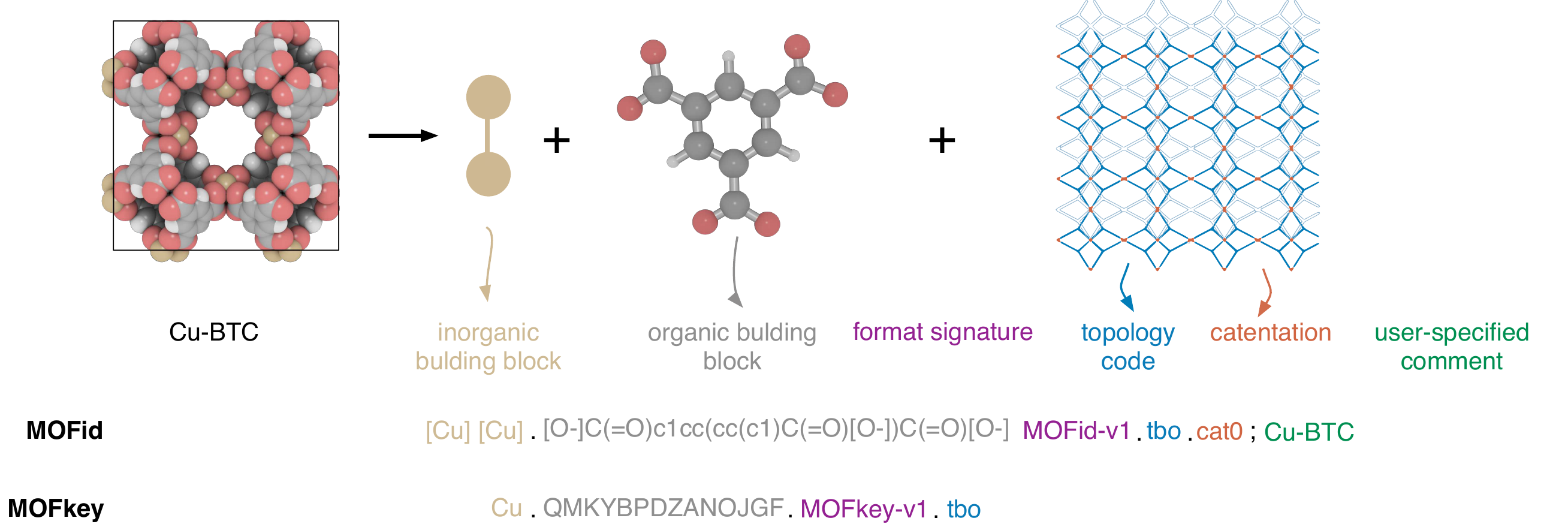}
    \caption{Building principle of the MOFid and MOFkey identifiers for HKUST-1. Bucior et al.\ use a \gls{smiles} derived format in the MOFid and whereas the MOFkey is inspired by the InChIkey format, which is a hashed version of the InChi fingerprint, which is more standardized than \gls{smiles}.  Figure adopted from Bucior et al,\protect{\cite{bucior_identification_2019}}}\label{fig:mofid}
\end{figure}

A decomposition of \glspl{mof} into the building blocks and encoding of the linker using \gls{smiles} was proposed in the MOFid scheme from Bucior et al.\ (cf.\ Figure~\ref{fig:mofid}).\cite{bucior_identification_2019}
This scheme is especially interesting to generate unique names for \glspl{mof} and in this way to simplify data-mining efforts.
For example, Park et al.\ had to use a six-step process to identify whether a string represents the name of a \gls{mof} in their text-mining effort,\cite{park_text_2018} and then one still has to cope with non-uniqueness problems (e.g., Cu-BTC vs.\ HKUST-1).
One main problem of such fingerprinting approaches for \glspl{mof} is that they require the assignment of bonds and bond orders, which is not trivial for solid structures,\cite{walsh_electron_2017} and especially for experimental structures that might contain disorder or incorrect protonation.

The most popular fingerprints for molecular systems are implemented and documented in libraries like \texttt{RDKit},\cite{landrum_rdkit_2006} \texttt{PaDEL}\cite{yap_padel-descriptor_2011} or \texttt{Mordred}.\cite{moriwaki_mordred_2018}
For a more detailed introduction into descriptors for molecules we can recommend a review by Warr\cite{warr_representation_2011} and the \enquote{Deep Learning for the Life Sciences} book,\cite{ramsundar_deep_2019} which details on how to build \gls{ml} systems for molecules.

\subsection{Descriptors}
There are several requirements that an ideal descriptor should fulfill to be suitable for \gls{ml}:\cite{ghiringhelli_big_2015, faber_crystal_2015}
\begin{itemize}

    \item A descriptor should be \textbf{invariant} with respect to transformations that preserve the target property (cf.\ Figure~\ref{fig:invariances}).

          \begin{figure}
              \centering
              \includegraphics[width=\textwidth]{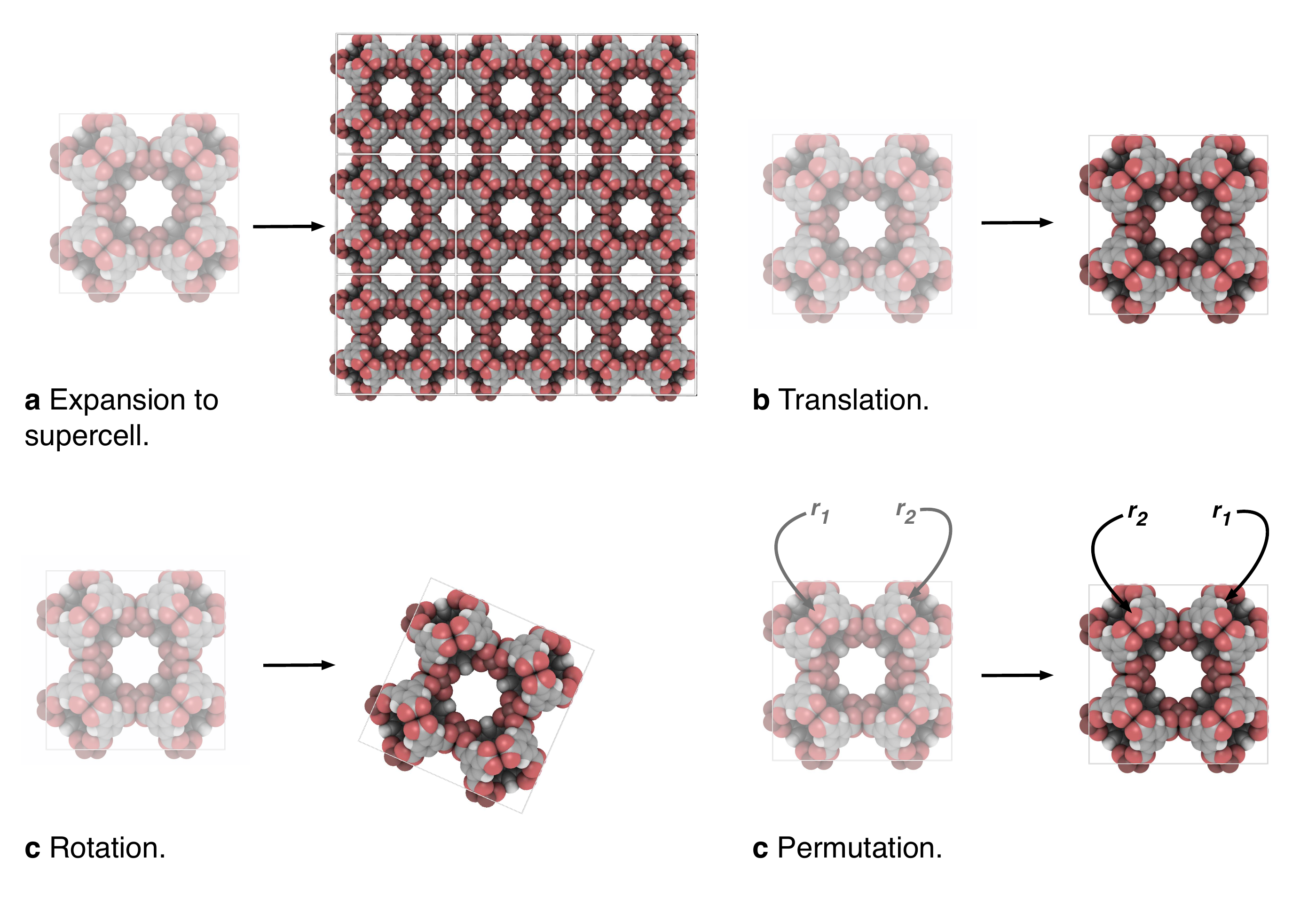}
              \caption{Illustration of transformations of crystal structures to which an ideal descriptor should be invariant.
                  Structures drawn with iRASPA.\protect{\cite{dubbeldam_iraspa_2018}}}\label{fig:invariances}
          \end{figure}

          For crystal structures, this means that the representations should respect periodicity, translational, rotational and permutation symmetry (i.e., the numbering of the atoms in the fingerprint should not influence the prediction).
          Similarly, one would want \textbf{equivariances to be conserved}.
          Equivariant functions transform in the same way as their arguments, as it is, for example, the case for the tensorial properties like the force (negative gradient of energy) or the dipole moment, which both translate the same way as the positions.\cite{noe_machine_2019, glielmo_accurate_2017}

          Respecting those symmetries is important from a physics perspective as (continuous) symmetries are generally linked to a conserved property (cf.\ Noether's theorem, e.g., rotational invariance corresponds to conservation of angular momentum).
          Conceptually, this is different from classical force field design where one usually focuses on correct asymptotic behavior.
          In \gls{ml}, the intuition is to rather use symmetries to preclude completely nonphysical interactions.

          As discussed above, one could in principle also attempt to include those symmetries using data augmentation techniques, but it is often more robust and efficient to \enquote{hard-code} them on the level of the descriptor.
          Notably, the introduction of the invariances on the descriptor level also removes alignment problems, when one would like to compare two systems.

    \item A descriptor should be \textbf{unique} (i.e., non-degenerate).
          This means that each structure should be characterized by one unique descriptor and that different structures should not share the same descriptor.
          When this is not the case, the model will produce prediction errors that cannot be removed with the addition of data.\cite{moussa_comment_2012}
          Von Lilienfeld et al.\ nicely illustrate this in analogy to the proof of the first Hohenberg-Kohn theorem trough \textit{reductio ad absurdum}.\cite{lilienfeld_fourier_2015}
          This uniqueness is automatically the case for invertible descriptors.
    \item A descriptor should allow for \textbf{(cross-element) generalization}.
          Ideally, one does not want to be limited in system size or system composition. Fixed vector or matrix descriptors, like the Coulomb matrix (see section~\ref{sec:distance_matrix_based}), can only represent systems smaller or equal to the dimensionality of the descriptor. Also, one sometimes finds that the linker type\cite{borboudakis_chemically_2017}  or the monomer type is used as a feature. Obviously, such an approach does not allow for generalization to new linkers or monomer types.

          The cross-element generalization is typically not possible if different atom types are encoded as being orthogonal (e.g., by using a separate \gls{nn} for each atom type in a \gls{hdnpp} or by grouping interactions by the atomic numbers, e.g., \gls{bob}, partial \gls{rdf}). To introduce generalizability across atom types one needs to use descriptors that allow for a chemically reasonable measure of similarity between atom types (and trends in the periodic table). What an appropriate measure of similarity is depends on the task at hand, but an example for a descriptor that can be relevant for chemical reactivity or electronic properties is the electronegativity.

    \item A descriptor should be \textbf{efficient} to calculate.
          The cardinal reason for using supervised \gls{ml} is to make simulations more efficient or to avoid expensive experiments or calculations.
          If the descriptors are expensive to compute, \gls{ml} no longer fulfills this objective and there is no reason to add a potential error source.

    \item A descriptor should be \textbf{continuous}: For differentiability, which is needed to calculate, e.g., forces, and for some materials design applications\cite{gomez-bombarelli_automatic_2018} it is desirable to have continuous descriptors.
          If one aims to use the force in the loss function (force-matching) of a gradient descent algorithm, at least second order differentiability is needed.
          This is not given for many of the descriptors which we will discuss below (like global features as statistics of elemental properties) and is one of the main distinctions of the symmetry functions from the other, often not localized, tabular descriptors which we will discuss.
\end{itemize}
Before we discuss some examples in more detail, we will review some principles that we should keep in mind when designing the columns of the feature matrix.

\begin{figure}
    \centering
    \includegraphics[scale=.7]{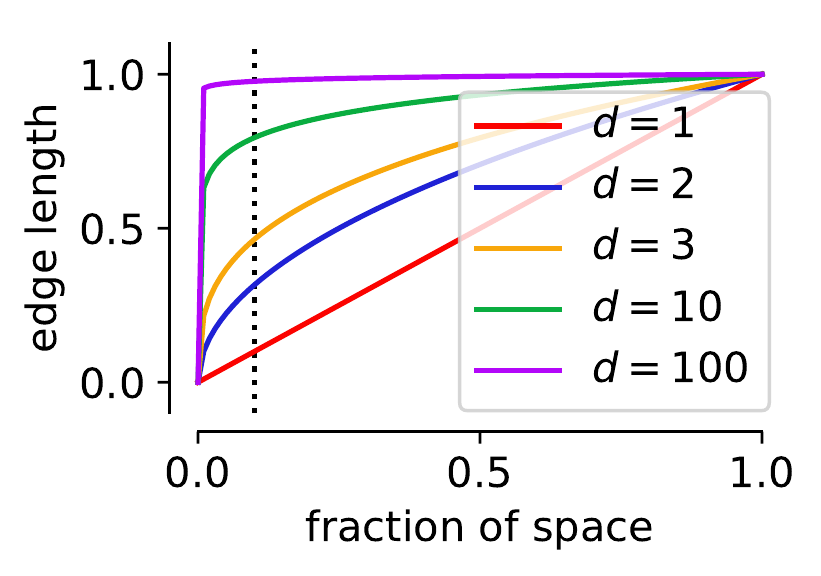}
    \caption{Illustration of the empty space phenomenon (the curse of dimensionality).
        For this illustration we consider the data to be uniformly distributed in a \(d\) dimensional unit cube.
        The edge length of a hypercube corresponding to a fraction \(q\) of the total volume is \(q^{1/d}\), which we plotted here for different \(d\).
        The dotted line in the figure represents \SI{10}{\percent} of the volume, for which we would nearly need to consider the full edge length in 100-dimensional space.
        This means that locality is lost in high dimensions, which can be problematic for algorithms that use the local neighborhood for their reasoning.}\label{fig:empty_space_phenomenon}
\end{figure}

\paragraph{The Curse of Dimensionality}\label{sec:curse_dimensionality}
One of the main paradigms that guide the development of materials descriptors is the so-called \emph{curse of dimensionality}, which describes that it is often hard to find decision boundaries in a high-dimensional space as the data often no longer covers the space.
For example, in 100 dimensions nearly the full edge length is needed to capture \SI{10}{\percent} of the total volume of the 100-dimensional hypercube (cf.\ Figure~\ref{fig:empty_space_phenomenon}).
This is also known as empty space phenomenon, and describes that similarity-based reasoning can fail in high dimensions given that also the nearest neighbors are no longer close in such high-dimensional spaces.\cite{domingos_few_2012}
Often, this is also discussed in terms of Occam's razor: \enquote{Simpler solutions are more likely to be correct than complex ones.}
This not only reflects that learning in high-dimensional space brings its own problems but also that simplicity, which might be another way of asking of explainability, for itself is a value (due to its aesthetics) we should strive for.\cite{domingos_role_1999}
More formally, this is related to the minimum descriptor length principle\cite{rissanen_modeling_1978} which views learning as a compression process and in which the best model is the smallest one in terms of itself and the data (this idea is rooted in Solomonoff’s general theory of inference\cite{Solomonoff1964}).\cite{grunwald_tutorial_2004, grunwald_minimum_2007}

\paragraph{Chemical Locality Assumption}\label{sec:locality}
Many descriptors that we discuss below are based on the assumption of chemical locality, meaning that the total property of a compound can be decomposed into a sum of contributions of local (atom-centered) environments:
\begin{equation}
    \text{property} (\text{descriptor}) = \sum_{i}^{\text{atoms}} \text{model}_i \left(\text{descriptor}_i\right).\label{eq:locality_approx}
\end{equation}
This approximation (cf.\ eq.~\ref{eq:locality_approx}) is often used in models describing the \gls{pes}.

The locality approximation is usually justified based on the nearsightedness principle of electronic matter, which says that a perturbation at a distance has little influence on the local density.\cite{prodan_nearsightedness_2005}
And this \enquote{nearsighted} approach also guided the development of many-body potentials like embedded atom methods, linear-scaling \gls{dft} methods or other coarse-grained models in the past (also here the system is divided into subsystems).\cite{kohn_density_1996, galli_large_1992}

The division into subsystems can also be a feat for training of \gls{ml} models, as one can learn on fragments to predict larger systems, as it has been done for example for a \gls{hdnpp} for MOF-5.\cite{gastegger_machine_2017}
Also, this approach makes it easier to incorporate size extensivity, i.e., to ensure that the  energy of a  system composed of the subsystems \(A + B\) is indeed the sum of the energies of \(A\) and \(B\).\cite{zhang_end--end_2018}

But such an approach might be less suited for cases like gas adsorption where both the local chemical environment (especially for chemisorption) but also the pore shape, size, and accessibility play a role---i.e., one wants pore-centered descriptors rather than atom-centered descriptors.
For this case global, \enquote{farsighted}, descriptors of the pore size and shape, like pore limiting diameters, accessible surface areas,\cite{first_computational_2011, first_mofomics_2013, ongariAccurateCharacterizationPore2017} or persistent homology fingerprints,\cite{moosavi_geometric_2020} can be better suited.
This is important to keep in mind as target similarity, i.e., how good we can the property of interest (e.g., the \gls{pes} or the gas adsorption properties), is one of the main contributions to the error of \gls{ml} models.\cite{huang_communication_2016}
Also, one should be aware that typically cutoffs of \SI{6}{\angstrom} around an atom are used to define the local chemical environments.
In some systems, the physics of the phenomenon is, however, dominated by long-range behavior\cite{stohr_quantum_2019} that cannot be described within the locality approximation.
Correctly describing such long-range effects is one of the main challenges of ongoing research.\cite{grisafi_incorporating_2019}

Importantly, a model that assumes atom-centred descriptors is invariant to the order of the inputs (permutational invariance).\cite{gastegger_high-dimensional_2015}
Interestingly, classical force fields do not show this property. The interactions are defined on a bond graph and the exchange of an atom pair can change the energy.\cite{noe_machine_2019, braams_permutationally_2009}

\subsection{An Overview of the Descriptor Landscape}
\begin{figure}
    \centering
    \includegraphics[width=\textwidth]{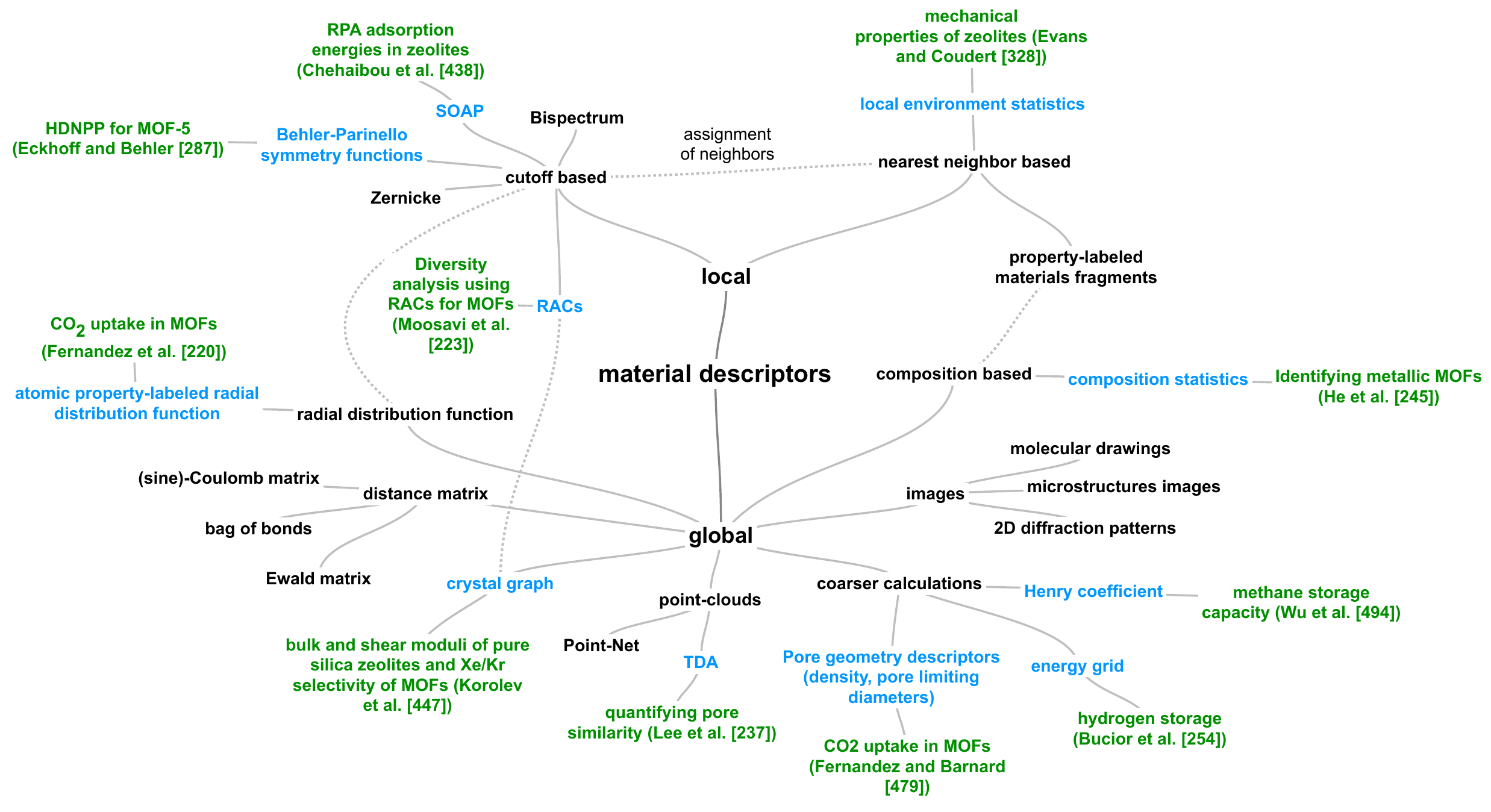}
    \caption{Non-exhaustive overview over the landscape of descriptors for solids. In blue, we highlighted descriptors for which we are aware of an application in the field of porous materials, for which we give an example in green.}\label{fig:desciptor_mindmap}
\end{figure}

In Figure~\ref{fig:desciptor_mindmap} we show an overview of the space of material descriptors.
We will distinct two main classes of descriptors; local ones, that only describe the local (chemical) environment and global ones which describe the full structures at once.

Nearly as vast as the descriptor landscape is the choice of tools that are available to calculate these descriptors.
Some notable developments are \texttt{matminer} package,\cite{ward_matminer_2018} which is written in Python, the \texttt{DSCribe} package, which has a Python interface, but where  the computationally expensive routines are written in C/C++ and \texttt{AMP}, which also has a Python interface and where the expensive fingerprinting can be performed in Fortran.\cite{khorshidi_amp_2016}
The von Lilienfeld group is currently also implementing efficient Fortran routines in their \texttt{QML} package.\cite{anders_s._christensen_qmlcode/qml_2017}
Other packages like \texttt{CatLearn},\cite{hansen_atomistic_2019} which has also functionalities for surfaces, or \texttt{QUIP},\cite{bartok_gaussian_2015} \texttt{aenet}\cite{artrith_implementation_2016} and \texttt{simple-nn}\cite{lee_simple-nn_2019}, \texttt{ai4materials}\cite{ziletti_ai4materials_2020} also contain functions for fingerprinting of solid systems.
For the calculation of features based on elemental properties, i.e., statistics based on the chemical composition, the Magpie package is frequently used.\cite{ward_general-purpose_2016}

\begin{figure}
    \centering
    \includegraphics[width=.4\textwidth]{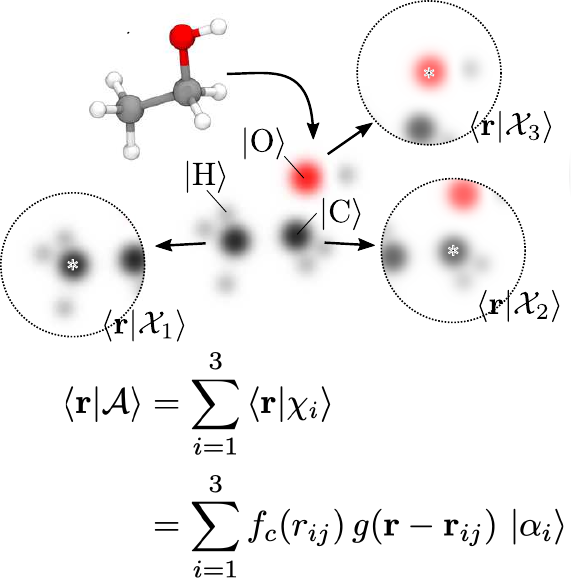}
    \caption{Illustration of the concept of featurization using symmetry functions.
        There are atom centered local environments that we can represent with abstract kets \(\ket{\chi}\), expressed in the basis of Cartesian coordinates \(\bra{\mathbf{r}}\). The Figure is a modified version of an illustration from Ceriotti and co-workers.\protect{\cite{willatt_atom-density_2019}}}\label{fig:symmetry_functions}
\end{figure}

\paragraph{General Theme of Local and Global Fingerprints}
In the following, we will also see that many fingerprinting approaches are just a variation of the same theme, namely many-body correlation functions, which can be expressed in Dirac notation as
\begin{equation}
    \braket{\mathbf{r} | \chi_j^{(1)}} = \sum_i  \, g^{(2)} (\mathbf{r} - \mathbf{r}_{ij}) \ket{\alpha_i}.\label{eq:dirac}
\end{equation}
This shows that the abstract atomic configuration \(\ket{\chi_j^{(v)}}\), in terms of the \((v+1)\)-body correlation, can be described with a cross-correlation function (\(g^{(2)}\) being equivalent to the radial distribution function) and information about the elemental identity of atom \(i\), \(\ket{\alpha_i}\) (see Figure~\ref{fig:symmetry_functions}).
And it also already indicates why the term \enquote{symmetry functions} is often used for functions of this type.
Descriptors based on eq.~\ref{eq:dirac} are said to be symmetrized, e.g., invariant to translations of the entire structure (symmetrically equivalent positions will give rise to the same fingerprint).

Some fingerprints take into account higher orders of correlations (like triples in the bispectrum) but the idea behind most of them is the same---they are just projected onto a different basis (e.g., spherical harmonics, \(\bra{n lm}\), instead of the Cartesian basis \(\bra{\mathbf{r}}\)).\cite{willatt_atom-density_2019, drautz_atomic_2019}
Notably, it was recently shown that also three-body descriptors do not uniquely specify the environment of an atom, but Pozdnyakov et al. also showed that in combination with many neighbors, such degeneracies can often be lifted.\cite{pozdnyakov_completeness_2020}

Different flavors of correlation functions are used for both local and global descriptors, and the different flavors might converge differently with respect to the addition of terms in the many-body expansion (going from two-body to the inclusion of three-body interactions and so on).\cite{bartok_representing_2013}
Local descriptors are usually derived by multiplying a version (projection onto some basis) of the many-body correlation function with a smooth cutoff function such as
\begin{equation}
   f_{c}\left(r_{i j}\right)=\left\{\begin{array}{ll}
0.5 \times\left[\cos \left(\frac{\pi r_{ij}}{r_{c}}\right)+1\right] & \text { for } r_{i j} \leq r_{c} \\
0 & \text { for } r_{i j}>r_{c}
\end{array}\right.\label{eq:cutoff_func}
\end{equation}
where \(r_\text{cut}\) is the cutoff radius, which determines the set of \(i\) the summation in equation~\ref{eq:dirac} runs over.

We will start our discussion with local descriptors that use such a cutoff function (cf.\ eq.~\ref{eq:cutoff_func}) and which are usually employed when atomic resolution is needed.

In some cases, especially when only the nearest neighbors should be considered, Voronoi tessellations are used to assign which atoms from the environment should be included in the calculation of the fingerprint.
This approach is based on the nearest neighbor assignment method that was put forward by O'Keeffe.\cite{okeeffe_proposed_1979}

\subsubsection{Local Descriptors}
\paragraph{Instantaneous Correlation Functions via Cutoff Functions}
For the training of models for \gls{pes}, flavors of instantaneous correlation functions have become the most popular choices, and are often used with kernel methods (cf.\ section~\ref{sec:kernel_methods}) or \gls{hdnpp} (cf.\ section~\ref{sec:ann}).

The archetypal examples of this type are the atom-centered symmetry functions suggested by Behler and Parinello, where the two-body term has the following form
\begin{equation}
    G_{i}^{2}=\sum_{j \neq i}^{\text {all}} \exp\left[-\eta\left(r_{i j}-R_{s}\right)^{2} \right] f_{c}\left(r_{i j}\right),\label{eq:behler}
\end{equation}
which is a sum of Gaussians and the number of neighbors that are taken into account in the summation is determined by the cutoff function \(f_c\) (cf.\ eq.~\ref{eq:cutoff_func}).
Behler and Parinello also suggest a three-order term, which takes all the internal angles for triplets of atoms, \(\theta_{ijk}\), into account.
This featurization approach has been the driver of the development of many \glspl{hdnpp} (cf.\ section~\ref{sec:ann}).

One should note that these fingerprints contain a set of hyperparameters that should be optimized, like the shift \(R_s\) or the width of the Gaussian \(\eta\), for which usually at set of different values is used to fingerprint the environment.
Also, similar to molecular simulations, the cutoff \(r_c\) is a parameter that should be carefully set to ensure that the results are converged.

Fingerprints of this type (cf.\ eq.~\ref{eq:dirac}) are translational invariant, because they only depend on internal coordinates and rotational invariant, because they only depend  on internal angles (in case of the $v=3$ correlation).
The permutation invariance is due to the summation (which does not depend on the order) over all neighbors \(i\), in eq.~\ref{eq:behler} (and also in the locality approximation itself, cf.\ eq.~\ref{eq:locality_approx}).

An alternative approach for fingerprinting in terms of symmetry functions has been put forward by Csányi and co-workers.\cite{bartok_gaussian_2010}
They started by proposing the bispectrum descriptor which is based on expanding the atomic density distribution (with Dirac delta functions for \(g\) in equation~\ref{eq:dirac}) in spherical harmonics.
This allows, as advantage over the Behler-Parinello symmetry functions, for systematic improvements via the addition of spherical harmonics.

This corresponds to a projection of the atomic density onto a four-dimensional sphere and representing the location in terms of four-dimensional spherical harmonics.\cite{bartok_representing_2013, behler_perspective_2016}
This descriptor was improved with the \gls{soap} methodology, which is a smooth similarity measure of local environments (covariance kernel, which we will discuss in section~\ref{sec:kernel_methods}) by writing \(g(r)\) in eq.~\ref{eq:dirac} using atom-centered Gaussians as expansions with sharp features (Dirac delta functions in the bispectrum) are slowly converging.

Given that \gls{soap} is a kernel, this descriptor found the most application in kernel-based learning (which we will discuss below in more detail, cf.\ section~\ref{sec:kernel_methods}), as it directly defines a similarity measure between environments (overlap between the smooth densities), which has recently extended to tensorial properties.\cite{grisafi_symmetry-adapted_2018} This enabled Wilkins et al.\ to create models for the polarizability of molecules.\cite{wilkins_accurate_2019}

\paragraph{Voronoi Tessellation Based Assignment of Local Environments}
In some cases the partitioning into Wigner-Seitz cells using Voronoi tessellation is used instead of a cutoff function. These Wigner-Seitz cells are regions which are closer to the central atom than to any other atom.
The faces of these cells can then be used to assign the nearest neighbors and to determine coordination numbers.\cite{okeeffe_proposed_1979}
Ward et al.\ used this method of assigning neighbors to construct local descriptions of the environment that are not sensitive to small changes that might occur during a geometry relaxation.\cite{ward_including_2017}
These local descriptors can be based on comparing elemental properties, like the electronegativity, of the central atom to its neighbors
\begin{equation}
    \delta_p = \frac{\sum_n A_n \left\|p_n - p_i \right\|}{\sum_n A_n },
\end{equation}
where \(A_n\) is the surface area of the face of the Wigner-Seitz cell and \(p_i\)  and \(p_n\) are the properties of central and neighboring atoms, respectively.

\begin{figure}
    \centering
    \includegraphics[width=\textwidth]{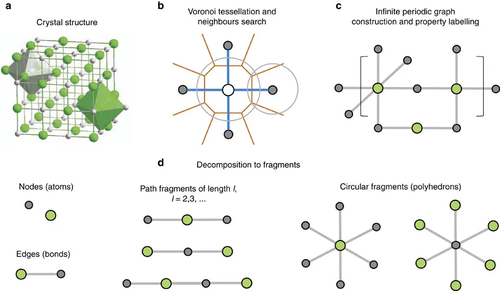}
    \caption{Schema illustrating the construction of \gls{plmf}. The concept behind this descriptor is that for crystal structure (a) the nearest neighbors are assigned using Voronoi tessellation (b) and then used to construct a crystal graph that can be colored with properties, which is then decomposed into subgraphs (d).  Figure reprinted from Isayev et al.\protect{\cite{isayev_universal_2017}}}\label{fig:property_labeled_materials_fragements}
\end{figure}
A similar approach was also used in the construction of \gls{plmf} which were proposed by Isayev et al.\cite{isayev_universal_2017}
There, a crystal graph is constructed based on the nearest-neighbor assignment from the Voronoi tessellation, where the nodes represent atoms that are labeled with a variety of different (elemental) properties.
Then, the graph is partitioned into sub-graphs and the descriptors are calculated using differences in properties between the graph nodes (neighboring atoms) (cf.\ Figure~\ref{fig:property_labeled_materials_fragements}).

The Voronoi decomposition is also used to assign the environment in the calculation of the orbital field matrix descriptor, which is the weighted sum of the one-hot encoded vector of the electron configuration.\cite{lam_pham_machine_2017}
One hot-encoding is a technique that is frequently used in language processing and that represents the feature vector of \(n\) possibilities with zeros (feature not present) and ones (feature present).
In the original work, the sum and average of the local descriptors were used as descriptors for the entire structure and also suggested to gain insight into the importance of specific electronic configurations using a decision tree analysis.

Voronoi tesselation is the dual problem of Delaunay triangulation which attempts to assign points into tetrahedrons (in three dimensions, in two dimensions into triangles, etc.) which circumspheres contain no other point in its interiors. The Delaunay tesselation found use in the analysis of zeolites, where the geometrical properties of the tetrahedrons, like the tetrahedrality or the volume, have been used to build models that can classify zeolite framework types.\cite{yang_identifying_2009, carr_machine_2009}

Overall, we will see that a common approach to generate global, fixed length, descriptors is that one calculates statistics (like the mean, standard deviation or maximum or minimum) of base descriptors, that can be based on elemental properties for each site.

\subsubsection{Global Descriptors}
\paragraph{Global Correlation Function}\label{sec:rdf}
As already indicated, some properties are less amenable to decomposition into contributions of local environments and might be better described using the full, global correlation functions.
These approaches can be seen, completely analogous to the local descriptors, as approximations to the many-body expansion, for example for the energy
\begin{equation}
    E = \sum_{i=1}^N E^{(1)} (\mathbf{r}_i) + \sum_{i<j}^N E^{(2)}(\mathbf{r}_i, \mathbf{r}_j) + \sum_{i<j<k}^N E^{(3)} (\mathbf{r}_i, \mathbf{r}_j, \mathbf{r}_k) + \dots
\end{equation}
As we discussed in the context of the symmetry functions for local environments, we can choose where we truncate this expansion (two-body pairwise distance terms, three-body angular terms \dots) to trade-off computational and data efficiency (more terms will need more training data) against uniqueness.
Similar to the symmetry functions for local chemical environments, different projections of the information have been developed.
For example, the \gls{bob} representation\cite{hansen_machine_2015} bags different off-diagonal elements of the Coulomb matrix into bags depending on the combination of nuclear charges and has then been extended to higher-order interactions in the \gls{baml} representation.\cite{huang_communication_2016}
A main motivation behind this approach, which has been generalized in the \gls{mbtr} framework,\cite{huo_unified_2017} is to have a more natural notion of chemical similarity than the Coulomb repulsion terms.
One problem with building bags is that they are not of fixed length and hence need to be padded with zeros to make them applicable for most \gls{ml} algorithms.

An alternative method to record pairwise distances, that is familiar to chemists form \gls{xrd}, is the \gls{rdf}, $g^{(2)}(r)$.
Here, pairwise distances are recorded in a binned fashion in histograms. This representation inspired  Schuett et al.\ to build a \gls{ml} model for the \gls{dos}.\cite{schutt_how_2014}
They use a matrix of partial \glspl{rdf}, i.e., a separate \gls{rdf} for each element pair---similar to how the element pairs were recorded in different bags in the \gls{bob} representation and quite similar to Valle's crystal fingerprint\cite{valle_crystal_2010} in which modified \glspl{rdf} for each element pair are concatenated.

Von Lilienfeld et al.\ took inspiration in the plane-wave basis sets of electronic structure calculations, which remove many problems that local (e.g., Gaussian) basis sets can cause, e.g., Pulay forces and basis set superposition errors, and created a descriptor that is a Fourier series of atomic \glspl{rdf}.
Most importantly, the Fourier transform removes the translational variance of local basis sets---which is one of the main requirements for a good descriptor.\cite{lilienfeld_fourier_2015}
The Fourier transform of the \gls{rdf} also is directly related to the \gls{xrd} pattern which has found widespread use in \gls{ml} models for the classification of crystal symmetries.\cite{park_classification_2017,vecsei_neural_2018,ziletti_insightful_2018}

For the prediction of gas adsorption properties property labeled \glspl{rdf} have been introduced by Fernandez et al.\cite{fernandez_atomic_2013}
The property labeled \gls{rdf} is given by
\begin{equation}
    \text{RDF}^\text{P} = f \, \sum_{i,j} P_i P_j  \exp\left[-B {\left(r_{ij} - R\right)}^2 \right],
\end{equation}
where \(P_i\) and \(P_j\) are elemental properties of atom \(i\) and \(j\) in a spherical volume of radius \(R\). \(B\) is a smoothing factor and \(f\) is scaling factor.
It was designed based on the insight that for some type of adsorption processes, like \ce{CO2} adsorption, not only the geometry but also the chemistry is important.
Hence, they expected that stronger emphasis on e.g.\ the electronegativity might help the \gls{ml} model.

\paragraph{Structure Graphs}\label{sec:structure_graphs}
Encoding structures in the form of graphs, instead of using explicit distance information, has the advantage that the descriptors can also be used without any precise geometric information, i.e., a geometry optimization is usually not needed.
In structure graphs, the atoms define the nodes and the bonds define the edges of the graph.
The power of such descriptors was demonstrated by Kulik and co-workers in their work on transition metal complexes.
They introduced the \gls{rac} functions\cite{janet_resolving_2017} (which is a local descriptor that correlates some atomic heuristics, like the atom type, on the structure graph) and used it to predict for example metal-oxo formation energies,\cite{nandy_machine_2019} or the success of electronic structure calculations.\cite{duan_learning_2019} Recently, they also have been adapted for \glspl{mof}.\cite{moosavi_understanding_2020}

For crystals, Xie and Grossmann built a \gls{gcnn} that directly learns from the crystal structure graph (cf.\ section~\ref{sec:graph_model}) and could predict a variety of properties such as formation energy or mechanical properties as the bulk moduli for structures from the Materials Project.\cite{xie_crystal_2018, xie_hierarchical_2018}
This architecture also allowed them to identify chemical environments that are relevant for a particular prediction.

\paragraph{Distance-Matrix Based Descriptors}\label{sec:distance_matrix_based}
Another large family of descriptors is built around different encodings of the distance matrix.
Intuitively, one might think that a representation such as the \(z\)-matrix, which is popular in quantum chemistry and is written in terms of internal coordinates, might be suitable as input for a \gls{ml} model.
And indeed, the \(z\)-matrix is translational and rotational invariant due to the use of internal coordinates---but it is not permutational invariant, i.e., the ordering matters.
This was also a problem with the original formulation of the Coulomb matrix which encodes structures using the Coulomb repulsion of atomic charges (proton count \(Z\)) on the off-diagonal and rescaled atomic charges on the diagonal:\cite{faber_crystal_2015}
\begin{equation}
    x_{ij} =
    \begin{cases}
        0.5\, Z_i^{2.4}                                                & i =j      \\
        Z_i\, Z_j \, \phi \left(\|\mathbf{r}_i - \mathbf{r}_j\|\right) & i \neq j,
    \end{cases}\label{eq:couloumb_matrix}
\end{equation}
as one structure could have many different Coulomb matrices, depending on where one starts counting.
The Coulomb matrix shares this problem with the older Weyl matrix,\cite{weyl_classical_1946} which is a \(N \times N\) matrix composed of inner products of atomic positions, and in this way also an overcomplete set.
To remedy this problem it was suggested to use sorted Coulomb matrices or the eigenvalue spectrum (but this violates the uniqueness criterion as there can be multiple Coulomb matrices with the same eigenspectrum).
Also, to be applicable to periodic systems, eq.~\ref{eq:couloumb_matrix} needs to be modified.

To deal with electrostatic interactions in molecular simulations, one usually uses the Ewald-summation technique which splits one non-converging infinite sum into two converging ones.
This trick has also been used to deal with the infinite summations which would occur if one attempted to use eq.~\ref{eq:couloumb_matrix} for periodic systems---the corresponding descriptor is known as the Ewald sum matrix.\cite{faber_crystal_2015}
The sine-Coulomb matrix is a more \textit{ad hoc} solution to apply the Coulomb matrix to periodic systems.
Here, the off-diagonal terms are calculated using a modified potential \(\phi\) that introduces periodicity using a sine over the product of the lattice vectors and the vector between the two sites \(i\) and \(j\).\cite{faber_crystal_2015}

\paragraph{Point Cloud Based}\label{sec:persistent_homology}
In object recognition much success has been achieved by representing objects as point clouds.\cite{maturana_voxnet_2015, qi_pointnet_2016}
This can also be applied to materials science, where solids can be represented as point clouds by sampling the structures with \(n\) points.
This point cloud can be then further processed to generate an input for a (supervised) \gls{ml} algorithm.
Such processing is often needed because most algorithms cannot to deal with irregular data structures, like point clouds, wherefore the data is often mapped to a grid.

\begin{figure}
    \centering
    \includegraphics[width=\textwidth]{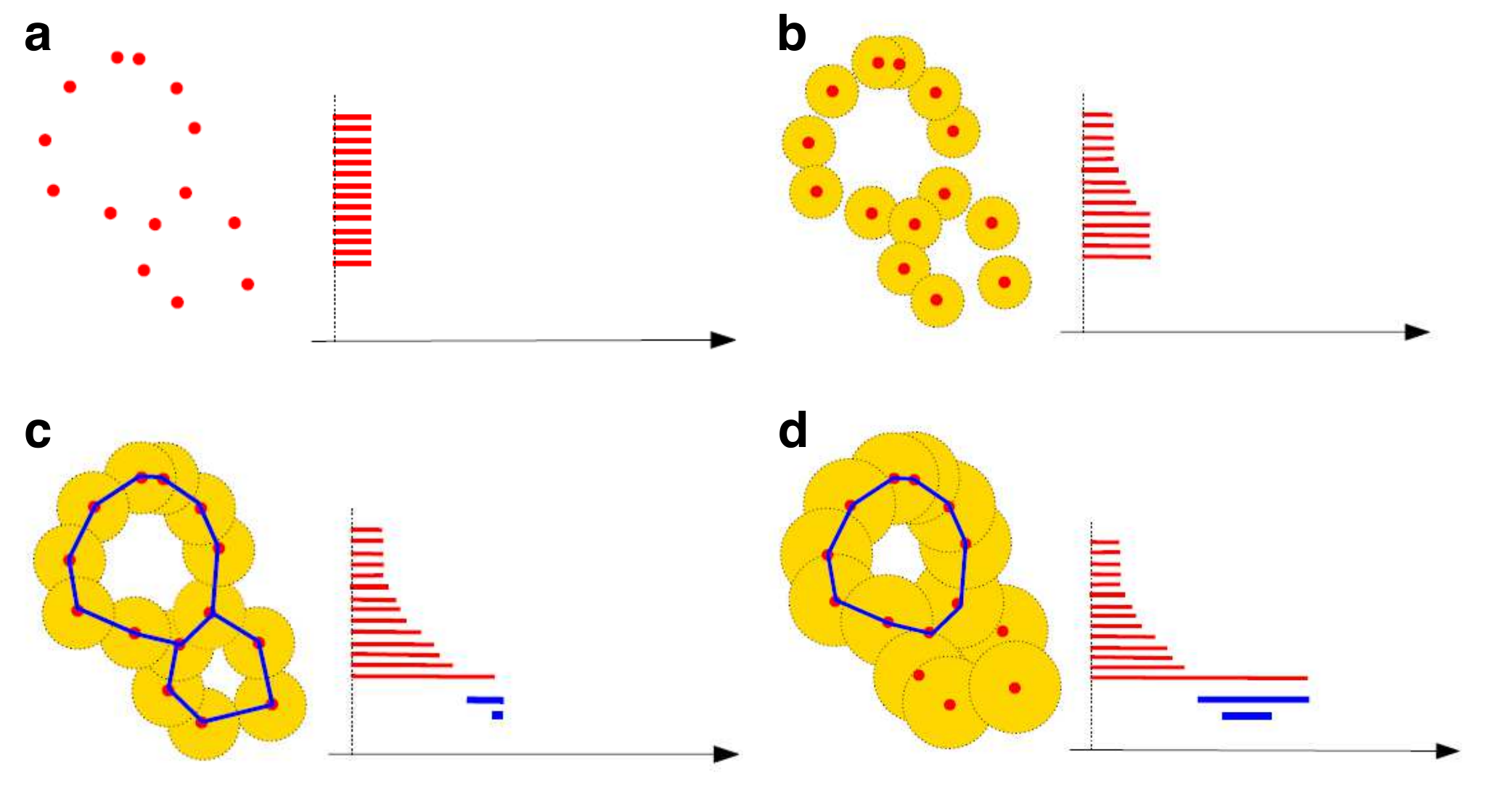}
    \caption{Illustration of the filtration of the distance function of a cloud of points.
        For the birth of each point, we create an interval (bar) in the persistence diagram.
        As we increase the radius of the points, some components die (and merge) as the circles start to overlap. The persistence diagram takes track of this by putting an end to the interval (b).
        As the radius of the circle further increases, we form new, one-dimensional, connected components (the holes, blue in c) and all the intervals associated with single points come to an end.
        The only interval that never dies is due to the union of all points.
        The figure is a modified version of the illustration from Chazal and Michel.\protect{\cite{chazal_introduction_2017} }}\label{fig:filtration}
\end{figure}

\subparagraph{Topological Data Analysis}\label{sec:tda}
A fruitful approach to generate features from point clouds is to use the persistence homology analysis rooted in \gls{tda}.\cite{weinberger_what_2011, chazal_introduction_2017}
Here, the underlying topological structures are extracted using a process called filtration. In a filtration one uses using a sequence of growing spaces, e.g., using balls of growing radii, to understand how the topological features change as a function of the radius.
A persistence diagram records when a topological feature is created or destroyed.
This is shown in Figure~\ref{fig:filtration} where at some radius the first circles start to overlap, which is reflected in the end of a bar in the persistence diagram. Then, the circles form two holes (c), which is reflected with the birth of new bars that die with increasing radius, when the holes disappear (d).

Using this technique has recently become even easier with the \texttt{scikit-tda} suite of packages,\cite{nathaniel_saul_scikit-tda/scikit-tda_2019} which gives an easy-to-use Python interface to the C++ \texttt{Ripser} library\cite{tralie_ripser.py_2018} and functions to plot persistent images\cite{adams_persistence_2015} and diagrams.

\begin{figure}
    \centering
    \includegraphics[width=\textwidth]{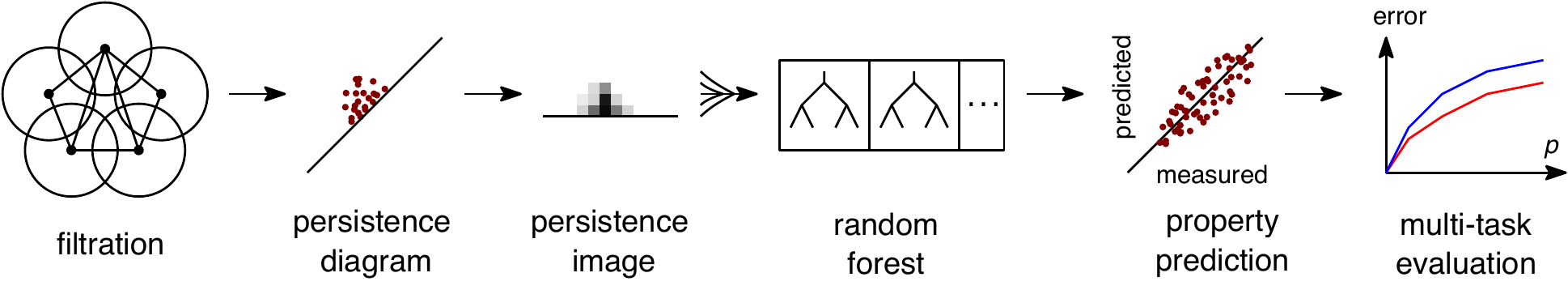}
    \caption{Illustration of the scheme used to predict gas adsorption properties using persistent images.
        A filtration is used to create a persistence diagram (as illustrated in Figure~\ref{fig:filtration}). This is then transformed into a persistence image that is used to train a \gls{rf} model to predict the methane uptake. Figure redrawn based on ref. \citep{krishnapriyan_robust_2020}}\label{fig:persistent_images}
\end{figure}

Unfortunately, most \gls{ml} algorithms only accept fixed length inputs, wherefore the persistent homology barcodes cannot directly be used as descriptors.
To work around this limitation, Lee and co-workers\cite{zhang_machine_2019} used a strategy that is similar to the general strategy for creating fix-length global descriptors that we discussed above, namely by computing statistics of the persistent homology barcodes (cf.\ section~\ref{sec:applications}). \\
Alternative finite-dimensional representation are persistence images,\cite{adams_persistence_2015} which have recently employed by  Krishnapriyan et al.\ to predict the methane uptakes in zeolites between \SIrange{1}{200}{\bar} (cf.\ Figure~\ref{fig:persistent_images}).\cite{krishnapriyan_robust_2020}
In persistence images, the birth-death pairs \((b,d)\), which are shown in persistence diagrams, are transformed into birth-persistence pairs \((b, d-b)\) which are spread using a Gaussian.
The images are then created by binning the function of \((b, d-b)\).  Krishnapriyan et al.\ then used \glspl{rf} to learn from this descriptor, but it might also be promising to investigate the use of transformations of the homology information that can be learned during training (e.g., using \glspl{nn}, see section~\ref{sec:messagepassing}).\cite{hofer_learning_2019}

The capabilities of \gls{tda} have been demonstrated in the high-throughput screening of the nanoporous materials genome.\cite{lee_high-throughput_2018,lee_quantifying_2017}
Here, the \texttt{zeo++} code has been used to analyze the pore structure of zeolites (using Voronoi tessellations), which then could be sampled to create point clouds that were used as an input for a persistent homology analysis, which output was summarized in persistence diagrams (\enquote{barcodes}).
The similarity between these persistence diagrams was then used to rank the materials, i.e., if the persistence diagram of one structure is similar to a high-performing structure, it is likely to also perform well. As Moosavi, Xu et al.\ recently showed, the similarity between barcodes can also be used to build kernels for \gls{krr} which then can be used to predict the performance for methane storage applications.\cite{moosavi_geometric_2020}

\subparagraph{Neural-Network Engineered Features}\label{sec:nn_feat_tda}
A promising alternative to \gls{tda} is to use specific \gls{nn} architectures such as PointNet that can directly learn from point cloud inputs.\cite{qi_pointnet_2016}
DeFever et al.\ used the PointNet for a task similar to object recognition: the classification of local structures in trajectories of molecular simulations.\cite{defever_generalized_2019}
Interestingly, the authors also demonstrated that one can use PointNet to create hydrophilicity maps, e.g., for self-assembled monolayers and proteins.

\paragraph{Coarse Tabular Descriptors}
Our discussion so far guided us from atomic-level descriptors to more coarse, global descriptors.
In this section, we will explore some more examples of such coarse descriptors.
Those coarse descriptors are frequently used in top-down modeling approaches, where a model is trained on experimental or high-level properties.
Obviously, such coarse, high-level descriptors are  not suited to describe properties with atomic resolution, e.g., to describe a \gls{pes}, but they can be efficient to model, for example, gas adsorption phenomena.

\subparagraph{Based on Elemental Properties}
Widely used in this context are compositional descriptors that encode information about the chemical elements a compound is made up of.
Typically, one finds that simple statistics such as sums, differences, minimums, maximums or covariance of elemental properties such as electronegativity or covalent radii are calculated and used as feature vectors.
There has been some success with using such descriptors for perovskites\cite{balachandran_predictions_2018, bartel_new_2019}, half-Heussler compounds\cite{legrain_materials_2017}, analysis of topological transitions,\cite{acosta_analysis_2018} the likelihood of substitutions\cite{singh_phenomenology_1982, hautier_data_2011} as well as the conductivity of \glspl{mof}.\cite{he_metallic_2018}
Generally, one can expect such descriptors to work if the target property is directly related to the constituent elements.
A prime example of this concept are perovskites for which there are empirical rules, like the Goldschmidt tolerance factor, that relate the radii of the ions to the stability, wherefore it is reasonable to expect that one can build meaningful \gls{ml} models for perovskite stability, that outperform the empirical rules, with ion radii as features.

\subparagraph{Cheap Calculations Crude Estimates of Target and Experimental Features}
Especially for our case-study problem, the gas adsorption in porous materials, tabular descriptors that are based on cheap calculations (e.g., geometry analysis, energy grids) are most commonly used.
As gas adsorption requires that the pore properties are \enquote{just right} it is natural to calculate them and use them as features.\cite{fernandez_large-scale_2013, gulsoy_analysis_2019, fanourgakis_robust_2019, bobbitt_molecular_2019}
Especially, since we know that target similarity governs the error of \gls{ml} models.\cite{huang_communication_2016}
Typically, such descriptors, as the \gls{psd},\cite{pinheiro_characterization_2013}, and accessible surface areas or pore volumes, can be computed with programs as \texttt{Zeo++},\cite{willems_algorithms_2012} \texttt{Poreblazer}\cite{sarkisov_computational_2011} or \texttt{MOFomics}/\texttt{ZEOMICS}.\cite{first_mofomics_2013, first_computational_2011}

A cheaper calculation was also used by Bucior et al.\ to construct descriptors. On a coarse grid they computed the interactions between the adsorbate and the framework, summarized this data in histograms and then used these histograms to construct \gls{ml} models for the adsorption of \ce{H2}.\cite{bucior_energy-based_2019}
This is the related to the approach Zhang and Ling put forward to use \gls{ml} on small datasets.\cite{zhang_strategy_2018}
They suggest including crude estimates of the target property into the feature set.
As an example, they included force-field derived bulk moduli to predict bulk moduli on \gls{dft} level of theory.
This idea is directly related to \(\Delta\)-ML and co-kriging approaches which we will discuss below in more detail.

Especially when one uses a large collection of tabular features it can be useful to curate feature dictionaries, which describe what the feature means and why it is useful---to aid collaboration and model development.

\subparagraph{Using Building Blocks as Features}
For materials like \gls{mof}, \gls{cof}, or also polymers that are constructed by self-assembly of simpler building blocks, one can attempt to directly use the building blocks as features.
Here, one typically one-hot encodes the presence of building blocks with ones and the absence with zeros. Therefore, there will be as many columns in the feature matrix as there are building blocks.
Due to the nature of this encoding, such a model cannot generalize to new building blocks.
This featurization was for example used by Borboudakis et al.\ who one-hot encoded linker and metal node types to learn gas adsorption properties of \glspl{mof} from a small database.\cite{borboudakis_chemically_2017}
Recently, Fanourgakis et al.\ reported a more general approach in which they use statistics over atom types (e.g., minimum, maximum and average of triple bonded carbon per unit cell), that would usually be used to set up force field topologies, as descriptors for \gls{rf} models to predict the methane adsorption in \glspl{mof}.\cite{fanourgakis_universal_2020}

\subsection{Feature Learning}
\subsubsection{Feature Engineering}
A key insight is that the \enquote{raw} features are often not the best inputs for a \gls{ml} model.
Therefore, it can be useful to transform the features.
This is also what every chemist or modeler already intuitively knows: Some phenomena like the dependence of the activation energy on the diffusion constant are better visible after a logarithmic transformation.
Sometimes it is also more meaningful to look at ratios, like the Goldschmidt tolerance ratio, rather than at the raw values.

The term feature engineering describes this process where new features are formed via the combination and/or mathematical transformation of raw features.
And this is one of the main avenues for domain knowledge to enter into the modeling process.
One approach to automate this process is to automatically try different mathematical operations and transformation functions as well as combinations of features.
Unfortunately, this leads to an exponential growth of the number of features and the modeler now faces the problem to select the best features to avoid the curse of dimensionality (cf.\ section~\ref{sec:curse_dimensionality}), which is not a trivial problem.
In fact, the featurization process is equivalent to finding the optimal basis set for the description of a physical problem.

\subsubsection{Feature Selection}
\begin{figure}
    \includegraphics[width=\textwidth]{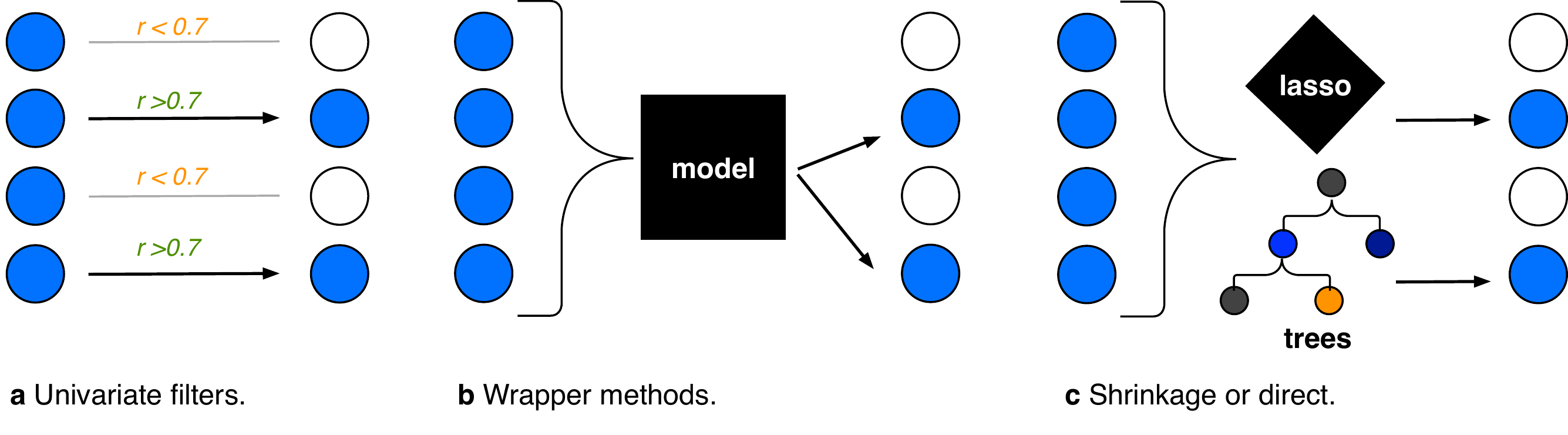}
    \caption{Overview of different feature selection strategies.
        Figure redrawn based on an illustration by Janet et al.\protect{\cite{janet_resolving_2017}}}\label{fig:feature_selection}
\end{figure}

For some phenomena one would like to develop \gls{ml} models but it might not be \textit{a priori} clear which descriptors one should use to describe the phenomenon, e.g., because it is a complex multi-scale problem.
Intuitively, one might try all possible combinations of descriptors that one can come up with to find the smallest, most informative set of features to avoid the curse of dimensionality.
But this approach is deemed to fail as it is a \gls{np} hard problem. This means that a candidate solution for this problem can be verified in polynomial time, but that the solution itself can probably not be found in polynomial time.
Hence, approximations or heuristics are needed to allow us to make the problem computationally tractable.
One generally distinguishes three approaches to tackle this problem:
First, simple filters can be used to filter out features (e.g., based on correlation with the target).
Second, iterations in wrapper methods (pruning, recursive feature elimination) can be used to find a good subset, or one can attempt to directly include the objective of minimizing the dimensionality in the loss function.\cite{guyon_introduction_2003, saeys_review_2007, janet_resolving_2017, imbalzano_automatic_2018}

\paragraph{Filter Heuristics}
Given a large set of possible features one can use some heuristics to compact the feature set.
A simple filter is to use the correlation, mutual information\cite{vergara_review_2014} or fitting errors for single features as surrogates and only use the features that show the highest correlation or mutual information with the target or the ones for which a simple model shows the lowest error.
Obviously, this approach is unable to capture interaction effects between variables.

Another heuristic that can be used to eliminate features is to eliminate those that do not show a lot of variance (\texttt{VarianceThreshold} in \texttt{sklearn}). The intuition here is that (nearly) constant features cannot help the model to distinguish between labels.

This is to some extent similar to \gls{pca} based feature engineering, where one tries to find the linear combinations of features that describe most of the variance and then only keeps those principal components.
This approach has the drawback that arbitrary linear combinations are not necessarily physically meaningful and that explaining the variance does not necessarily mean being predictive.

\paragraph{Wrapper Approaches}
Often, one also finds stage-wise feature selection approaches.\cite{kursa_feature_2010}
Either by weight pruning, i.e., by fitting the model on all features and then removing those with low weights or by \gls{rfe}.
\Gls{rfe} starts by fitting a model on all features and then iteratively removes the least important features until a desired number of features is reached.
This iterative procedure is needed because the feature importance can change after each elimination, but it is computationally expensive for moderately sized feature sets.
The opposite approach, i.e., the iterative addition of features is known as \gls{rfa} and is often used in conjunction with \gls{rf} feature importance, which is used to decide which features should be included.
This approach was for example used in a work by Kulik and co-workers in which they built models to predict metal-oxo formation energies, which are relevant for catalysis.
In doing so, they found that they can reduce the size feature set from ca.\ 150 to 22 features using \gls{rf}-\gls{rfa} which led to reduction of the \gls{mae} on the test set from \SI{9.5}{\kilo cal\per\mole}  to \SI{5.5}{\kilo  cal\per\mole}.\cite{nandy_machine_2019}

\paragraph{Direct Approximations: LASSO/Compressed Sensing}\label{sec:lasso}
As an alternative to iterative approaches, there are efforts to use objective functions that directly describe both modeling goals: First, to find a model that minimizes the error and, second, to find a model that minimizes the number of variables (following Occam's razor, cf.\ section~\ref{sec:curse_dimensionality}).
In theory, this can be achieved by adding a regularization term \(\| \mathbf{w} \|^p = \left(\sum_{i=1}^n w_i^p \right)^{1/p}\) to the loss function and attempting to find the coefficients \(\mathbf{w}\) that minimize this loss function.
In the limit \(p=0\), there is nothing won as it is the \gls{np} hard problem of minimizing the number of variables, we mentioned above.\cite{ghiringhelli_learning_2017}
Hence, the \(l_1\) norm (also known as Taxicab or Manhattan norm), i.e., the case \(p=1\), is often used as an approximation (to relax the \(l_0\) condition).\cite{hastie_statistical_2015}
This has the advantage that the optimization is now convex and that the edges of the regularization region tend to favor sparsity (cf.\ Figure~\ref{fig:lasso_ridge} and accompanying discussion for more details).
The minimization of the \(l_1\) known is known in statistics as the \gls{lasso} and widely used to avoid overfitting (regularization), by penalizing high weights (cf.\ section~\ref{sec:regularization}).\cite{hastie_statistical_2015}
Compressed sensing\cite{nelson_compressive_2013} uses this idea to recover a signal with only a few sensors while giving conditions on the design matrix (with materials in the rows and the descriptors in the columns) for which the \(l_0\) and the \gls{lasso} solution will likely coincide.
An in-depth discussion of the formalism of feature learning using compressed sensing is given by Ghiringhelli et al.\cite{ghiringhelli_learning_2017}
This approach works well in materials science as many physical problems are sparse and it also works well with noise, which is also common to physical problems.\cite{nelson_compressive_2013}
Ghiringhelli et al.\ applied this idea to materials science but also highlighted that a procedure based only on the \gls{lasso} has difficulties in selecting between correlated features and dealing with large feature spaces.\cite{ghiringhelli_big_2015}
With \gls{sisso} Ouyang et al.\ add a \gls{si} layer before the \gls{lasso}.\cite{ouyang_sisso_2018}
This \gls{si} layer pre-selects a subspace of features that show the highest correlation with the target and that can then be further compressed using the \gls{lasso}.
This approach, for which open-source code was published,\cite{ouyang_rouyang2017/sisso_2019} allowed Scheffler and co-workers to construct massive sets of \num{E9} descriptors using combinations of algebraic functions applied primary features, like the atomic radii, and to discover new tolerance factors for the stability of perovskites\cite{bartel_new_2019} or to predict new quantum spin-Hall insulators using interpretable descriptors.\cite{acosta_analysis_2018}

Another approach to the feature selection problem uses projected gradient descent to locally approximate the minimization of the \(l_0\) norm.\cite{xiang_simultaneous_2014}
It is efficient as it uses the gradient and it achieves sparsity by, stepwise, setting the smallest components of the weights vector \(\mathbf{w}\) to be zero (cf.\ Listing~\ref{lst:projected_gradient_descent} for pseudocode).\cite{keys_iterative_2017, jain_iterative_2014}
\begin{lstlisting}[language=Python, caption={Pseudo-code for iterative hard thresholding (also known as projected gradient descent).}, label=lst:projected_gradient_descent]
i = 0
weights_i = intitialize_weights()
k = 5 # hyperparameter that needs to be tuned

while halting condition not fulfilled:
    # select a stepsize for gradient descent
    eta[i] = chose_stepsize() 
    # perform the gradient descent step
    chi[i] = weights_i - eta[i] * gradient(lossfunction(weights[i])) 
    # select the largest component
    weights[i+1] = select_k_large_components(chi[i], k) 
    # update the counter
    i = i + 1
\end{lstlisting}
A modified version was also used Pankajakshan et al.\cite{pankajakshan_machine_2017, kumar_machine_2019}
They combined this feature selection method with clustering (to combine correlated features) and created a representative feature for each cluster, which they then used in the projected gradient algorithm to compress the feature set. Additionally, they also employed the bootstrap technique to make their selection more stable.

The bootstrapping step is also the key to another method known as stability selection.
Here, the selection algorithm (e.g., the \gls{lasso}) is run on different bootstrapped samples of the dataset and only those features that are important in every bootstrap are selected, which can help to counter chance correlation.\cite{meinshausen_stability_2008}
This is currently being implemented as randomized \gls{lasso} in the \texttt{sklearn} Python framework.

\subsubsection{Data Transformations}
An additional problem with features is that their distribution or the scale on which they are on (e.g., due to the choice of units) might not be appropriate for \gls{ml}.
One of the most important reasons to transform data is to improve interpretability.
Some features are more natural to think about on a logarithmic scale (e.g., the concentration of protons is known as \(\text{pH} =  - \lg_{10} \ce{H3O^+}\) in chemistry and also the Henry coefficient is naturally represented on logarithmic scale), or reciprocal scale (e.g., temperature in the case of Arrhenius activation energy analysis).
In other cases, the underlying algorithm will profit from transformations, e.g., if it assumes a particular distribution for the data (e.g., the archetypal linear regression assumes a normal distribution of the residuals).
The most widely used transformations are power transformations like the Box-Cox (defined as \((x^\lambda -1)/\lambda\) for \(\lambda > 0\), \(\ln x\) for \(\lambda =0\), where  \(\lambda\) can be used to tune the skew),\cite{box_analysis_1964} the inverse hyperbolic sine\cite{burbidge_alternative_1988, friedline_transforming_2015} or the Yeo-Johnson transformation which all aim to make the data more normally distributed.
The Box-Cox transformation, or a simple logarithmic transformation (\(\lg x\)), are the most popular techniques, but the inverse hyperbolic sine and the Yeo-Johnson transformation have the advantage that they can also be used on negative values.

\begin{figure}
    \centering
    \includegraphics[width=.7\textwidth]{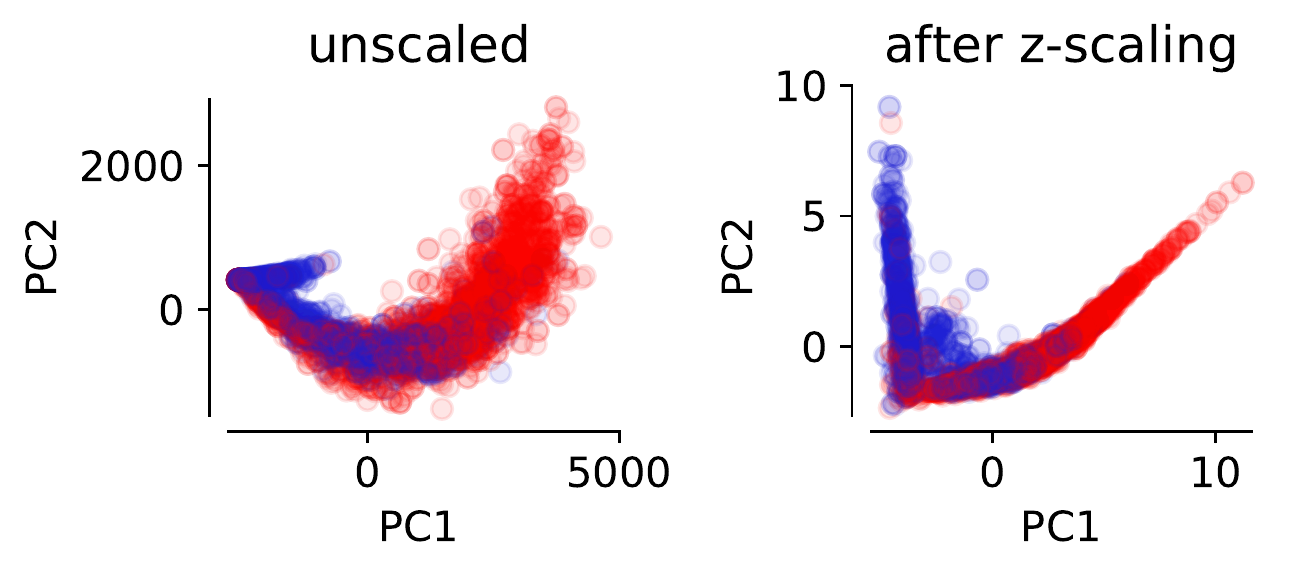}
    \caption{Influence of scaling, here standard scaling (z-scaling), of features on the dimensionality reduction using \gls{pca}.
        For this example we used the data from Boyd et al.\ and performed the \gls{pca} on the feature matrix of pore properties descriptors and plot the data in terms of the first two principal components (PC1, and PC2).
        We then color code the structures with above-median \ce{CO2} uptake (red) different from those with below-median \ce{CO2} uptake (blue) and plot the points in random order.
        It is observable that the separation after scaling is clearer.}\label{fig:pca_scaling}
\end{figure}

\paragraph{Normalization and Standardization}
In the following, we will show that many algorithms perform interference by calculating distances between examples.
But in the physical world, our features might have different scales, e.g., due to the arbitrary choice of units.
Surface areas might be recorded as numbers in the order of \num{E3} and void fractions as numbers on the order of \num{E-3}.
For \gls{ml} one wants to remove such influences from the model, as illustrated in Figure~\ref{fig:pca_scaling}.
Also, optimization algorithms will have problems if different directions in feature space have different scales. This is intuitive if we look at the gradient descent update step, where the values of the features, \(x_i\), are directly involved and for which reason some weights might update faster than others (using a fixed learning rate \(\eta\)).

The most popular choices to remedy these problems are min-max scaling and standard scaling (\(z\)-score normalization).
Min-max scaling transforms features to a range between zero and one (by subtracting the minimum and dividing by the range), and in this way minimizes the effect of outliers.
In contrast to that, the standard scaling transforms feature distributions to distributions centered around zero and unity variance by subtracting the mean and diving by the standard deviation.
Note that by using this transformation we do not bind the range of features, which can be important for some analyses like \gls{pca}, which work on the variance of the data.

In case there are many outliers or strong skew, it might be more reasonable to scale data based on robust estimators of centrality and spread, like subtracting the median and dividing by the interquartile range (this is implemented as \texttt{RobustScaler} in \texttt{sklearn}).

It is important that those transformations need to be applied to training and test data---but using the distribution parameters \enquote{learned} from the training set.
If we computed those parameters also on the test set we would risk data leakage, i.e., provide information about the test data to the model.

\paragraph{Decorrelation}\label{sec:correlation}
Often, one finds oneself in a position where the initial feature set contains multiple variables that are highly correlated with each other, like gravimetric and volumetric pore volumes or surface areas.
Usually, it is better to remove those correlations.
The reasoning behind this is that multicolinearity usually means that there is data redundancy, which violates the minimum description length principle we discussed above (cf.\ section~\ref{sec:curse_dimensionality}).
In particular severe cases, it can make the predictions unstable (and also the feature selection as we discussed above) and in general it undermines causal interference as it is not clear which of the correlated variables is the reason for a particular prediction.\cite{dormann_collinearity_2013, cronin_pitfalls_2003}

Widespread ways to estimate the severity of multicolinearity is to use pair-correlation matrices or the \gls{vif}, which estimates how much of the variance is inflated by colinearity with other features.\cite{roy_chapter_2015, james_introduction_2013}
It does this by predicting all the features using the remaining features \(\text{VIF} = 1/(1-R_i^2)\), where \(R_i\) is the coefficient of determination for the prediction of feature \(i\).
A \gls{vif} of ten would mean that the variance is ten times larger than it would be for fully orthogonal features.

%% file: main/3_learning.tex
\section{How To Learn: Choosing a Learning Algorithm}\label{sec:learning_algorithms}

\begin{figure}
    \centering
    \includegraphics[width=\textwidth]{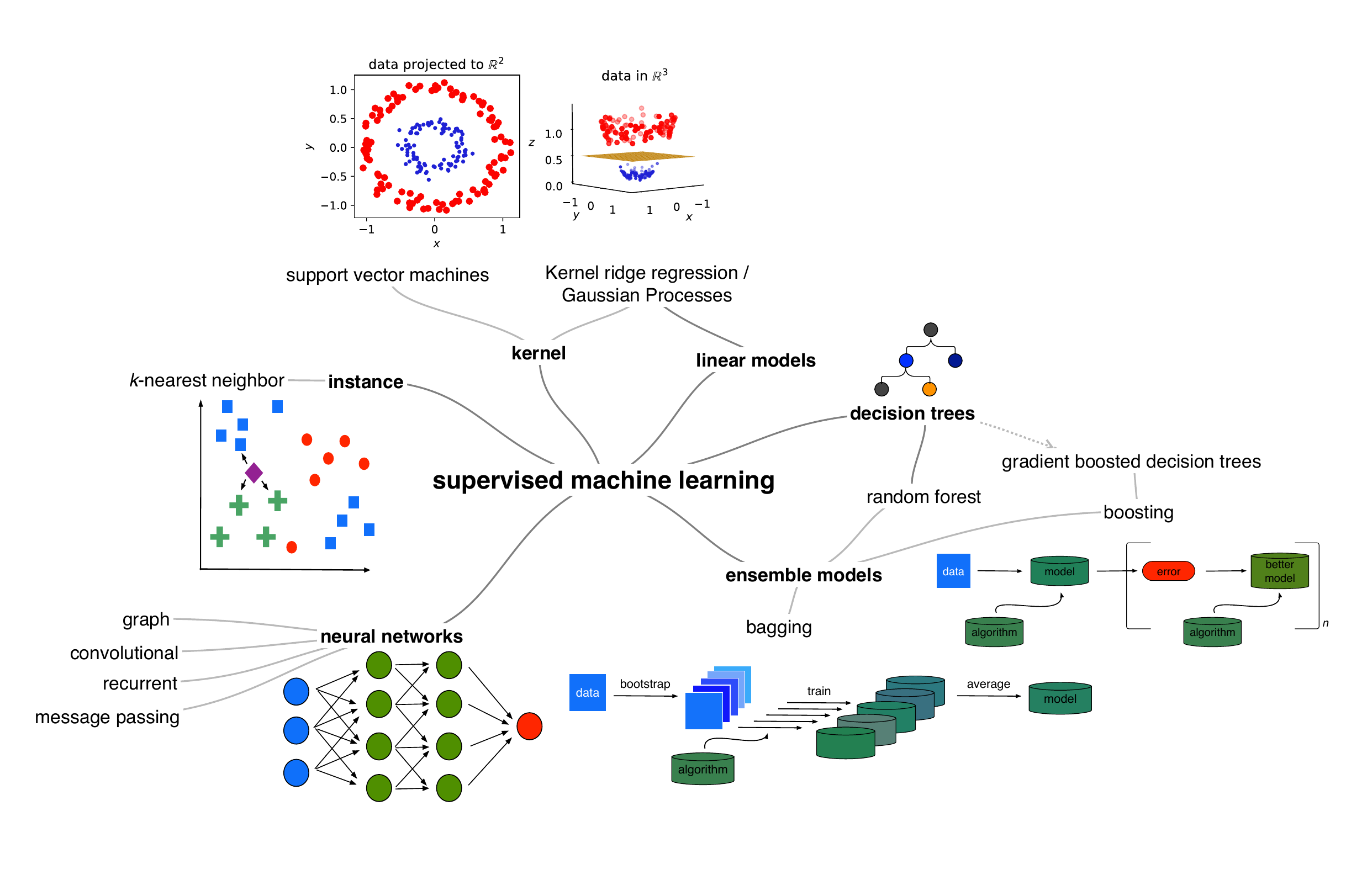}
    \caption{Overview of the supervised \gls{ml} algorithm landscape. We do not distinguish between classification and regression as many of the algorithms can be formulated both for regression and classification problems.}\label{fig:learner_landscape_mindmap}
\end{figure}

After data selection (cf.\ section~\ref{sec:datasource} and featurization (cf.\ section~\ref{sec:featurization}) one can proceed to training a \gls{ml} model.
But also here, there are a lot of choices one can make.
In Figure~\ref{fig:learner_landscape_mindmap} we give a non-exhaustive overview of the learning algorithm landscape.

In the following, we discuss some rules of thumb that can help to choose the appropriate algorithm for a given problem and discuss the principles of the most popular ones.
Typically, we will not distinguish between classification and regression as many algorithms can be formulated both for regression and classification problems.

\begin{figure}
    \centering
    \includegraphics[]{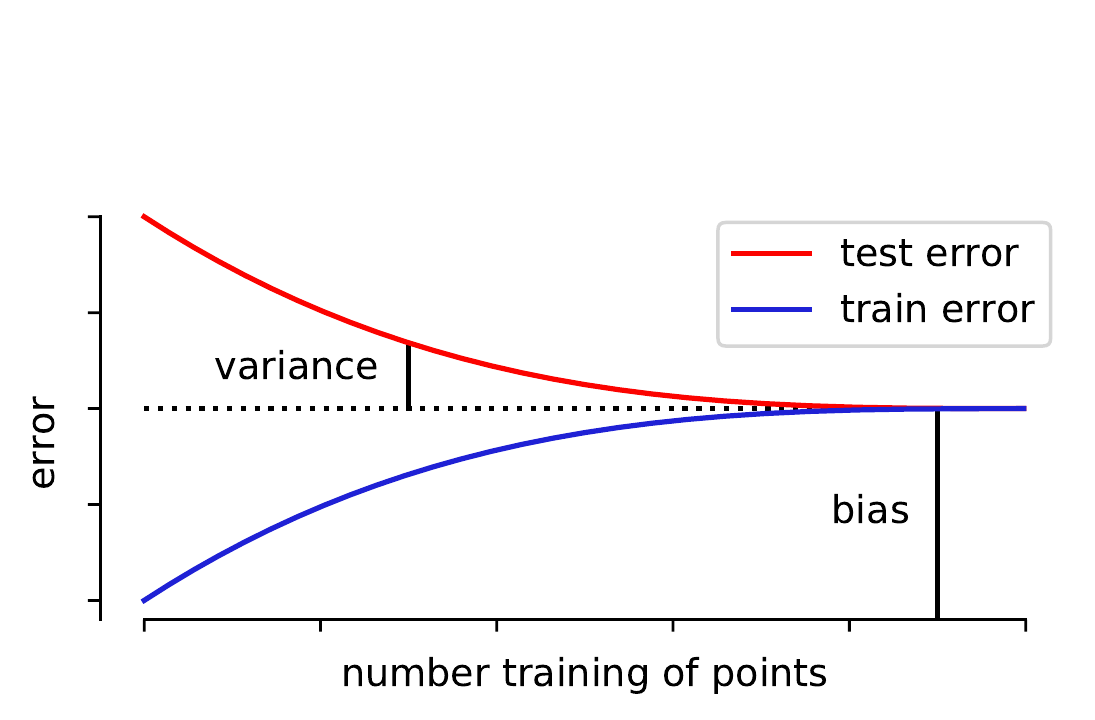}
    \caption{Train and test error as a function of the number of training points and the definition of bias and variance; bias being the error that remains on the training set for an infinite number of training points, and variance the error due to the finite size of the training set.}\label{fig:bv_explaination}
\end{figure}

\paragraph{Principles of Learning}
One of the main principles of statistical learning theory is the bias-variance decomposition (cf.\ eq.~\ref{eq:bias_variance}), which describes that the total error can be described as the sum of squared bias, variance and an irreducible error (Bayes error)
\begin{equation}
    \text{error} = \text{bias}^2 + \text{variance} + \text{Bayes error},\label{eq:bias_variance}
\end{equation}
and can easily be derived by rewriting of the cost function for the mean square error.\cite{mehta_high-bias_2019}
The variance of a model describes the error due to finite training size effects, i.e., how much the estimation fluctuates due to the fact that we need to use a finite number of data points for training and testing (cf.\ Figure~\ref{fig:bv_explaination}).
The bias is the difference between the prediction and the expectation value; it is the error we would obtain for an infinite number of training points (cf.\ Figure~\ref{fig:bv_explaination}).
In this case, the bias represents the limit of expressivity for our model, e.g., that the order of the polynomial is not high enough to describe the problem that should be modeled. But this error could in principle be removed by choosing a better model.
All the remaining error, which cannot be removed by building a better model, is for example due to noise in the training data. For this reason, this term is called irreducible error (also known as Bayes error).
\begin{figure}
    \centering
    \includegraphics{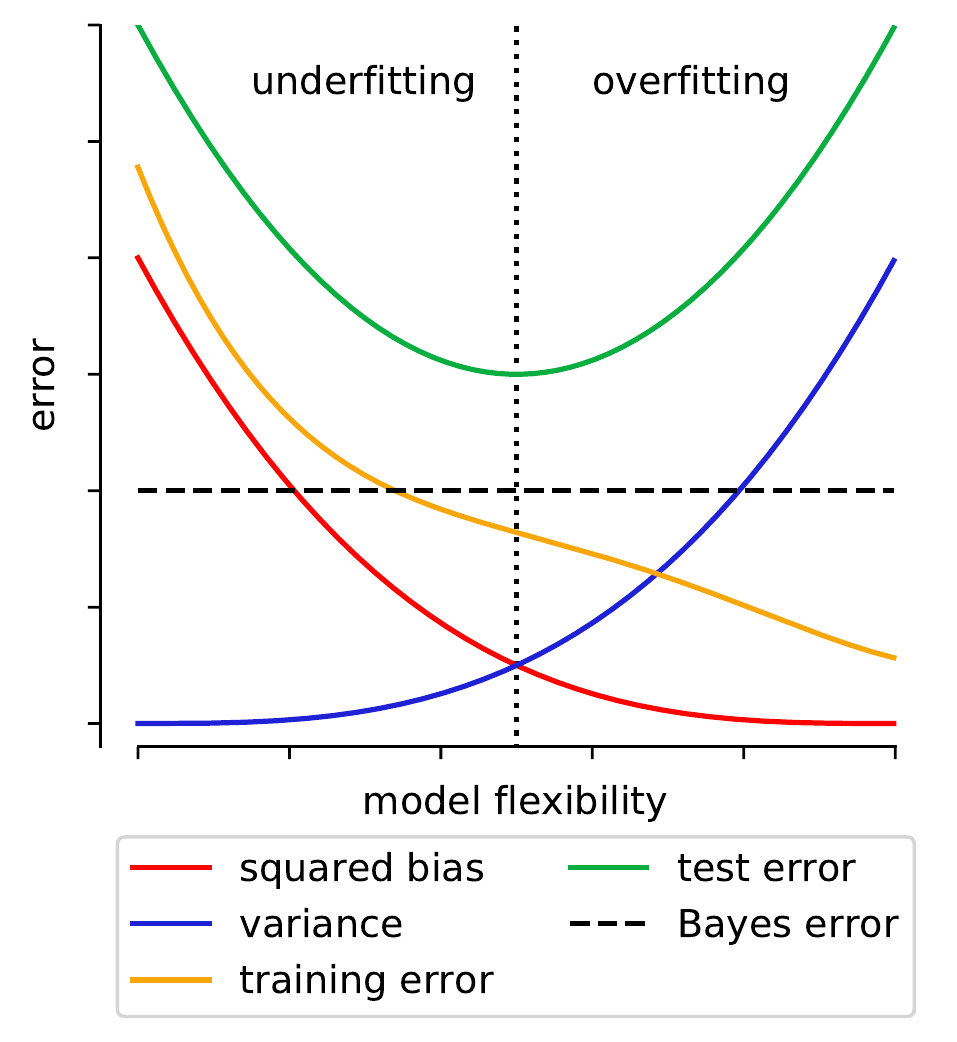}
    \caption{Bias, variance, training, and test error as well as Bayes error (irreducible error) as function of the model flexibility. }\label{fig:bv_tradeoff}
\end{figure}

This trade-off between bias and variance is directly linked to model flexibility.
A highly flexible model, which is also often less interpretable, like a high-order polynomial, tends to have a high variance whereas a simple model, such as a regularized linear regression, tends to have a high bias (cf.\ Figure~\ref{fig:bv_tradeoff}).
In practice, it is often useful to first create a model that overfits, hence has close to zero training error, and in this way ensure that the expressivity is high enough to model the phenomenon.
Then, one can use techniques which we will describe in section~\ref{sec:learning_well} to reduce overfitting.\cite{ng_machine_2018}

The classical bias variance-trade-off curve (cf.\ Figure~\ref{fig:bv_tradeoff}) suggests that there is a \enquote{sweetspot} (dotted line) in which the test error is minimal.
One current research question in \gls{dl} is why one still can achieve good testing error with highly overparameterized models, i.e., models for which the number of  parameters is larger than the number of training points.\cite{geiger_jamming_2019, allen-zhu_learning_2018}
Belkin et al.\ suggest that \enquote{modern}, overparameterized, models do not work in the regime described by the bias-variance trade off curve in Figure~\ref{fig:bv_tradeoff}.
Rather, they suggest a double descent curve where following a jamming transition, when we reach approximately zero train error (the interpolation threshold), the error decreases with the number of parameters.\cite{belkin_reconciling_2018}
Belkin et al.\ hypothesize that this is due to the larger function space that is accessible to more complex models which might allow them to find interpolating functions that are simpler (and hence better approximations according to Occam's razor, cf.\ section~\ref{sec:curse_dimensionality}).

In the following, we give an overview of the most popular learning techniques.
We see \glspl{nn} mostly suited for large, unstructured, datasets, data sources, e.g.\  images or spectra, or feature sets which are not yet highly preprocessed (e.g., directly using the coordinates and atom identities)---as \glspl{nn} can also be used to create features (representation learning), which in the chemical science is often used in a \enquote{message passing} approach (cf.\ section~\ref{sec:messagepassing}).\cite{gilmer_neural_2017}

\subsection{Lots of (Unstructured) Data (Tall Data)}
In (computational) materials science a large array of data is created every day and some of it is even deposited in a curated form on repositories.
Still, most of it does not contain highly engineered features.
To learn from such large amounts of data \gls{nn} are one of the most promising approaches.
The field of \gls{dl}, which describes the use of deep \glspl{nn}, is too wide to be comprehensively reviewed, wherefore we just give an overview of the basic building principles of the most popular building blocks.

\subsubsection{Neural Networks}
Classical, feed-forward, \gls{nn} approximate a function \(f\) using a chain of matrix evaluations
\begin{equation}
    f(\mathbf{X}) = g^{(L)}\left(\mathbf{W}^{(L)} \cdots g^{(2)}\left(\mathbf{W}^{(2)}g^{(1)}\left(\mathbf{W}^{(1)}\mathbf{X} \right)\right)\right),
\end{equation}
where \(\mathbf{X}\) is the input vector, \(g\) are activation functions---non-linear functions such as sigmoid functions or the \gls{relu}---and the \(\mathbf{W}\) are the weight matrices the neural network learns using the data.
\(L\) is here the number of layers, and the most popular and promising case is when there are multiple nonlinear layers.  This is known as \gls{dl}. The multiplication with the weight matrix is a linear transformation of the data, the bias corresponds to a translation and the activation function enables us to introduce non-linearities.

One of the most frequently cited theorems in the \gls{dl} community is the universal approximator theorem which states that, under given constraints, a single hidden layer of finite width is able to approximate any continuous function (on a set of \(\mathbb{R}\)).
What is perhaps more surprising is that those models work, that we can train them on random labels without any convergence problems,\cite{zhang_understanding_2016} and that they still generalize---these questions are active areas of research in computer science.

One of the strengths of neural networks is that they scale really well since training them does not involve an expensive matrix inversion (which scales with \(\mathcal{O}(n^3)\)) and since they can be trained efficiently in batch mode with stochastic gradient descent, where only a small part of the complete data needs to be loaded into memory.
The large expressivity of deep networks combined with the benign scaling makes them the preferred choice for massive (unstructured) datasets, whereas classical statistical learning methods might be the preferred choice for small datasets of structured data.\cite{makridakis_statistical_2018}

\paragraph{High-Dimensional Neural Network Potential}\label{sec:ann}
\begin{figure}
    \centering
    \includegraphics[width=\textwidth]{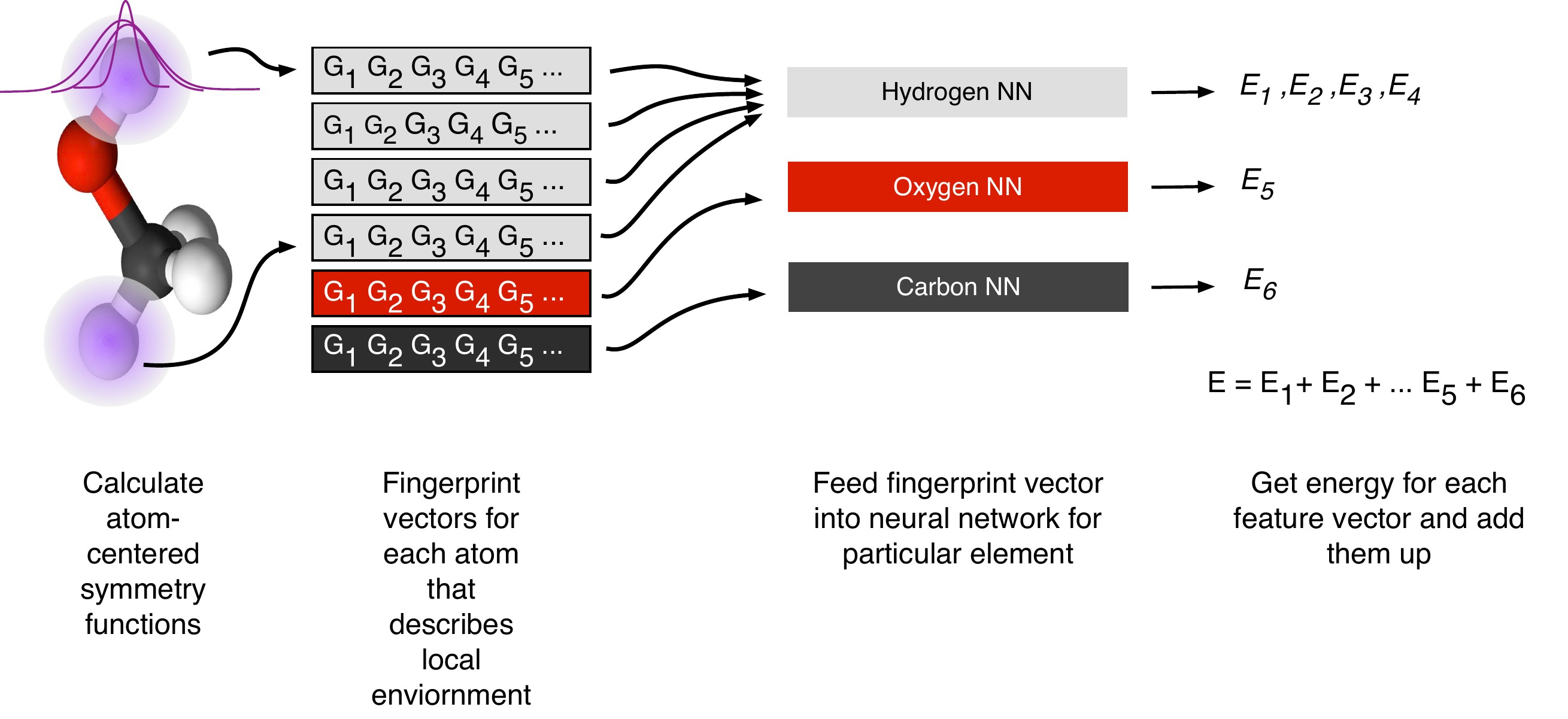}
    \caption{Schematic representation of the architecture of a \gls{hdnpp} (Behler-Parinello scheme) at the example of methanol. The local environment around each atom is described with symmetry functions (pink, Gaussians). Each symmetry function can probe different length scales and will return one value. The values can then be concatenated into one fingerprint vector. This fingerprint vector can then be fed into a \gls{nn} corresponding to one particular element, i.e., we will feed the four fingerprints for the four hydrogens into the same neural network but will receive different outputs due to the different fingerprints.
        The predictions can then be added up to calculate the energy of the entire system.}\label{fig:hdnpp}
\end{figure}
One of the cases where neural networks shine in the field of chemistry are high-dimensional neural networks that can be used to \enquote{machine learn} potential energy surfaces---as it has recently been done for MOF-5 (cf.\ section~\ref{sec:applications}),\cite{eckhoff_molecular_2019} and which can be used to access time or length scales that are not accessible with ab initio techniques at accuracies that are not accessible with force fields.
One prime example is the ANI-1X potential, which is a general-purpose potential that approaches coupled-cluster theory accuracy on benchmark sets.\cite{s.smith_ani-1_2017, s_smith_approaching_2019}
And due to the nature of molecular simulation in which there is a lot of correlations between the properties at different time steps, and hence data redundancy, they are an ideal application for \gls{ml}.\cite{rupp_machine_2015}

\Gls{nn} models for potential energy surface have already been proposed more than two decades ago.
But due to the architecture of those models, it was difficult to scale them to larger systems and the models did not incorporate fundamental invariances of the potential.\cite{blank_neural_1995}
This has been overcome with the so-called \gls{hdnpp} (also known as Behler-Parinello scheme, cf.\ Figure~\ref{fig:hdnpp}).
Each atom of the structure will be represented by a fingerprint vector (using symmetry functions) that describes its chemical environment within a cutoff radius (cf.\ chemical locality approximation in section~\ref{sec:locality}).
For each element, a separate \gls{nn} is trained (cf.\ Figure~\ref{fig:hdnpp}) and each atomic fingerprint vector is fed into its corresponding \gls{nn} that predicts an energy.
The total energy is then the sum of all atomic contributions (cf.\ eq.~\ref{eq:locality_approx}).
This additive approach is scalable by construction (nearly linear with system size) and the invariances with respect to rotation and translation are introduced on the level of the symmetry functions.
Also, the weight sharing (one \gls{nn} for many environments of a particular element) makes this approach efficient and allows for generalization (similar to the sharing of filters in \gls{cnn} which we will discuss in section~\ref{sec:cnn}).
One additional advantage of such models is that they are not only efficient and accurate, but that they are also reactive (again due to the locality assumption combined with the fact that no functional form is assumed)---which most classical force fields are not.
For more technical details, we recommend reviews from Behler.\cite{behler_representing_2014, behler_constructing_2015}

\paragraph{Message-Passing Neural Networks/Representation Learning}\label{sec:messagepassing}
\begin{figure}
    \centering
    \includegraphics[width=.7\textwidth]{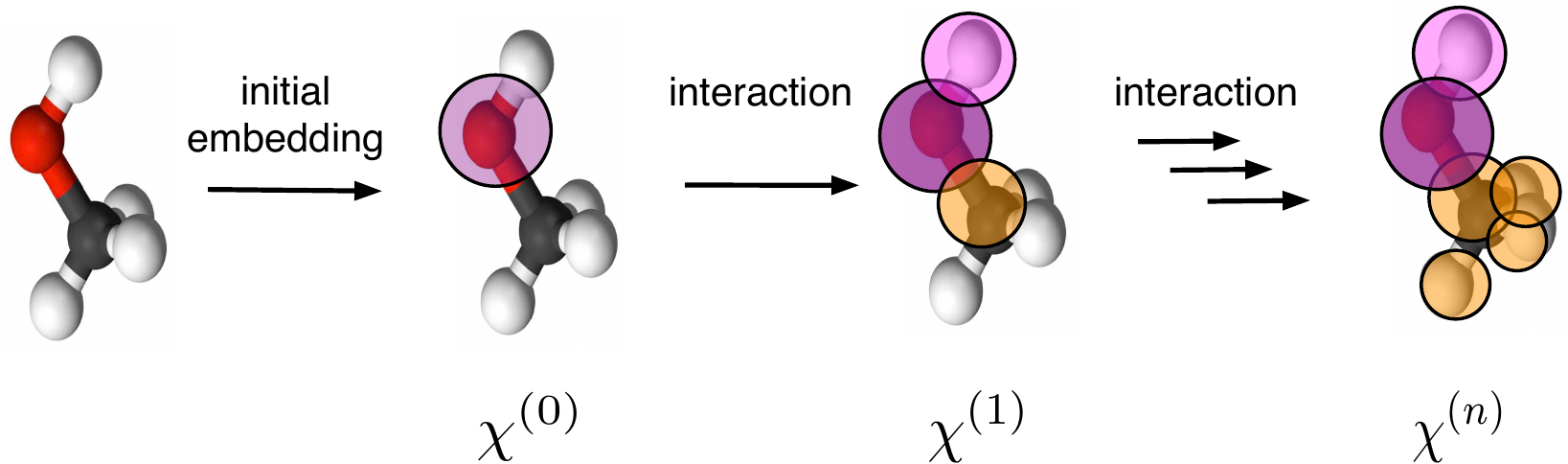}
    \caption{Schematic illustration of the idea behind the message-passing architecture. Following the initial embedding of the molecule each environment \(\chi\) represents one atom. Successive interactions in the message passing architecture refine the local chemical environments \(\chi\) by taking into consideration the interaction between the neighboring environments.}\label{fig:messagepassing}
\end{figure}
In message-passing neural networks, the input can be nuclear charges and positions, which are also the variables of the Schrödinger equation.
A \gls{dnn} then constructs descriptors that are relevant for the problem at hand (representation learning). The idea behind this approach is to build descriptors \(\chi\) by recursively adding interactions \(v\) with more and more complex neighboring environments at a distance \(d_{ij}\) (cf.\ Figure~\ref{fig:messagepassing})
\begin{equation}
    \chi_{i}^{(t+1)}  = \chi_i^{(t)} + \sum_{j<i} \mathbf{v}\left(\chi_j^{(t)}, d_{ij} \right).
\end{equation}
This approach is for example used in \gls{dtnn},\cite{schutt_quantum-chemical_2018} SchNet,\cite{schutt_schnet_2018}, SchNOrb,\cite{schutt_unifying_2019} \gls{hip}-\gls{nn},\cite{nebgen_transferable_2018} and PhysNet.\cite{unke_physnet_2019}
A detailed discussion of this architecture type is provided by Gilmer et al.\cite{gilmer_neural_2017}

\paragraph{Images or Spectra}\label{sec:cnn}
\begin{figure}
    \centering
    \includegraphics[width=\textwidth]{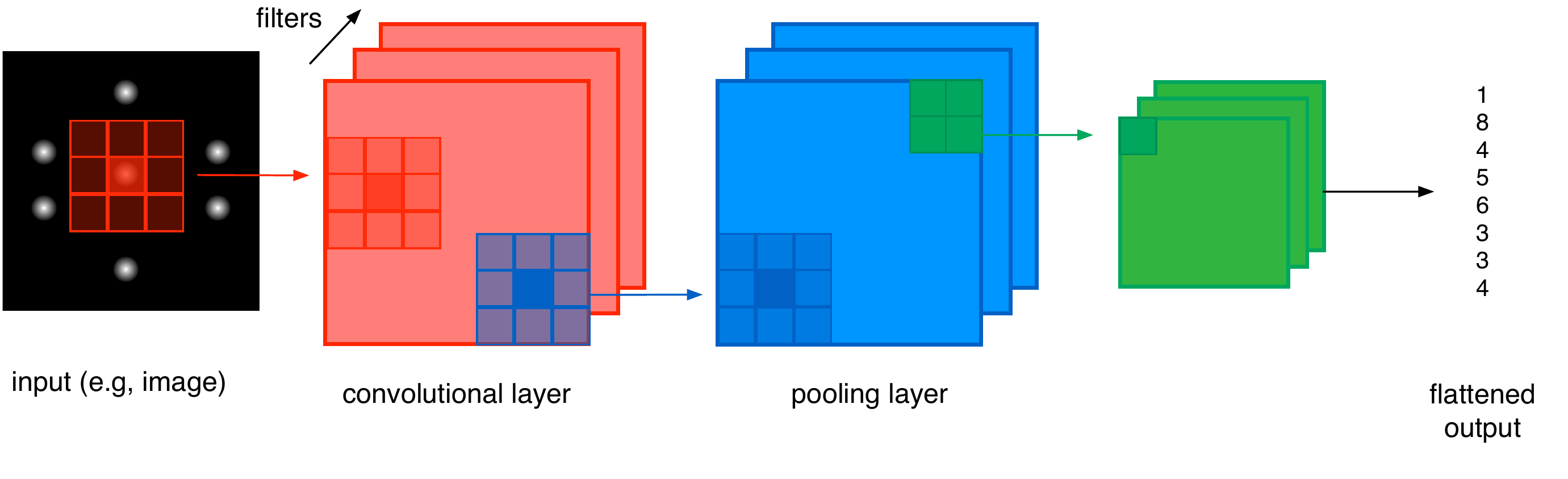}
    \caption{Example of the use of a \gls{cnn}.
        One slides convolution layers (red) over an image, which for example can be a two-dimensional diffraction pattern.\cite{ziletti_insightful_2018}
        Usually, one then uses a pooling layer to compress the matrices after convolution.
        After flattening, the output can be used for conventional hidden \gls{nn} layers.}\label{fig:cnn_schema}
\end{figure}
For learning from images or patterns, \gls{cnn} are particularly powerful.
They are inspired by the concept of receptive fields in biological processes, where each neuron responds only to activation in a specific region of the visual field.

\Glspl{cnn} work by sliding a filter matrix over the input to extract higher-level features (cf.\ Figure~\ref{fig:cnn_schema}).
An example of how such filters work is the set of the Sobel filter matrices, which can be used as edge detectors:
\begin{equation}
    \mathbf{G}_x =
    \begin{pmatrix}
        1 & 0 & -1 \\
        2 & 0 & -2 \\
        1 & 0 & -1
    \end{pmatrix}.\label{eq:sobel}
\end{equation}
The middle column, which is centered on the cell (pixel) on which the filter is used, is filled with zeros and the column left and right to it have opposite signs.
In case there is no edge, the values on the left and the right of the pixel will be equal.
But in case there is an edge, this is no longer the case and the matrix multiplication will give a result that highlights the edge.
By sliding the \(\mathbf{G}_x\) matrix horizontally over an image one can hence highlight horizontal edges. A collection of different filter layers are used to learn the different correlations between (neighboring) elements.
\Gls{cnn} apply, on each layer, a set of different filters that share weights (similar to the way in which different atoms of the same element share weights in \gls{hdnpp}).
Usually, convolutions are used together with pooling layers that compress the matrix by, again, sliding a filter matrix, which for example takes the maximum or the average in a \(2 \times 2\) block of the matrix, over the matrix (cf.\ Figure~\ref{fig:cnn_schema}).
This leads to approximate translational invariance as the maximum pixel after the convolution will still be extracted by a maximum pooling layer if the translation was not too large (since the pooling effectively filters out small translations).

\Glspl{cnn} tend to generalize well and are computationally efficient due to the weight sharing between the different filter layers for each convolutional layer.
Not surprisingly, ample works attempted to use \glspl{cnn} to analyze spectra.
Ziletti et al.\ used this approach to classify crystal structures based on two-dimensional diffraction patterns.\cite{ziletti_insightful_2018}
Others used them to perform classification based on steel microstructures,\cite{azimi_advanced_2018} or a representation based on the periodic table, where the positions of the elements of full-Heussler compounds were encoded and the authors hoped to implicitly leverage the information encoded in the structure in the periodic table using the \gls{cnn}.\cite{zheng_machine_2018}

\paragraph{Case Study: Predicting the Methane Uptake in COFs Using a Dilated CNN}\label{sec:xrd_cnn}
For this case study, we use the \gls{xrd} pattern as a geometric fingerprint of the structure as it fulfills many of the criteria for an ideal descriptor: it is cheap and invariant to symmetry operations like an expansion of the unit cell.
But the way in which information is encoded in the fingerprint makes it not suitable for all learners: one could try using it in kernel machines to do similarity-based reasoning---similar to what von Lilienfeld and co-workers have done with radial distribution functions.\cite{lilienfeld_fourier_2015}
However, one could also try to create a \enquote{pattern recognition} model---this is where \glspl{cnn} are powerful.
Importantly, the patterns do not only span a small range, like neighboring reflexes, but are composed of both nearby and far-apart reflexes (due to the symmetry selection rules).
For this reason, conventional convolution layers might be not ideal.
We use dilated convolutions to exponentially increase the receptive field: Dilated convolutions are basically convolutions with holes and in our model for which we increase the hole size from layer to layer.
To avoid overfitting, we use spatial dropout, which is especially well suited for convolutional layers (cf.\ section~\ref{sec:cnn}) and which randomly deactivates some neurons.
\begin{figure}
    \centering
    \includegraphics[width=\textwidth]{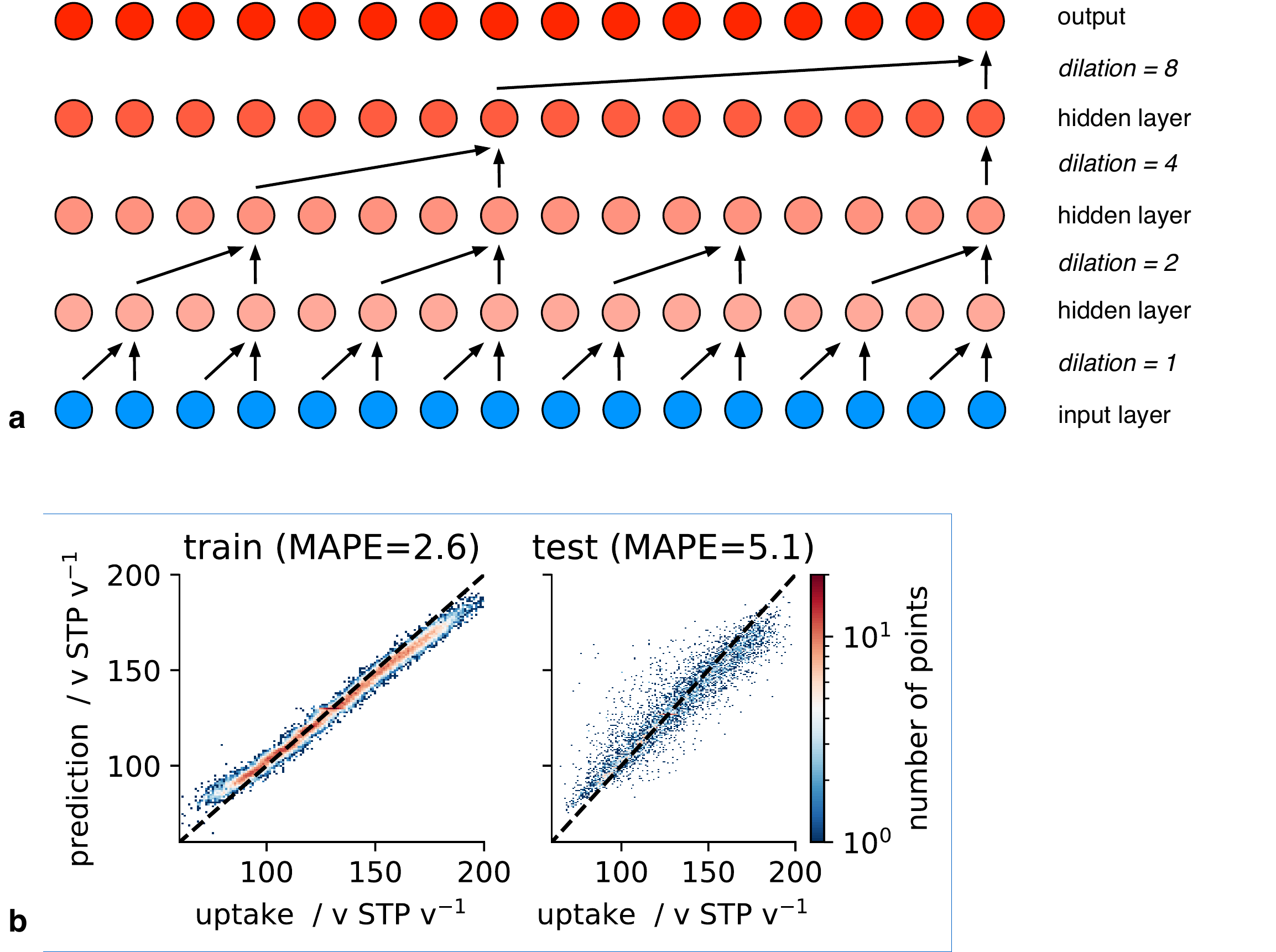}
    \caption{Using a dilated \gls{cnn} to predict the methane uptake of \glspl{cof} assembled by Mercado et al.\cite{mercado_silico_2018}
        For this example, we use dilated convolutions to extract correlations from the \gls{xrpd} pattern (a). We can then pass the output to some hidden layers to predict the methane uptake (b). We overfit to the training set, but can also get decent performance on the test set without major tuning of the model. MAPE is an acronym for the mean absolute percentage error.}\label{fig:dilational_cnn}
\end{figure}
From Figure~\ref{fig:dilational_cnn} we see that such a model is indeed able to predict the deliverable capacity for methane in \glspl{cof} based on the \gls{xrd} pattern.

\paragraph{Sequences}\label{sec:rnn}
\begin{figure}
    \centering
    \includegraphics[width=.75\textwidth]{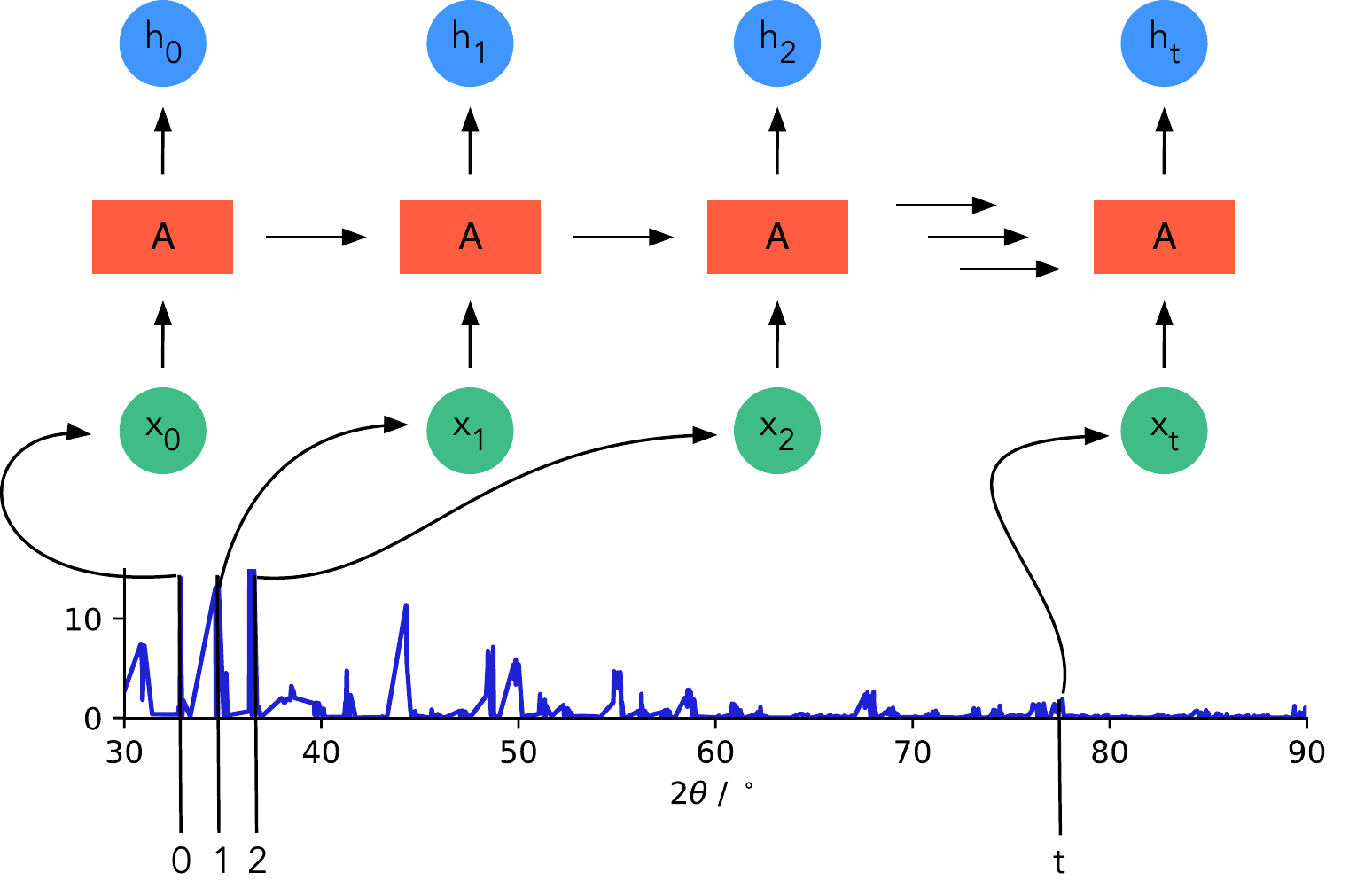}
    \caption{Schematic illustration of the building principle of a \gls{rnn}.
        A \gls{nn} \(A\), uses some input \(x\), like a peak of a \gls{xrpd} pattern, to produce some output \(h\).
        Importantly, some information is passed from one \gls{nn} to the next.}\label{fig:rnn}
\end{figure}
\Glspl{rnn} are frequently used for the modeling of time-series data as they, in contrast to classical feed-forward models, have a feedback loop that gives the network a \enquote{memory} which it can use to recognize information that is encoded in the sequence itself (cf.\ Figure~\ref{fig:rnn}).
This fitness for temporal data was for example used by van Nieuwenburg to classify phases of matter based on their dynamics, which in their case was a sequence of magnetizations.\cite{van_nieuwenburg_learning_2018}
Similarly, Pfeiffenberger and Bates used a \gls{rnn} to find improved protein conformations in \gls{md} trajectories for protein structure prediction.\cite{pfeiffenberger_predicting_2018}

Another approach to model sequences is to use autoregressive models, which also incorporate reference to \(p\) prior sequence points
\begin{equation}
    X_t - \phi_1X_{t-1} - \phi_2X_{t-2} - \cdots - \phi_p X_{t-p} = \epsilon_t,
\end{equation}
where \(\phi_p\) are the parameters of the model and \(\epsilon\) is white noise.
This approach has for example been used by Long et al.\ to model the degradation of lithium-ion batteries based on their capacity as a function of the number of charge/discharge cycles.\cite{long_improved_2013}

\paragraph{Graphs}\label{sec:graph_model}
As indicated above (cf.\ section~\ref{sec:structure_graphs}), graphs are promising descriptors of molecules and crystals as they can provide rich information without the need for precise geometries.
But learning from the graph directly requires special approaches.
\begin{figure}
    \centering
    \includegraphics[width=.7\textwidth]{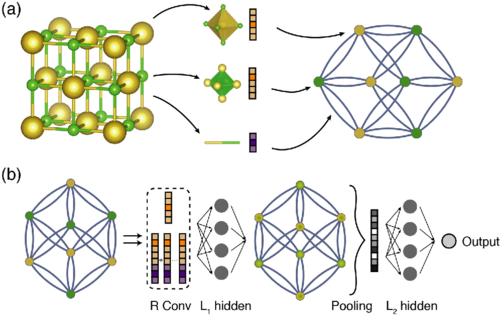}
    \caption{Schematic illustration of the crystal graph \gls{cnn} developed by Xie and Grossman.\cite{xie_crystal_2018}
        Reprinted from Xie and Grossman,\cite{xie_crystal_2018} Copyright (2018) by the American Physical Society.
        After representing the crystal structure as a graph (a) by using the atoms as nodes and the bonds as edges, the graph can be fed into a graph \gls{cnn} (b).
        For each node of the graph, \(K\) convolutional layers and \(L_1\) hidden layers are used to create a new graph that is then, after pooling, send to \(L_2\) hidden layers.}\label{fig:crystalgraphnn}
\end{figure}
Similar to message passing neural networks, Xie and Grossman developed convolution operations on the structure graph that let an edge interact iteratively with its neighbors to update the descriptor vector (cf.\ Figure~\ref{fig:crystalgraphnn}) and in this sense is a special case of the message-passing \glspl{nn} (cf.\ section~\ref{sec:messagepassing}).\cite{xie_crystal_2018}
Again, this approach has been shown to be promising in the molecular domain before it has been applied to crystals.\cite{kearnes_molecular_2016}

\subsection{Limited Amount of (Structured) Data (Wide Data)}
Especially for structured data, conventional \gls{ml} models, like kernel-based models, can often perform equally or better than neural networks---especially when the amount of data is limited.
In any case, it is generally useful to implement the simplest model possible first, to have a baseline and also to ensure that the infrastructure (getting the data into the model, calculating metrics \dots) works before starting to implement a more complex architecture.

\subsubsection{Linear and Logistic Regression}\label{sec:linear_model}
The most widely known regression method is probably linear regression.
In its ordinary form, it assumes a normal distribution of residuals, but we want to note that also generalized versions are available that work for other distributions.
One significant advantage of linear regression is that it is simple and interpretable.
One can directly inspect the weights of the model to understand how predictions are made and it has been the workhorse of cheminformatics.
Even though the simple architecture limits the expressivity of the model this is also a feat as one can use it for initial debugging, feedback loops, and to get some initial baseline results.

\subsubsection{Kernel Methods}\label{sec:kernel_methods}
\begin{figure}
    \centering
    \includegraphics[width=.7\textwidth]{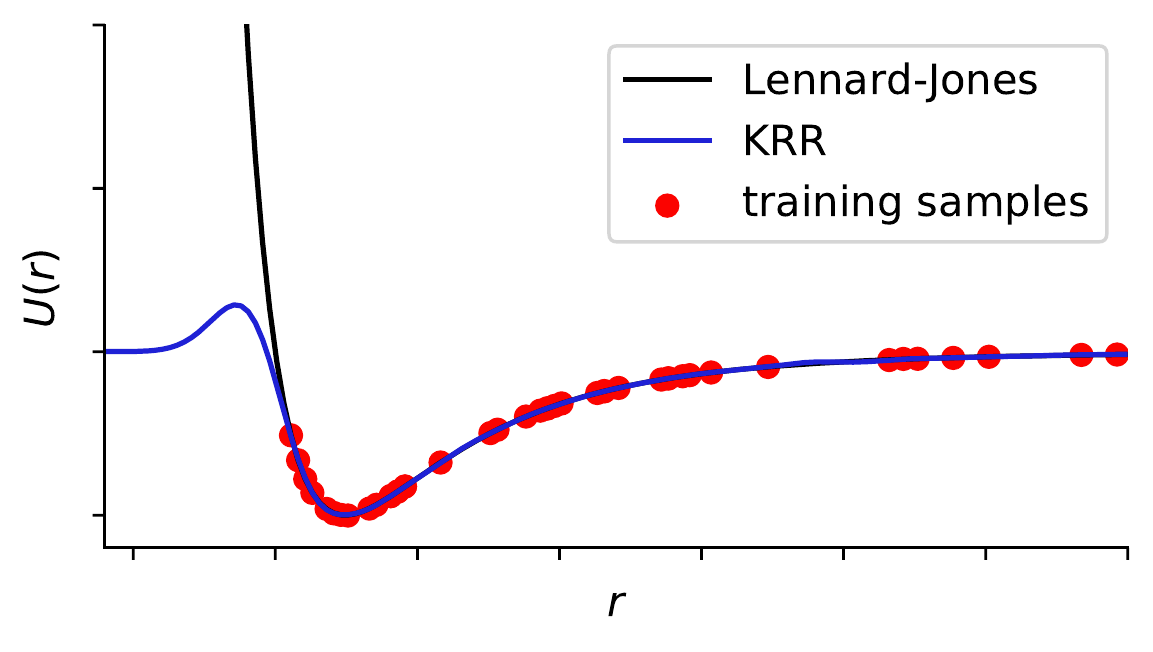}
    \caption{\gls{krr} to learn the Lennard-Jones (12,6) potential.
        We randomly sampled 80 points on the potential, then tuned the hyperparameters of the kernel and then predicted for all points.
        The model fails completely to model the strong repulsion due to the lack of training examples in that region.}\label{fig:lj_krr}
\end{figure}
\begin{figure}
    \centering
    \includegraphics{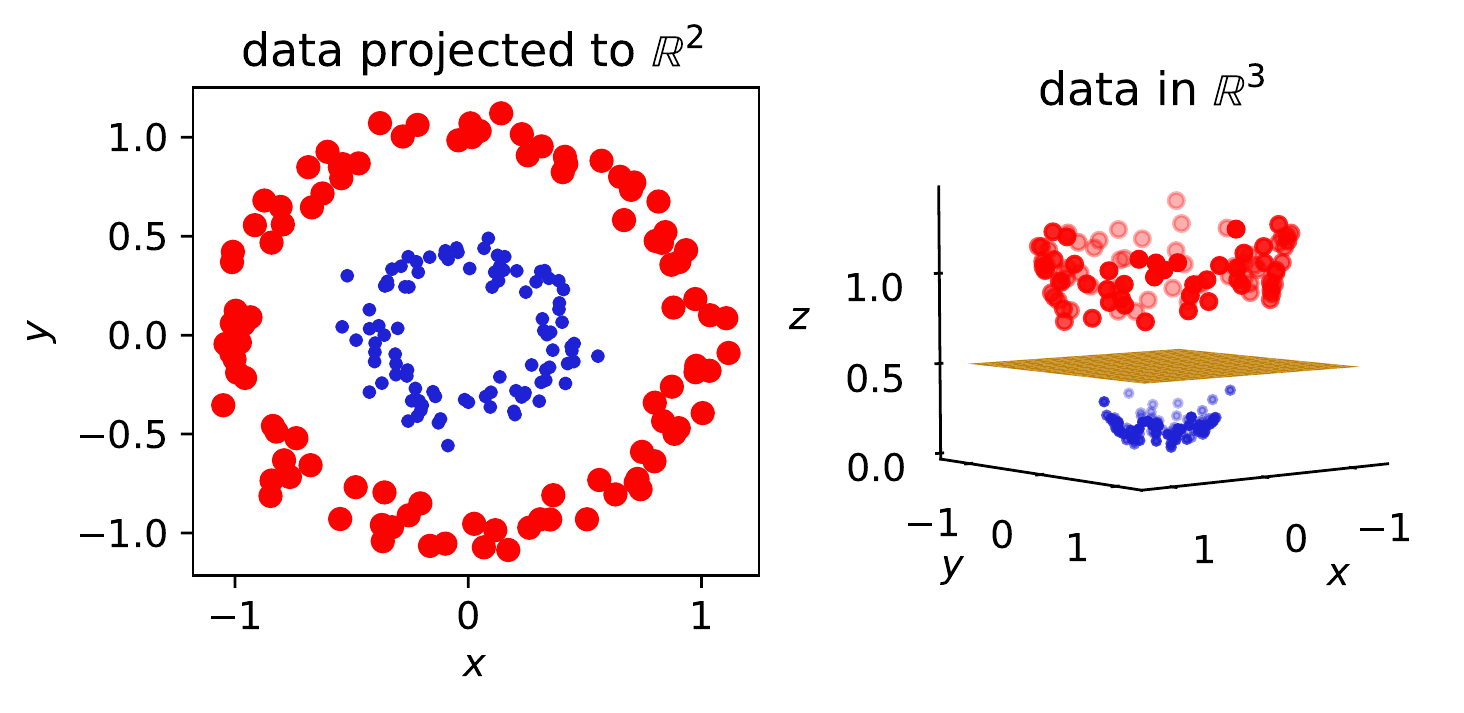}
    \caption{Visualization of one idea behind the kernel trick---mapping to higher-dimensional spaces to make problems linearly separable.
        In two dimension, the data (two different classes, colored in red and blue, respectively) is not linearly separable, but after applying the kernel \(K(x,y) = x \cdot y  + \| x \| ^2 = \| y \| ^2\) we can draw a plane to separate the classes (three-dimensional plot on the right).}\label{fig:kernel_trick}
\end{figure}

One of the most popular learning techniques in chemistry is \gls{krr}.
The core idea behind kernel methods is to improve beyond linear methods by implicitly mapping into a higher-dimensional space which allows treating non-linearities in a systematic and efficient way (cf.\ Figure~\ref{fig:kernel_trick}).
A naive approach for introducing non-linearities would be to compute all monomials of the feature columns, e.g., \(\phi(x_1, x_2) = (x_1^2, x_1x_2, x_2x_1, x_2^2)\).
But this can become computationally infeasible for many features.
The kernel trick avoids this by using kernel functions, i.e., inner products in some feature space.\cite{smola_tutorial_2004}
If they are used, the computation scales no longer with the number of features but with the number of data points.

There are strict mathematical rules that govern what a function needs to fulfill to be a valid kernel (Mercer's theorem),\cite{smola_tutorial_2004} but the most popular choices for kernel functions are the Gaussian (\(K(x,x^*) = \exp\left(\gamma \|x-x^* \|^2 \right)\)) or the Laplacian (\(K(x,x^*) = \exp\left(\gamma \|x-x^* \| \right)\)))  kernels, which width (\(\gamma\)) controls how local the similarity measure is.

The general intuition behind a kernel is to not consider the isolated data points but rather the similarity between a query point \(x\), for which we want to make a prediction, and the training points \(x^*\) (landmarks, which are usually multi-dimensional vectors), and to measure this similarity with inner products (as many algorithms can be rewritten in terms of dot products).
At the same time, one then uses this similarity measure to work implicitly in a higher-dimensional space where the data might be easier separable.
That is, it is most useful to think about predictions with \gls{krr} using the following equation
\begin{equation}
    \underbrace{y(x)}_{\text{prediction}} = \sum_i \underbrace{a_i}_{\text{weight}} \, \overbrace{K(\underbrace{x_i^*}_{\text{landmark}}, \underbrace{x}_{\text{query point}})}^{\text{kernel}},
\end{equation}
or in matrix form, we write
\begin{equation}
    \mathbf{y} = \mathbf{K}\, \mathbf{a} \quad  \Leftrightarrow \quad \mathbf{a} = \mathbf{K}^{-1}\,\mathbf{y}.
\end{equation}
But this equation assumes that \(\mathbf{K}^{-1}\) can be found, which might not be the case if there is no \(\mathbf{K}\) or more than one \(\mathbf{K}\) that satisfies the equation (i.e., it is an ill-posed, unstable or non-unique, problem).
For this reason, one typically adds a regularization term \(\lambda \mathbf{I}\), with \(\mathbf{I}\) being the identity matrix (we will explore the concept of regularization in more depth and from another viewpoint in section~\ref{sec:learning_well}) which acts as a high-pass filter, i.e., it filters out the noise and makes the inversion more stable and the solution smoother.
One then solves
\begin{equation}
    \mathbf{a} = {\left(\mathbf{K} + \lambda \mathbf{I}\right)}^{-1} \, \mathbf{y}.\label{eq:krr_equation}
\end{equation}
The most widely known algorithms which use this kernel trick are \glspl{svm} and \gls{krr}.
They are equivalent except for the loss function and the fact that the \gls{krr} is usually solved analytically.
The \glspl{svm} use a special loss function, the \(\epsilon\)-insensitive loss, where errors smaller than \(\epsilon\) are not considered.
The \gls{krr}, on the other hand, uses the ridge loss function, which penalizes high weights and which we will discuss in section~\ref{sec:regularization} in more detail.

One virtue of kernel learning is the mathematical framework which it provides.
It allows deriving a scheme in which data of different fidelity can be combined to predict on the high-fidelity level---a concept that was used to learn using a lot of \gls{gga} data (PBE functional) to predict hybrid functional level (HSE06 functional) band gaps.\cite{pilania_multi-fidelity_2017}
We will explore this concept, that can be promising for the \gls{ml} of electronic properties of porous materials with large unit cells, in more detail in section~\ref{sec:transfer}.

Also, kernels pave an intuitive way to multitask predictions; by using the same kernel for different regression tasks and predicting the coefficients for the different tasks at the same time, Ramakrishnan and von Lilienfeld could predict many properties from only one kernel (computing the Kernel is usually the expensive step as it involves a matrix inversion which scales cubically).\cite{ramakrishnan_many_2015}
Due to the relative ease of use of kernel methods and their mathematical underpinning, they are the workhorse of many of the quantum \gls{ml} works.\cite{rupp_fast_2012, bartok_machine_2017}
Also, kernel methods are useful for the development of new descriptors as they are much more sensitive to the quality of the descriptor than \gls{nn} or tree-based models as they are similarity-based.
This is, a kernel-based method will likely fail if two compounds that are distant in property space are close in fingerprint space.

\subsubsection{Bayesian Learning}\label{sec:bayesian}
\begin{figure}
    \centering
    \includegraphics[width=.6\textwidth]{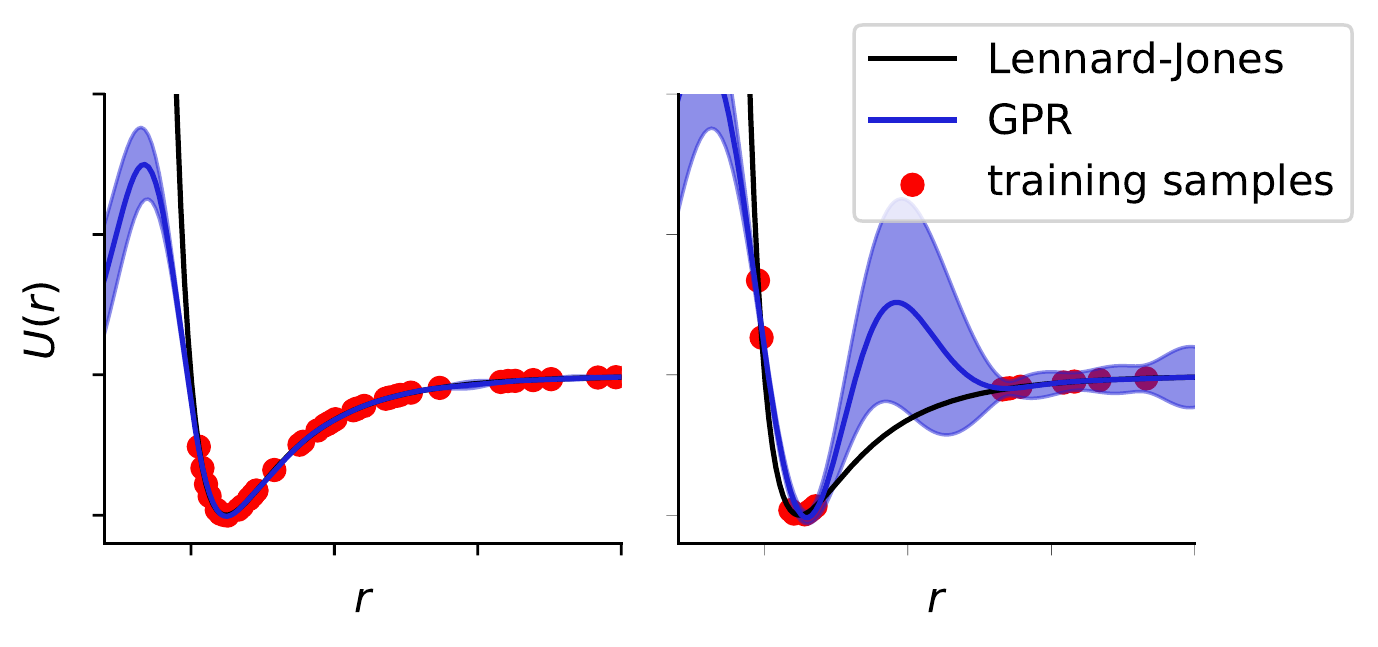}
    \caption{Using \gls{gpr} to learn the Lennard-Jones potential (same as in Figure~\ref{fig:lj_krr}).
        Here, we trained two different \gls{gpr} models: First, on the same 80 points we used for Figure~\ref{fig:lj_krr}, and then one for a bad training set with \enquote{holes}, i.e., areas from which we did not sample training points.
        Again, we tuned the hyperparameters of the kernel and then predicted for all points. We can observe that, similar to our \gls{krr} results, our model cannot predict the strong repulsion due to the lack of training points.
        But, in contrast to the \gls{krr}, the \gls{gpr} gives us an estimate for the uncertainty that is larger when we lack examples in a particular region.}\label{fig:gp_krr}
\end{figure}
Up to now, we surveyed the models from a frequentist point of view in which probabilities are considered as long-run frequencies of events.
A more natural framework to look at probabilities is the Bayesian point of view.
Bayesian learning is built around Bayes rule\cite{tipping_bayesian_2004}
\begin{equation}
    \underbrace{P\left(\theta | D\right)}_\text{posterior} = \frac{\overbrace{P\left(D | \theta \right)}^\text{likelihood} \overbrace{P \left(\theta\right)}^\text{prior}}{\underbrace{P\left(D\right)}_\text{evidence}},
\end{equation}
which describes how the likelihood \(P\left(D | \theta \right)\) (probability of observing the data given the model parameters) updates prior beliefs \(P \left(\theta\right)\) after observing the data \(D\).
This updated distribution is the posterior distribution \(P\left(\theta | D\right)\) of model parameters \(\theta\).

Similar to molecular Monte-Carlo simulations one can use Markov chain Monte-Carlo to sample the posterior distribution \(P\left(\theta | D\right)\).
Several packages like \texttt{pymc3}\cite{salvatier_probabilistic_2016} and \texttt{Edward}\cite{tran_edward_2017} offer a good  starting point for probabilistic programming in Python.

The power of Bayesian modeling is that one can incorporate prior knowledge with the choice of the prior distribution and that it allows for a natural way to deal with uncertainties as the output, the posterior distribution \(P\left(\theta | D\right)\), is a distribution of model parameters. Furthermore, it gives us a natural way to compare models: The best model is the one with the highest evidence, i.e., probability of the data given the model.\cite{mackayProbableNetworksPlausible1995a}

An example of how prior knowledge can be incorporated is a work by Mueller and Ceder who incorporated physical insight to fit cluster expansions, which are simple but powerful models that express the property of system using single-site descriptors.
An archetypal example is the Ising model.
They used physically intuitive insights such as the distance of the prediction to a simple model, like a weighted average of pure component properties for the energy of an alloy, or that observation that similar cluster functions should have similar values, to improve the predictive power of such cluster expansions.
This is effectively a form of regularization, equivalent to Tikhonov regularization (cf.\ section~\ref{sec:regularization}).

\paragraph{Gaussian Process Regression}
Bayesian methods are most commonly used in the form of \gls{gpr},\cite{rasmussen_gaussian_2004} which drives the \glspl{gap}.\cite{bartok_gaussian_2015}
\Gls{gpr} is the Bayesian version of \gls{krr}, i.e., it also solves eq.~\ref{eq:krr_equation}.

\begin{figure}
    \centering
    \includegraphics{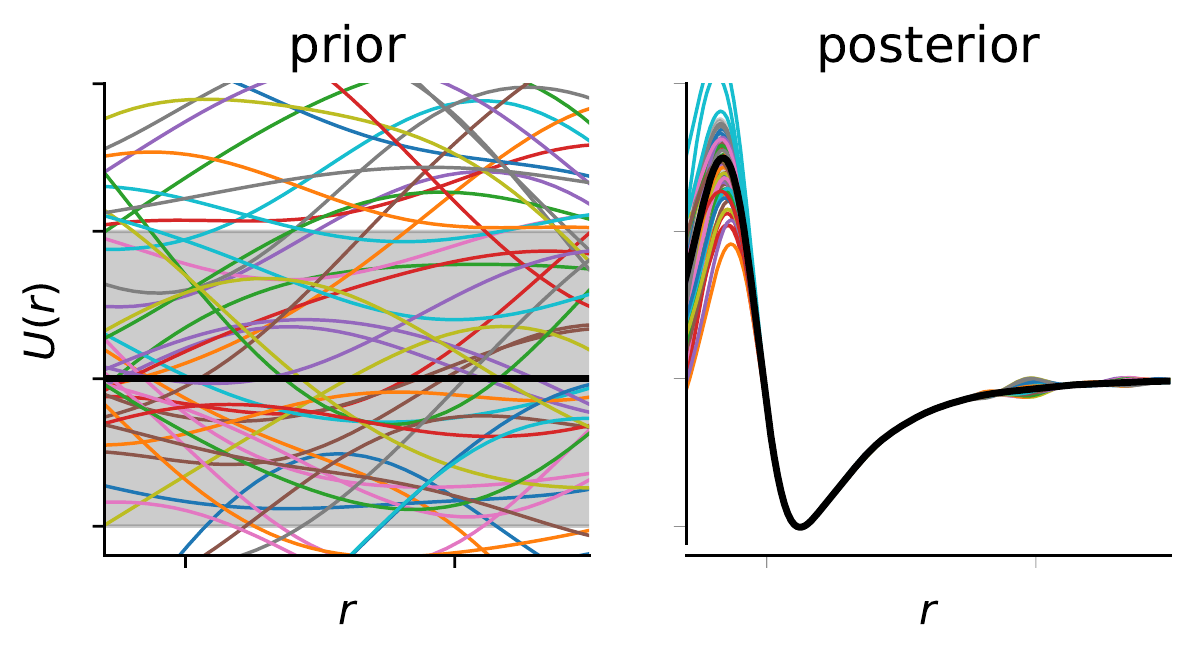}
    \caption{Samples from the prior and posterior distributions for the fit shown in Figure~\ref{fig:gp_krr} using the same scale for the axes.
        Here, we assume a zero mean (thick black line) for the prior but the mean in the posterior is no longer zero after the inference.
        The standard deviation is shown as a gray area. }\label{fig:prior_posterior}
\end{figure}

In \gls{gpr} one no longer uses a parametric functional form (like polynomials or a \gls{mlp}) to model the data but uses learning to adapt the distribution (\enquote{ensemble} of functions), where the initial distribution (the prior) reflects the prior knowledge.\cite{seeger_gaussian_2004}
That is, in contrast to standard (multi)linear regression one does not directly choose the basis functions but rather allows for a family of different possible functions (this is also reflected in the uncertainty band shown in Figure~\ref{fig:gp_krr} and the spread of the functions in Figure~\ref{fig:prior_posterior}).

We can think of the prior distribution as samples that are drawn from a multivariate normal distribution, that is characterized by a mean \(\boldsymbol{\mu}\) and a covariance \(\mathbf{C}\), i.e., we can write the prior probability as
\begin{equation}
    P(\mathbf{y}) =  \mathrm{Normal}(\boldsymbol{\mu},\mathbf{C}) \overset{\boldsymbol{\mu}=0}{\propto} \exp\left(-\frac{1}{2}\, \mathbf{y}^T\mathbf{C}^{-1}\mathbf{y}\right)
\end{equation}
Usually, one uses a mean of zero and the covariance matrix \(\mathrm{cov}(y(x), y(x^*))\) that describes the covariance of function values at \(x\) and \(x^*\)---i.e., it is fully analogous to the kernel in \gls{krr}.
But in \gls{krr} one needs to perform a search over the kernel hyperparameters (like the width of the Gaussian), whereas the \gls{gpr} framework allows learning the hyperparameters using gradient descent on the marginal likelihood, which is the objective function in \gls{gpr}.

Also, the regularization term has another interpretation in \gls{gpr}, as it can be thought of as noise \(\sigma_f\) in the observation
\begin{equation}
    \mathrm{cov}(y_i, y_j) = C_{ij} + \sigma_f \, \delta_{ij}
\end{equation}
with Kronecker delta \(\delta_{ij}\) (1 for \(i=j\), else 0).
Hence, the regularization also has a physical interpretation, whereas in \gls{krr} we introduced a hyperparameter \(\lambda\) that we need to tune.

But the most important practical difference is that the formulation in the Bayesian framework generates a posterior distribution and hence a natural estimate of the uncertainty of the prediction.
This is especially valuable in active learning settings (cf.\ section~\ref{sec:active_learning}) where one needs an estimate of the uncertainty to decide whether to trust the prediction for a given point or whether additional training data are needed.
This was for example successfully used by Jinnouchi et al.\ employing ab inito force fields derived in the \gls{soap}-\gls{gap} framework.\cite{jinnouchi_phase_2019}
During the molecular dynamics simulations of hybrid perovskites, they monitored the uncertainty of the predictions and then could switch to \gls{dft} in case the uncertainty was too high and refined the force field with this new training point.
Using this approach, which is implemented in \texttt{VASP 6}, they could access time scales that would require years of simulations with first principle techniques.

\subsubsection{Instance-Based Learning}\label{sec:instance}
Thinking in terms of distances to training examples, as we do in kernel methods, is also the key ingredient to the understanding of instance-based learning algorithms like \gls{knn} regression.
Here, the learner only memorizes the training data and the prediction is a weighted average of the training data.
For this reason, \gls{knn} regressors are said to be non-parametric---as they do not learn any parameters and only need the data itself to make predictions.

The difference between kernel learning and \gls{knn} is that in the case of kernel learning the prediction is influenced by all training examples and the nature of the locality is influenced by the kernel.
\Gls{knn}, on the other hand, only uses a weighted average of the \(k\) nearest training examples.
This limits the expressivity of the model but makes it easy to inspect and understand.
As it requires that examples that are close in feature space are also close in property space, there might be problems in the case of activity cliffs\cite{cruz-monteagudo_activity_2014} and per definition, such a model cannot extrapolate.
Still, such models can be useful---especially due to the interpretability.
For example, Hu et al.\ combined \gls{knn} with a Gaussian kernel weighting over the \(k\) neighbors to predict the capacity of lithium-ion batteries.\cite{hu_data-driven_2014}

An interesting extension of \gls{knn} for virtual high-throughput screenings was developed by Swamidass et al.
The idea here is to refine the weighting of the neighbors using a small \gls{nn}, which allows taking non-linearities into account.\cite{swamidass_influence_2009}
The advantages here are the short training time, the low number of parameters, and hence the low risk of overfitting and the interpretability, which is only slightly lower than for a vanilla \gls{knn}.


\subsubsection{Ensemble Methods}\label{sec:ensemble_models}
Ensemble models try to use the \enquote{wisdom of the crowds} by using a collection (an ensemble) of several weak base learners, which are often high-variance models like decision trees, to produce a more powerful predictor.\cite{dietterich_ensemble_2000, rokach_ensemble-based_2010}

The power of ensemble models is to reduce the variance (the error due to the finite sample, i.e., the instability of the model) while not increasing the bias of the model.
This works if the predictors are uncorrelated.\cite{mehta_high-bias_2019}
In detail, one finds that the variance is given by
\begin{equation}
    \text{variance} (x) = \rho{(x)}\sigma^2 + \frac{1 - \rho(x)}{M} \sigma^2,
\end{equation}
where \(M\) is the covariance matrix of the \(M\) predictors with variance \(\sigma\). The bias is given by
\begin{equation}
    \text{bias}^2 (x) = {(f(x) - \mu)}^2.
\end{equation}
These equations mean that for an infinite number of predictors (\(M \to \infty\)) with no correlations with each other (\(\rho = 0\)) we can completely remove the variance and the only remaining sources of error are the bias of the single predictor and the noise.
Hence, this approach can be especially valuable to improve unstable models with high variance.
One example for high-variance models are \glspl{dt} (also known as \gls{cart}) which build flow chart like models by splitting the data based on particular values of variables, i.e.,  based on rules like \enquote{density greater than \SI{1}{\gram\per\centi\meter\cubed}?}
Only one such rule is usually not enough to describe physical phenomena, wherefore usually many rules are chained.
But such deep trees can have the problem that their structure (splitting rules) is highly dependent on the training set, wherefore the variance is high.
One approach to minimize this variance is to build ensemble models.
Another motivation for ensemble models can be given based on the Rashomon\footnote{Rashomon is a Japanese movie in which one person dies and four persons witness the crime, and report the same facts at court but in a different story.} effect which describes that there are usually several models with different functional forms that perform similarly.
Averaging over them using an ensemble can resolve to some extent this non-uniqueness problem, and make models more accurate and stable.\cite{breiman_statistical_2001}

\begin{figure}
    \centering
    \includegraphics[width=\textwidth]{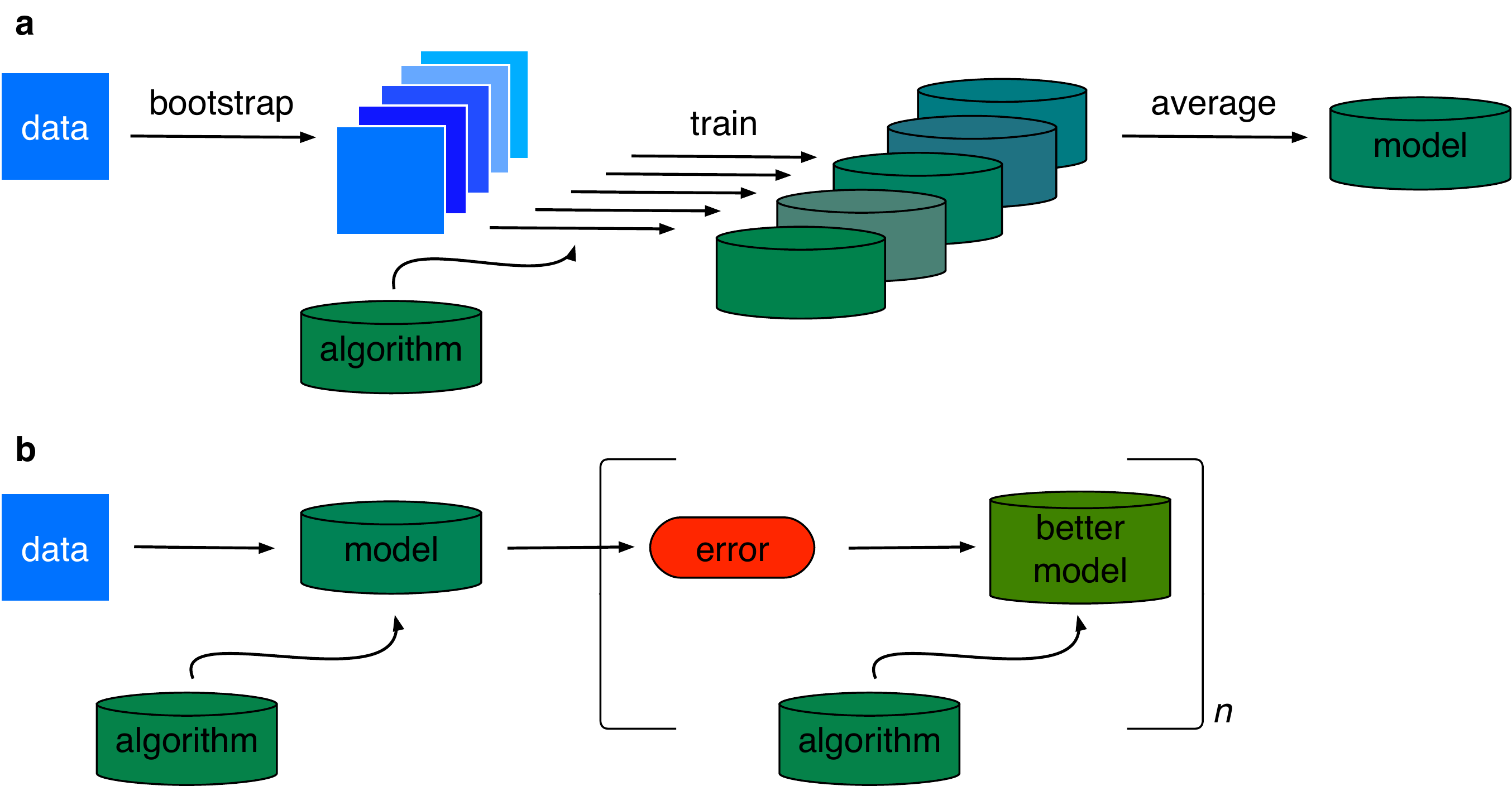}
    \caption{Schematic representation of the two most popular approaches for the creation of ensemble models, bagging (a) and boosting (b).}\label{fig:ensemble}
\end{figure}
There are two main approaches for the creation of ensemble models (cf.\ Figure~\ref{fig:ensemble}): The first one is called bagging (bootstrap aggregating) in which bootstraps of the training are fitted to a model and the predictions of all models are averaged to give the final prediction.
In \glspl{rf}, which are one of the most popular models in materials informatics, this idea is combined with random feature selection, in which the model is fitted only on a subset of randomly selected features.
ExtraTrees, are even more randomized by not using the optimal cut at different points in the decision tree but the best one from a random selection of possible cuts.\cite{geurts_extremely_2006}
Additionally, they also do not use bootstraps but the original training set.
In a benchmark of \gls{ml} models for the prediction of the thermodynamic stability of perovskites (based on composition features) Schmidt et al.\ found that ExtraTrees outperform random forest, neural networks, ridge regression and also adaptive boosting (which we will discuss in the following).\cite{schmidt_predicting_2017}

The other approach for the ensembling of models is boosting.
Here, models are not trained in parallel but iteratively, one after another, on the error of the previous model.
The most popular learners from this category are AdaBoost\cite{freund_decision-theoretic_1997} and \glspl{gbdt}\cite{friedman_greedy_2001} which are efficiently (and in a refined version) implemented in the XGBoost\cite{chen_xgboost_2016} and LightGBM\cite{ke_lightgbm_2017} libraries.
Given that \gls{gbdt} models are fast to train on datasets of moderate size, easy to use and robust, they are a good choice as a first baseline model on tabular descriptor data.\cite{caruana_empirical_2006, schapire_boosting_1998}
\Glspl{gbdt} were used in many studies on porous materials (cf.\ section~\ref{sec:applications}). For example, they were used by Evans et al.\ to predict mechanical properties of zeolites based on structural properties like \ce{Si-O-Si} bond lengths and angles as well as additional descriptors such as the porosity.\cite{evans_predicting_2017, gaillac_speeding_2020}

An approach that is different from bagging and boosting is model stacking.
In boosting and bagging one usually uses the same base estimator, like a \gls{dt}, whereas in stacking one combines different learners and can use a meta learner to make the final prediction based on the prediction of the different models. This approach was, for example,  successfully used by Wang, who could reduce the error in predicting atomization energies by \SI{38}{\percent}, compared to the best single learner, using a stacked model.\cite{wang_significantly_2018}

%% file: main/4_learn_well.tex
\section{How To Learn Well: Regularization, Hyperparameter Tuning, and Tricks}\label{sec:learning_well}

\subsection{Hyperparameter Tuning}\label{sec:hyperpar_tuning}
Almost all \gls{ml} models have several \enquote{knobs} that need to be tuned to achieve good predictive performance.
The problem is that one needs to evaluate the model to find the best hyperparameters---which is expensive because this involves training the model with the set of parameters and then evaluating its performance on a validation set.
This problem setting is similar to the optimization of reaction conditions, where the execution of experiments is time-consuming, wherefore akin techniques are used.

The most popular way in the materials informatics community is to use grid search, where one loops over a grid of all possible hyperparameter combinations.
Unfortunately, this is not efficient as all the information about previous evaluations remains unused and one has to perform an exponentially growing number of model evaluations.
It was shown that even random search is more efficient than grid search, but especially Bayesian hyperparameter optimization was demonstrated to be drastically more efficient.\cite{bergstra_random_2012,bergstra_algorithms_2011}
This approach is formalized in \gls{smbo}.
The idea behind \gls{smbo} is that a (Bayesian) model is initialized with some examples and then used to select new examples that maximize a so-called acquisition (or selection) function \(a\), which is used to decide which points to choose next---based on the surrogate model.
The task of the acquisition function is to balance exploration and exploitation, i.e., to choose a balanced ratio between points \(\mathbf{x}\) where the surrogate model is uncertain (exploration) and points where \(f\), the target, is maximized (exploitation).
The need for an uncertainty estimate (to be able to balance exploration and exploitation) and the ability to incorporate prior knowledge makes this task ideally suited for Bayesian surrogate models.
For example, \glspl{gp} are used to model the expensive function in the \texttt{spearmint}\cite{snoek_spearmint_2019} and \texttt{MOE (Metric Optimization Engine)}\cite{clark_moe_2019} libraries.
The \texttt{SMAC} library\cite{lindauer_smac3_2019} on the other hand uses ensembles of \gls{rf}, which are appealing as they naturally allow incorporating conditional reasoning.\cite{dewancker_bayesian_2001}
A popular optimization scheme is the \gls{tpe} algorithm, which is implemented in the \texttt{hyperopt} package\cite{bergstra_hyperopt_2013} and which has an interface to the  \texttt{sklearn}\cite{scikit-learn} framework with the \texttt{hyperopt-sklearn} package.\cite{komer_hyperopt-sklearn_2014}
The key idea behind the \gls{tpe} algorithm is to model the hyperparameter selection process with two distributions; one for the good parameters and one for the bad ones. In contrast to that, \glspl{gp} and trees model it as dependent on the entire joint variable configuration.
The Parzen estimator, which is a non-parametric method to estimate distributions, is used to build these distributions.
To encode conditional hyperparameter choices, the Parzen estimators are structured in a tree.

\subsection{Regularization}
Many problems in which we are interested in the chemical sciences and materials science are ill-posed.
In some cases, they are not smooth, in other cases, not every input vector is feasible (only a fraction of all imaginable compounds exist at standard conditions), and in other cases, our descriptors might not be as unique as we would want them to be, or we have to deal with noise in the data.
Moreover, we often have to cope with little (and wide) data which can easily lead to overfitting.
To remedy these problems one can use regularization techniques.\cite{chen_different_2002}

Particularly powerful regularization techniques are based on physical or chemical insights, like the reaction tree heuristic form Rhone et al., where they only consider reaction products that are close to possible outcomes of a rule-based reaction tree.\cite{rhone_predicting_2019}

In the following, we will discuss more conventional techniques that require no physical or chemical insight and that are applicable to most problems.

\subsubsection{Explicit Regularization: Adding an Term or Layer}\label{sec:regularization}
The most popular way to avoid overfitting is to add a term that penalizes high model weights (\enquote{large slopes}) to the loss function:
\begin{equation}
    L(\mathbf{w}) = \lambda \|\mathbf{w} \|^p.
\end{equation}
In most of the cases, one uses either the Manhattan norm (\(p = 1\)), which is known as the \gls{lasso} (\(l_1\)), or the \(p = 2\), which is known as ridge regularization.
As we discussed previously (cf.\ section~\ref{sec:lasso}), the \gls{lasso} yields sparse solutions which can be seen as a general physical constraint.
\begin{figure}
    \centering
    \includegraphics[width=\textwidth]{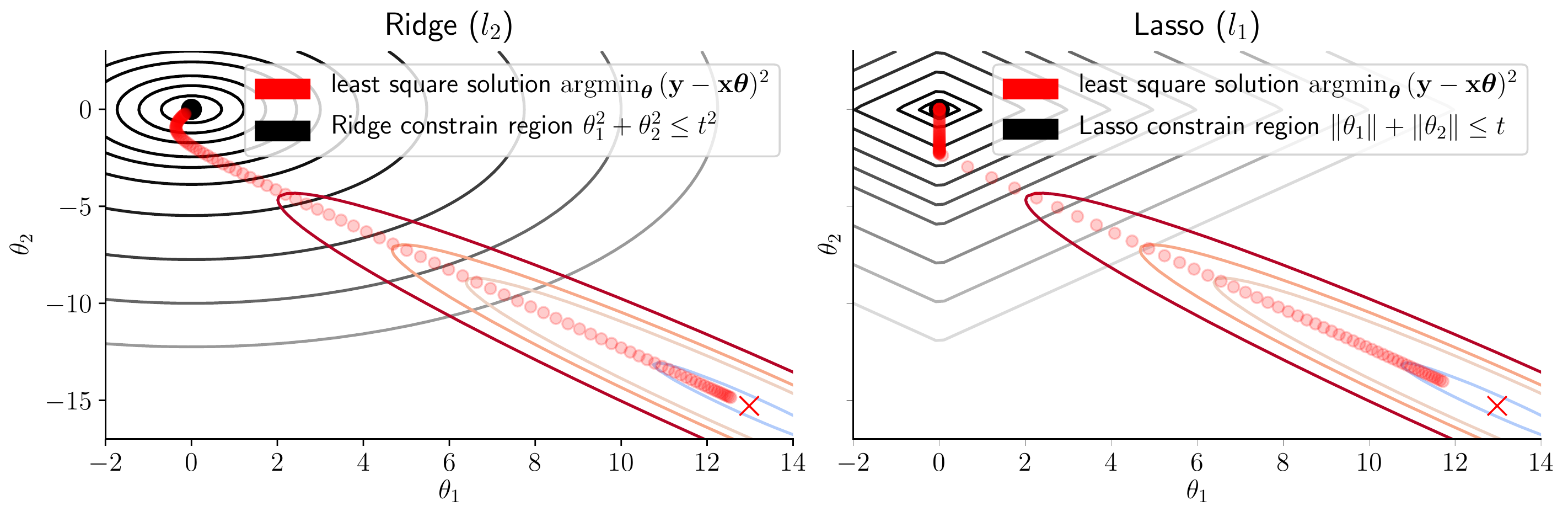}
    \caption{Visualization of the \(l_1\) and \(l_2\) constraints and the solution paths. The solution (dots) of the constrained optimization is at the intersection between the contours of the least square solution (red/blue colored ellipses indicating with the color the error for different parameter choices) and the regularization constraint region (black), which extent depends on \(\lambda \propto 1/t\).
        For \(\lambda = 0\), we recover the least square solution, for \(\lambda \to \infty\), the solution will lie at (0,0).
        If we increase \(\lambda\), the optimal solution will tend to be zero in one dimension at the vertex of the \gls{lasso} constrain region.
        For the ridge case, the smooth constrain region will lower the magnitude of the weights, but will not force them to exactly zero.
        Figure created based on an illustration in Tibshirani, Friedman and Tibshirani\protect{\cite{tibshirani_elements_2017}} and code by Sicotte.\protect{\cite{sicotte_ridge_2018}}}\label{fig:lasso_ridge}
\end{figure}
Since the ridge term shrinks high weights smoothly (there are no edges in the regularization hypercube, cf.\ Figure~\ref{fig:lasso_ridge}) it does not lead to sparse solutions but it can be seen as a way to enforce smoother solutions.
For example, we do expect potential energy surfaces to vary smoothly with conformational changes---a squiggly polynomial with high weights will hence be a bad solution that does not generalize.
Ridge regression can be used to enforce this when training models.
For both \gls{lasso} and ridge regression, we recover the original solution for \(\lambda \rightarrow 0\) and force it to zero for \(\lambda \to \infty\).

In \gls{dl} specific regularization layers are often used to avoid overfitting.
The most widely known technique, dropout, randomly disables some neurons from training.\cite{srivastava_dropout_2014}
As it is computationally cheap and can be implemented in almost any network architecture, it belongs to the most popular choices.

For trees, one usually uses pruning heuristics to limit overfitting.
One can either limit the number of splits or the maximum depth of the trees before fitting them or eliminate some leaves after fitting.\cite{esposito_comparative_1997}
This idea was also used in \glspl{nn}, e.g., by automatically deleting weights (also known as \gls{obd}).\cite{lecun_optimal_1990}
This procedure not only improves generalization but can also speed up inference (and training).\cite{molchanov_pruning_2016}

\subsubsection{Implicit Regularization: More Subtle Ways to Stop the Model From Remembering}
\begin{figure}
    \centering
    \includegraphics[scale=.9]{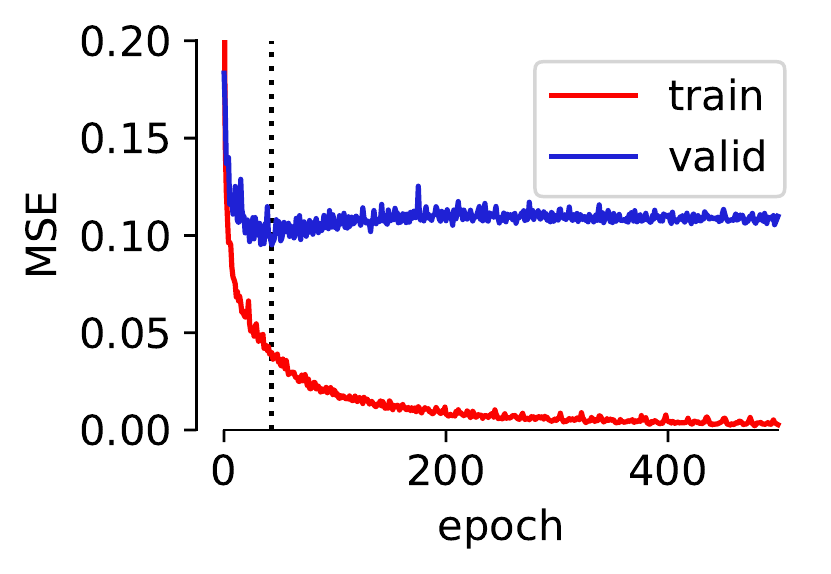}
    \caption{Example of early stopping.
        For this example, we trained a \gls{nn} (three hidden layers with \gls{relu} activation and 250, 100, and 10 neurons, respectively, followed by linear activation in the output layer) using the Adam optimizer,\cite{kingmaAdamMethodStochastic2014} to predict the \ce{CO2} uptake for structures in the database from Boyd et al.\protect{\cite{boyd_data_2019}} using \glspl{rac} and pore geometry descriptors as features.
        We can observe that after approximately 43 epochs (dotted vertical line) the training error still decreases, whereas the validation error starts to increase again.}\label{fig:early_stopping}
\end{figure}

But there are also other, more subtle ways, to avoid overfitting.
One of the simplest, most powerful and generally applicable techniques is early stopping.
Here, one monitors both the error on the training and a validation set over the training process and stops training as soon as the validation error no longer decreases (cf.\ Figure~\ref{fig:early_stopping}).\cite{goos_early_1998}
Another simple and general technique is to inject noise in the training process.\cite{noh_regularizing_2017, bishop_training_1995}

For the training of \gls{nn} batch normalization is widely used.\cite{ioffe_batch_2015}
Here, the input to layers of a \gls{dnn} is normalized in each training batch, i.e., the means and the variance are fixed in this way.
It was shown that this can accelerate training but it also acts as a regularizer as each training example no longer produces a deterministic value as it depends on which batch it is in.\cite{ioffe_batch_2015}

Similarly, the training algorithm itself, batched \gls{sgd}, was shown to induce implicit regularization due to its stochasticity as only a part of all training examples is used to approximate the gradient.\cite{lei_implicit_2018, hardt_train_2015-1}

In general, one finds that stochasticity is a theme underlying many regularization techniques.
Either through the addition of noise, by randomly dropping layers, or by making the prediction not fully deterministic by means of batch normalization.
This is in some sense similar to bagging as we also average over many slightly different models.\cite{goodfellow_deep_2016}

%% file: main/5_metrics.tex
\section{How to Measure Performance and Compare Models}\label{sec:metrics}

\begin{figure}
    \centering
    \includegraphics[width=\textwidth]{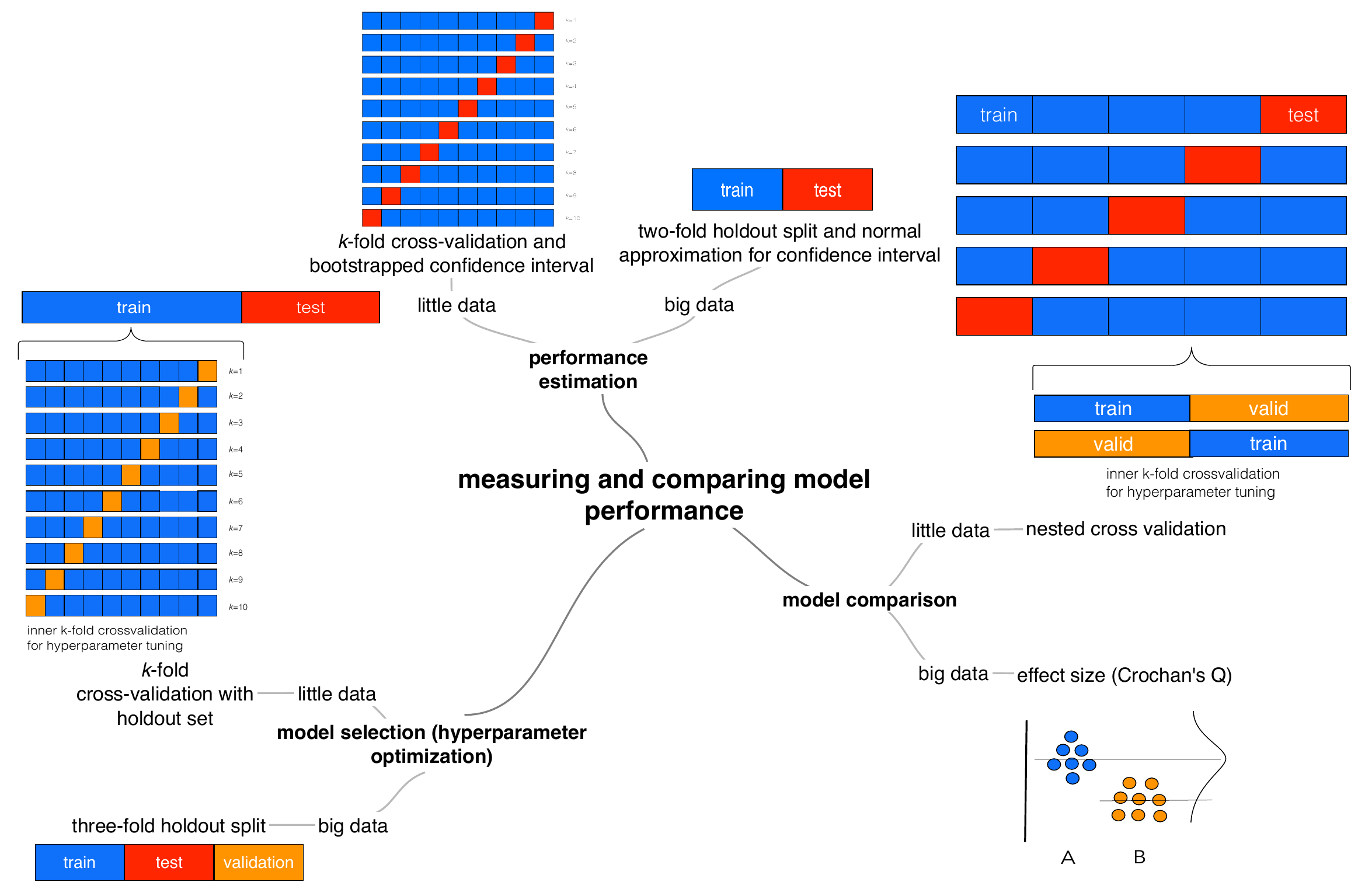}
    \caption{Model performance evaluation and comparison landscape, following the schema from Raschka.\protect{\cite{raschka_model_2018}}. Blue boxes represent training data, red ones test data. Validation data, for hyperparameter optimization, is shown with orange boxes. Differences between groups can be shown with Gardner-Altman plots where the data for each group are shown with dots and the effect size is shown with a bootstrapped confidence interval.}\label{fig:model_eval_landscape}
\end{figure}

In \gls{ml}, we want to create a model that performs well on unseen data for which we often do not know the underlying distribution when we train a model.
To optimize our models towards good performance on unseen data we need to develop surrogates for the performance on the unseen data (empirical error estimates).
An article by Sebastian Rascka gives an excellent overview (see Figure~\ref{fig:model_eval_landscape}) of different techniques for model evaluation and selection (the \texttt{mlxtend} Python library of the same author implements all the methods we discuss).\cite{raschka_model_2018}

\begin{figure}
    \centering
    \includegraphics[width=\textwidth]{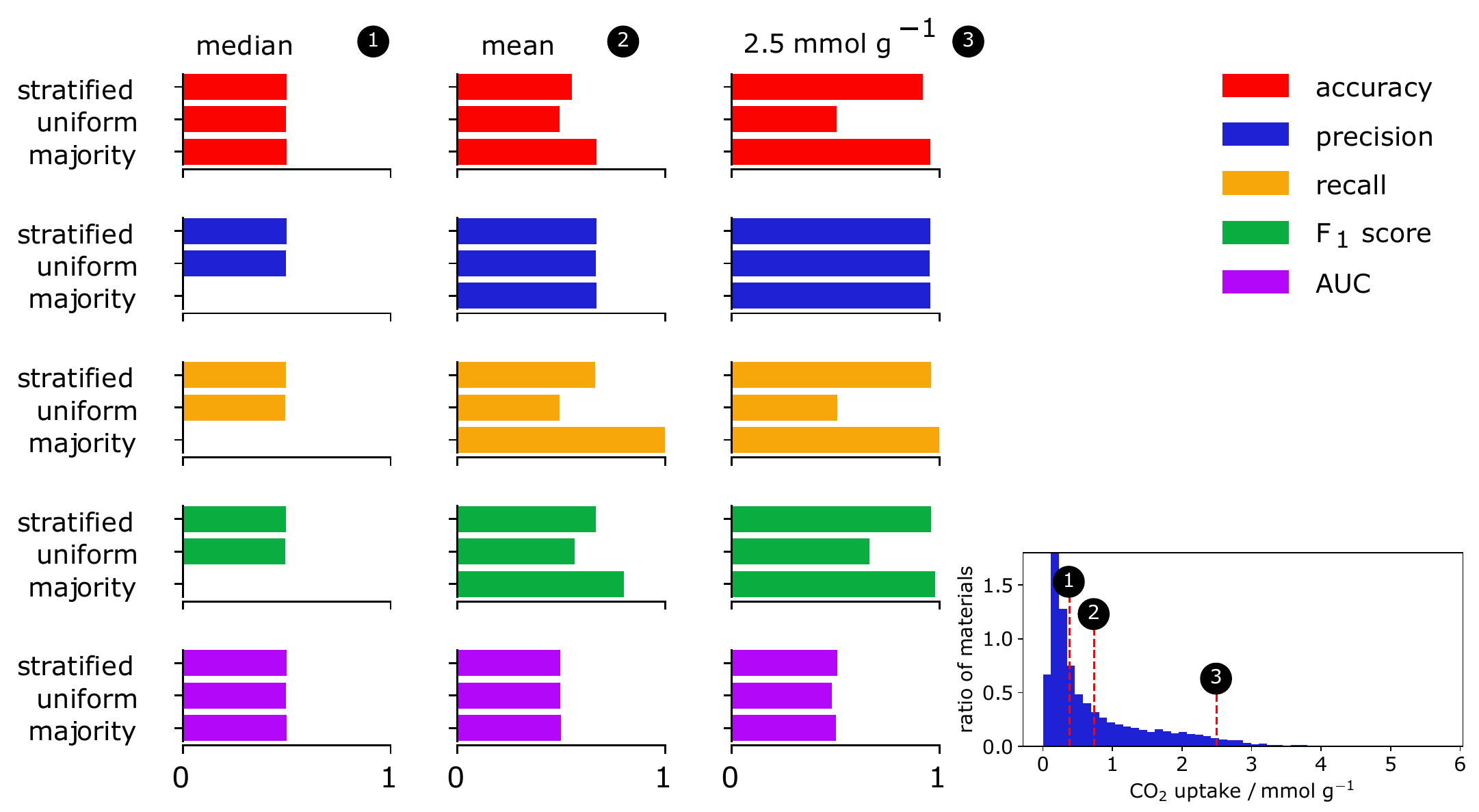}
    \caption{Influence of class imbalance on different classification metrics.
        For this experiment, we used different thresholds (median, mean \SI{2.5}{\milli\mole\per\gram}) for \ce{CO2} uptake to divide structures in \enquote{high performing} and \enquote{low performing} (see histogram inset). I.e., we convert our problem with continuous labels for \ce{CO2} uptake to a binary classification problem for which we now need to select an appropriate performance measure.
        We then test different baselines that randomly predict the class (uniform), i.e., sample from a uniform distribution, that randomly draw from the training set distribution (stratified) and that only predict the majority class (majority).
        For each baseline and threshold, we then evaluate the predictive performance on a test set using common classification metrics as the accuracy (red), precision (blue), recall (yellow), F\(_1\) score (green), and the \gls{auc} (pink).
        We see that by only reporting one number, without any information about the class distribution, one might be overly optimistic about the performance of a model, i.e., some metrics give rise to a high score even for only random guessing in the case of imbalanced distributions. For example, using a threshold of \SI{2.5}{\milli\mole\per\gram} we find high values for precision for all of our sampling strategies. Note that some scores are set to zero due to not being defined due to zero division.}\label{fig:classification_metrics}
\end{figure}

Often, one finds that models are selected, compared and evaluated based on only one single number, which is the \gls{mae} in many materials informatics applications.
But this might not be the optimal metric in all cases---especially since such global metrics depend on the distribution of data points (cf.\ Figure~\ref{fig:classification_metrics}) and in materials informatics we often do not only want a model that is \enquote{on average right} but one that can also reliably find the top performers.
Moreover, in some cases, we want to consider other parameters such as the training time, the feature set or the amount of training data needed.
Latter we can for example extract from learning curves in which a metric for the predictive performance, like the \gls{mae}, is plotted against the number of training points.\cite{cortes_learning_1993, amari_statistical_1993, huang_communication_2016}

The optimal (and feasible) model evaluation methodology depends on the amount of available data, the problem setting (e.g., if extrapolation ability is important) and the available computational resources.
We will discuss these trade-offs in the following.

\subsection{Holdout Splits and Cross-Validation: Sampling Without Replacement}
\begin{figure}
    \centering
    \includegraphics[width=.5\textwidth]{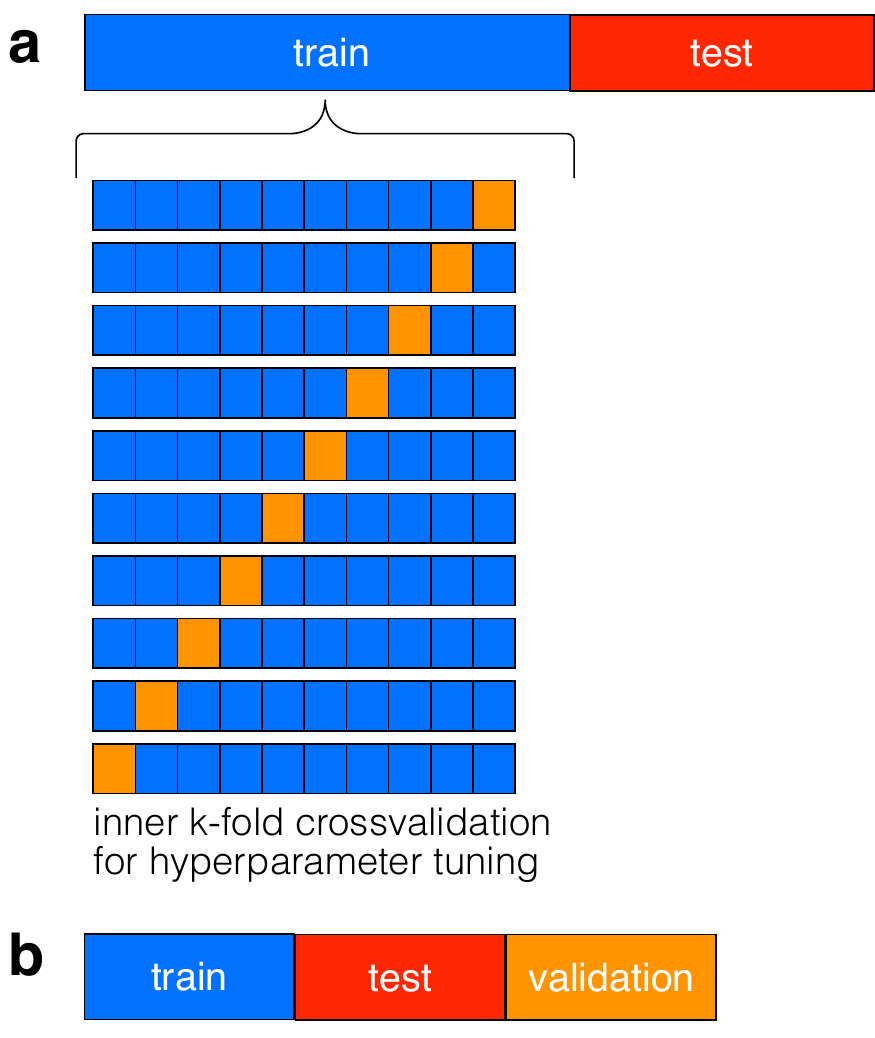}
    \caption{Comparison of model selection techniques for little and big data.
        For little data, one can use \(k\)-fold cross-validation with a separate test set (a) whereas the holdout method with three sets can be used for big data (b). In \(k\)-fold cross-validation the data is split into \(k\) folds and one loops over all the \(k\)-folds, using \(k-1\) folds as the training set and the \(k\)th fold for testing.}\label{fig:model_selection}
\end{figure}
The most common approach to measure the performance is to create two (or three) different data sets: the training set, on which the learning algorithm is trained on, the development (or validation set), which is used for hyperparameter tuning, and the test set, which is the ultimate surrogate for the performance on unseen data (cf.\ Figure~\ref{fig:model_selection}~b).
We do not use the test set for hyperparameter tuning to avoid data leakage, i.e., by tuning our hyperparameters on the test we might overfit to this particular test set.
The most common choice to generate these sets is to use a random split of the available data.

But there are caveats with this approach.\cite{raschka_model_2018}
First, and especially for small datasets, the number of training points is reduced (which introduces a pessimistic bias) in this way.
But at the same time, the test set must still be large enough to detect statistically significant differences (and avoid too much variance).
Second, one should note that random splitting can change the statistic, i.e., we might find different class ratios in the test set than in the training set, especially in case of little data (cf.\ the discussion for Figure~\ref{fig:pmof_stratification}).

The most common approach to deal with the first problem is \(k\)-fold cross-validation (cf.\ inner loop in Figure~\ref{fig:model_selection}~a), which is an ensemble approach to the holdout technique.
The idea here is to give every example the chance to be part of the training set by splitting the dataset into \(k\) parts, using one part for the validation and the remaining \(k-1\) parts for training and iterate this procedure \(k\) times.
A special case of the \(k\)-fold method is when the number of folds is equal to the number of data points, i.e., \(k=n\).
This case has a special name, \gls{loocv}, as it is quite useful for small datasets where one does not want to waste any data point, and it is also an almost unbiased estimator since nearly all data is used for the training.
But it comes with a high computational burden and a high variance (the training set merely changes but the test example can change drastically from one fold to the next).
Empirically, it was found that \(k=10\) provides a good trade-off between bias and variance for many datasets.\cite{kohavi_study_1995}
But, one needs to keep in mind that a pessimistic bias might not be a problem as in some cases, as in the model selection, we are only interested in relative errors of different models.

A remedy for the second problem of the holdout method (the change of the class distributions upon sampling) is stratification (cf.\ Figure~\ref{fig:pmof_stratification}), which is a name for the constraint that the original class proportions are kept in all sets.
To use this approach in regression one can bin the data range and apply stratification on the bins.

One caveat one should always keep in mind when using cross-validation is that the data splitting procedure must be applied before any other step of the modeling pipeline (filtering, feature selection, standardization, \ldots) to avoid data leakage.
The problem of performing for example feature selection before splitting the data is that feature selection is then performed based on all data (including the test data) which can bias which features are selected (based on the information from the test set)---which is an unfair advantage.

\subsection{Bootstrap: Sampling With Replacement}
An alternative to \(k\)-fold cross-validation is to artificially create new datasets by means of sampling with replacement, i.e., bootstrapping.
If one samples \(n\) examples from \(n\) data points with replacement, some points might not be sampled (in the limit of large data, only \SI{63.2}{\percent} will be sampled).\cite{efron_bootstrap_1986}
Those can be used as a \gls{loob} estimator of the generalization error and using 50--100 bootstraps, one also finds reliable estimates for confidence intervals (\textit{vide infra}).
Since only \SI{63.2}{\percent} of the examples are selected also this estimator is pessimistically biased and corrections like the 0.632(+) bootstrap\cite{efron_improvements_1997} have been developed to correct for this pessimistic bias. In practice, the bootstrap is more complicated than the \(k\)-fold cross-validation for the estimation of the prediction error, e.g., because the size of the test set is not fixed in the \gls{loob} approach.
Therefore, in summary, the 10-fold cross-validation offers the best compromise for model evaluation on modestly sized datasets---also compared to the holdout method which is the method of choice for large datasets (like for \gls{dl} applications).\cite{hawkins_assessing_2003}

\subsection{Choosing the Appropriate Regression Metric}
One of the most widely known metrics is the \(R^2\) value (for which several definitions exist, which are equal for the linear case).\cite{kvalseth_cautionary_1985}
The most basic definition of this score is the ratio between the variance of the predictions and the labels.
The problem is that in this way it can be arbitrarily low even if the model is correct and e.g., on Anscombe's quartet it has the same value for all datasets (cf.\ Figure~\ref{fig:anscombe}).
Hence, this metric should be used with great care.
The choice between the \gls{mae} and the \gls{mse} depends on how one wants to treat outliers.
If all errors should be treated equally one should choose the \gls{mae}, if large errors should get higher weights, one should choose the \gls{mse}.
Often, the square root of latter, the \gls{rmse}, is used to achieve a metric that is more easily interpretable.

To get a better estimate of the central tendency of the errors, one can use for example the median or trimean\cite{weisberg_central_1992} absolute error, which is a weighted average of the median, the first quartile, and the third quartile.

Especially in the process of model development it is valuable to analyze the cases with maximum errors by hand to develop ideas why the model's prediction was wrong.
This can for example show that a particular structure class is underrepresented---in which case it might be worth generating more data for this class or to try techniques for imbalanced learning (cf.\ section~\ref{sec:datasource}).
In other cases one might also realize that the feature set is inadequate for some examples or that features or labels are wrong.

\subsection{Classification}
\subsubsection{Probabilities That Can Be Interpreted as Confidence}
An appealing feature of many classification models is that they output probabilities and one might be tempted to interpret them as \enquote{confidence in the prediction}.
But this is not always possible without additional steps. Ensemble models, such as random forest for example tend to rarely predict high or low probabilities.\cite{niculescu-mizil_predicting_2005}
To remedy this, one can calibrate the probabilities using either Platt scaling or isotonic regression.
Platt scaling is a form of logistic regression where the outputs of the classifier are used as input for a sigmoid function and the parameters of the sigmoid are estimated using maximum likelihood estimation on a validation set.
In isotonic regression, on the other hand, one fits to a piecewise constant, stair-shaped, function which tends to be more prone to overfitting.
To study the quality of the probabilities that are produced by a classifier it is convenient to plot a reliability diagram in which the probabilities are divided into bins and plotted against their relative frequency.
A well-calibrated classifier should fall onto the diagonal of this plot.

\subsubsection{Choosing the Appropriate Classification Metric}
Especially in a case in which one wants to identify the few best materials, accuracy---although widely used---is not the ideal classification metric.
This is the case as accuracy is defined as the ratio of correct predictions over the total number of predictions and can, in the case of imbalanced classes, be maximized by always predicting the majority class---which certainly is not the desired outcome (cf.\ Figure~\ref{fig:classification_metrics}).
Popular alternatives to the accuracy are precision and recall:
\begin{align}
    \text{accuracy}  & = \frac{\text{true positive} + \text{true negative}}{\text{true positive} + \text{true negative} + \text{false positive} + \text{false negative}} \\
    \text{precision} & = \frac{ \text{true positives} }{ \text{true positives} + \text{false positives}}                                                                 \\
    \text{recall}    & =  \frac{\text{true positives}}{\text{true positives} + \text{false negatives}}.
\end{align}
The precision will be low when the model classifies many negatives as positives and the recall, on the other hand, will be low if the model misses many positive results.
Similar to accuracy these metrics have their issues, e.g., recall can be maximized by predicting only the positive class.
But as there is usually a trade-off between precision and recall, summary metrics have been developed. The \(F_1\) score tries to summarize precision and recall using a harmonic mean
\begin{equation}
    F_1 = \frac{2}{\frac{1}{\text{precision}} + \frac{1}{\text{recall}}} = 2 \, \frac{\mathrm{precision} \cdot \mathrm{recall}}{\mathrm{precision} + \mathrm{recall}},
\end{equation}
which is useful for imbalanced data.

Since the classification usually relies on a probability (or score) threshold (e.g., for binary classification we could treat all predictions with probability \(> 0.3\) as positive), \gls{roc} curves are widely used.
Here, one measures the classifier performance for different probability thresholds and plots the true positive rate [true positives / (true positives + false negatives)]  against the false positive rate \([1 - \mathrm{true\ negative} / (\mathrm{true\ negative} + \mathrm{false\ positive})]\).
A random classifier would fall on the diagonal of a \gls{roc} curve and the optimal classifier would touch the top left corner (only true positives).
This motivated the development of metrics that try to capture the full curve in only one number.
The most popular one is the \gls{auc},\cite{bradley_use_1997, jin_huang_using_2005} but also this metric is no silver bullet.
For example, care has to be taken when one wants to use the \gls{auc} as a model selection criterion.
For instance, the \gls{auc} will not carry information about how confident the models are in their predictions---which would be an important for model selection.\cite{lobo_auc_2008}

\begin{figure}
    \centering
    \includegraphics[scale=.9]{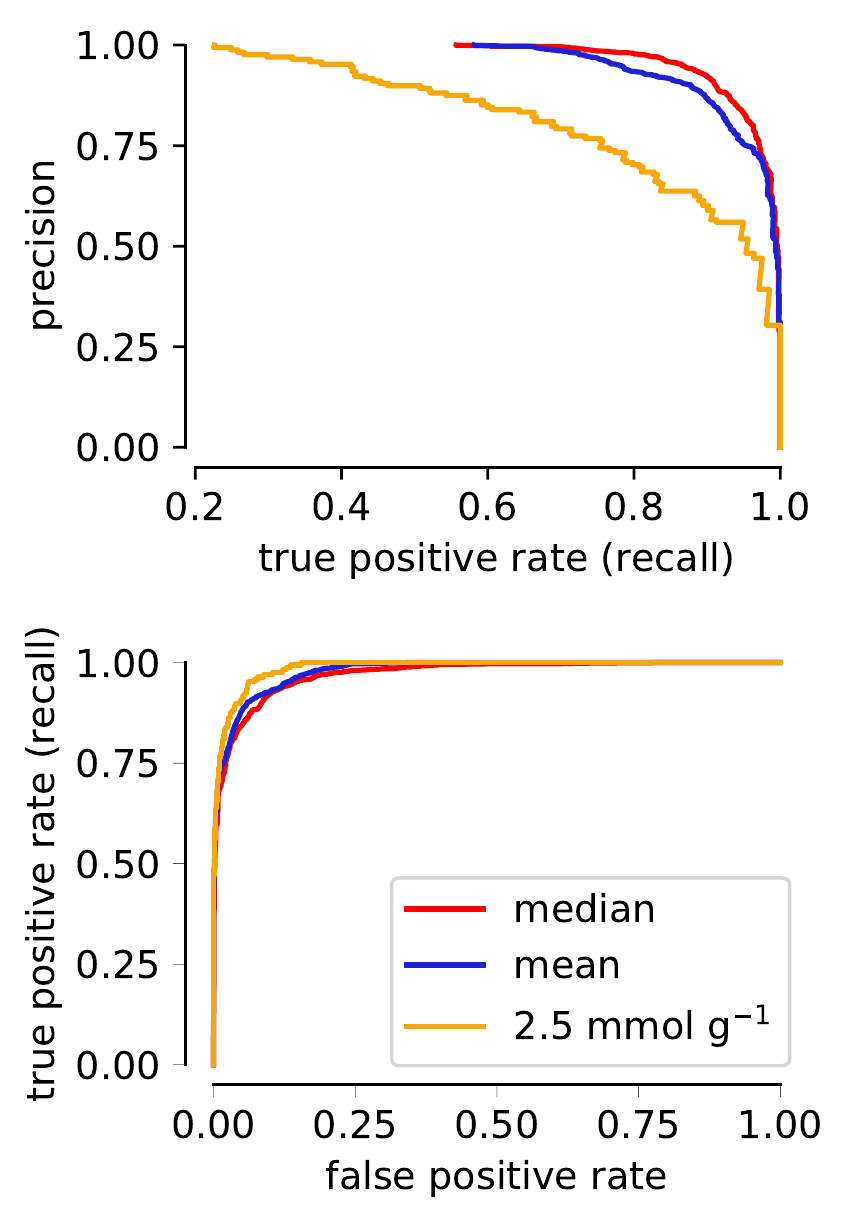}
    \caption{Comparison of precision-recall (top) and \gls{roc} (bottom)  curves  for different thresholds for the binary classification of \ce{CO2} uptake (same as for Figure~\ref{fig:pmof_stratification}).
        For this example, we fitted a \gls{gbdt} classifier on the dataset from Boyd et al.\protect{\cite{boyd_data_2019}}
        We can observe that for increasingly imbalanced class distributions (e.g., higher threshold for \enquote{high} performing MOFs, i.e., there are few of them) the difference between the shape of the precision-recall curve and the \gls{roc}, as well as the area under those curves, are more different.
        For imbalanced classes, the precision-recall curve (and the area under this curve) is a more sensible measure of model performance.
    }\label{fig:roc_vs_pr}
\end{figure}

Related to \gls{roc} curves are precision-recall curves.
They share the recall (true positive rate) with the \gls{roc} curves but plot it against the precision, which is, for a small number of positives, more sensitive to false positive predictions than the false positive rate. For this reason, we see an increasing difference between the \gls{roc} and the precision-recall curves with increasing class imbalance (cf.\ Figure~\ref{fig:roc_vs_pr}).\cite{wu_moleculenet_2018}

Usually, it is also useful to print a confusion matrix in which the rows represent the actual classes and the columns the predicted ones.
This table can be useful to understand between which classes misclassification happens and allows for a more detailed analysis than a single metric.
A particularly useful Python package is \texttt{PyCM} which implements  most of the classification metrics, including multi-class confusion matrices.\cite{haghighi_pycm_2018}

\subsection{Estimating Extrapolation Ability}
For some tasks, like the discovery of new materials, one wants models that can robustly extrapolate.
To estimate the extrapolation ability, specific metrics have been developed.
The \gls{lococv} technique proposed by Meredig et al.\ is an example of such a metric.\cite{meredig_can_2018}
The key idea is to perform clustering in the \(n\) cross-validation runs and leave one of the clusters out in the training set and then use this cluster as the test set.
Xiong et al.\ propose a closely related approach:
But instead of clustering the data in feature space they partition the data in target property space and use only a part of property space for training in a \(k\)-fold cross-validation loop and the holdout part for testing purposes.\cite{xiong_evaluating_2020}

Similar to that is the scaffolding splitting technique,\cite{wu_moleculenet_2018} in which the two-dimensional framework of molecules\cite{bemis_properties_1996} is used to separate structurally dissimilar molecules into training and test set.

\subsection{Domain of Applicability}\label{sec:domain_applicability}
In production, one would like to know if the predictions the model gives are reliable.
This question received particular attention in Cheminformatics\cite{sahigara_comparison_2012, varnek_machine_2012} with the emphasis of the \gls{reach} regulations on the reliability of \gls{qsar} predictions.\cite{weaver_importance_2008, tetko_critical_2008, gramatica_principles_2007}
Often, comparing the training and production distributions is a good starting point to understand if a model can work.
Here, one could first consider if the descriptor values of the production (test) examples fall into the range of the descriptors of the training examples (boundary box estimate).
This approach gives a first estimate if the prediction is made on solid ground, but it does not consider the distribution of the training examples, i.e., it might overlook \enquote{holes} in the training distribution.\cite{sahigara_comparison_2012}
But it is easy to implement and can, for example, be used during a molecular simulation with a \gls{nn} potential.
If a fingerprint vector outside the bounding box is detected, a warning could be raised (or the ab initio data can be calculated in an active learning setting).\cite{behler_representing_2014}

More involved methods often use clustering,\cite{stanforth_measure_2007} subgroup discovery,\cite{sutton_identifying_2019} and distances to the nearest neighbors of the test datum.
If this distance is greater than a threshold, which can be based on the average distance of the points in the training set, the model can be considered unreliable.
Again, the choice of the distance metric requires some testing.

More elaborate are methods based on the estimation of the probability density distribution of datasets and the evaluation of their overlaps.
These methods are closely related to \gls{kmm}---a method to mitigate covariate shift---which attempts to estimate the density ratio between test (production) and training distribution and then reweights the training distribution to more closely resemble the test (or production) distribution.\cite{quinonero-candela_covariate_2008}

\subsection{Confidence Intervals and Error Estimates}
The outputs of \gls{ml} models are random variables, with respect to the sampling, e.g., how the training and test set are created (cf.\ sections~\ref{sec:sampling} and~\ref{sec:learning_well})\cite{varoquaux_cross-validation_2018} and the optimization (one may end up in a different local minimum for stochastic minimization) and in some cases also with respect to the initialization.
Hence, one needs to be aware that there are errorbars around the predictions of any \gls{ml} model that one needs to consider when comparing models (cf.\ section~\ref{sec:comparing_models}), using the predictions, or simply to estimate the stability of a learning algorithm.

In addition, reliable error estimates are also needed to make predictions based on \gls{ml} models trustworthy.
Bayesian approaches automatically produce uncertainty estimates (cf.\ section~\ref{sec:bayesian}) but are not applicable to all problem settings.
In the following, we will review techniques that can be used to get error estimates in a model-agnostic way.


\subsubsection{Ensemble Approach}
Based on the insight that the outputs are random variables it seems natural to use an ensemble approach to calculate error bars.\cite{heskes_practical_1997}
One of the most popular ways to do this is to train the same model on different bootstraps of the dataset and then take the variance of this ensemble as a proxy for the error bars.
This is connected to two insights.
First, the training set is only one particular realization of a probability distribution  (which is the key idea behind the bootstrap), and second, the variance of the ensemble will be larger for cases in which the model is uncertain and has seen few training data.\cite{peterson_addressing_2017}

A related approach is to use to same data but to vary the architecture of the model, e.g., the number of hidden layers.
If the variance between the predictions in a particular part of chemical space is too large, this indicates that the models are still too \enquote{flexible} and need more training data in that particular region.\cite{behler_representing_2014} In contrast to the bootstrap approach, the ensemble surrogate can also be used in production, i.e., when we do not know the actual labels.

The fact that all ensemble or resampling approaches increase the computational cost motivated the development of other approaches for uncertainty quantification.

\subsubsection{Distance-based}
Most of the distance-based uncertainty surrogates are based on the idea that there is a relationship between the distance of a query example from the training set and the uncertainty of the prediction.
This is directly related to the concept of the domain of applicability, which we discussed above (cf.\ section~\ref{sec:domain_applicability}).
Although this approach may seem straightforward, there are caveats as the feature vector and the distance metric must be carefully chosen to allow for the calculation of a meaningful distance.
Also, this approach is not applicable to models that perform representation learning (cf.\ section~\ref{sec:messagepassing}).

This motivated Kulik and co-workers to develop uncertainty estimators that are cheaper than ensemble approaches and applicable to \gls{nn} in which feature engineering happens in the hidden layers.\cite{janet_quantitative_2019}
The idea of this approach is to use the distance in the latent space of the \gls{nn}, which is calibrated by fitting it to a conditional Gaussian distribution of the errors, as a surrogate for the uncertainty.

\subsubsection{Conformal Prediction} \label{sec:conformal_prediction}
A less widely known technique is conformal prediction, which is a rigorous mathematical framework that only assumes interchangeability (which is the case for \gls{iid} data, which is usually assumed for interpolative applications of \gls{ml}) and can be used for any learning framework with minimal cost.
Practically, given a test datum \(x_i\) and a significance level of choice \(\epsilon \in (0, 1)\), a conformal predictor calculates a prediction region \(\Gamma_i^{\epsilon} \subseteq Y\) that contains the ground truth \(y_i \in Y\) with a probability of \(1-\epsilon\).
\begin{figure}
    \centering
    \includegraphics[width=.8\textwidth]{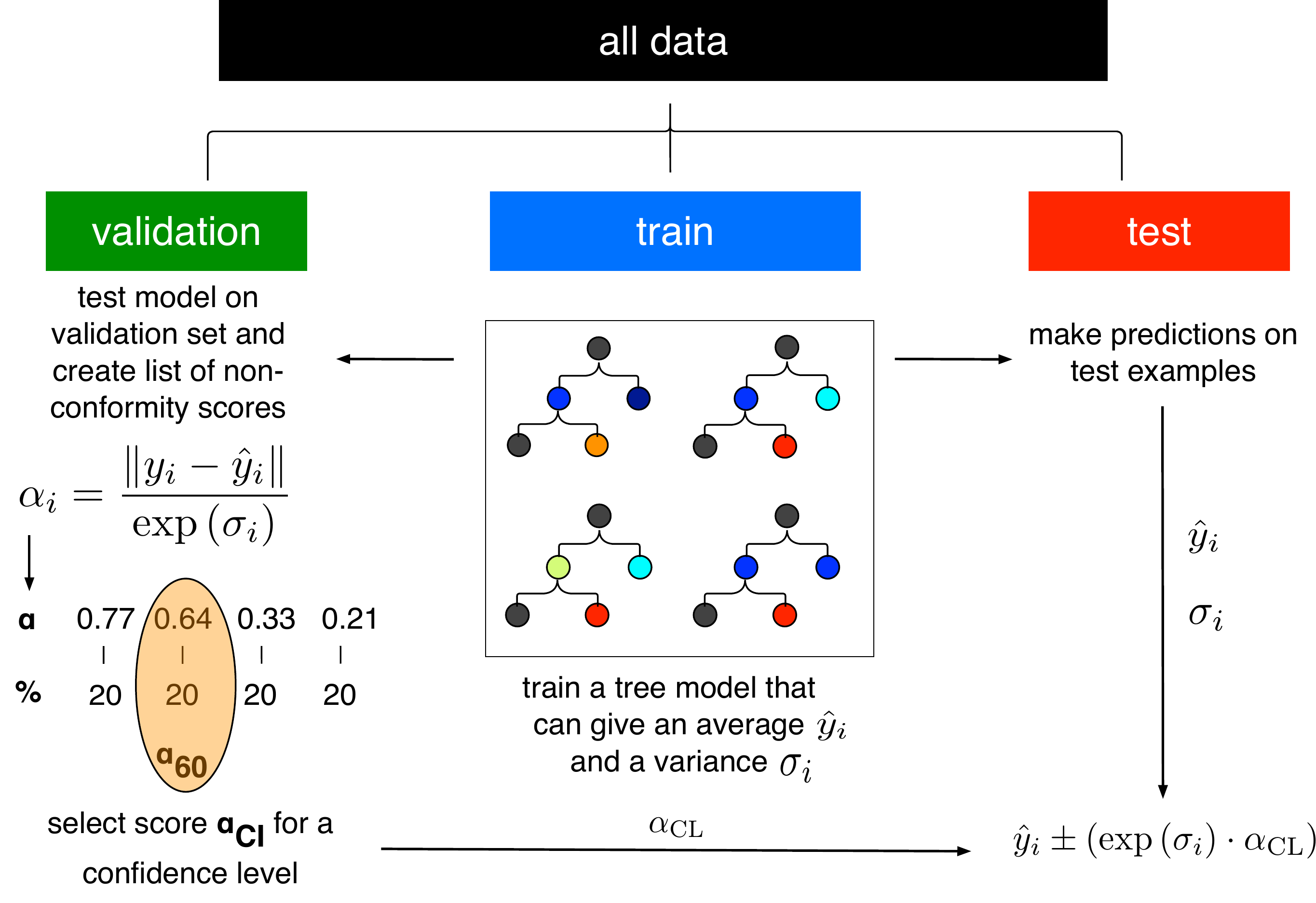}
    \caption{Example of inductive conformal prediction for the regression with tree models.}\label{fig:conformal}
\end{figure}
The idea behind this concept (cf.\ Figure~\ref{fig:conformal}) is to compute the nonconformity scores that measure the \enquote{uniqueness} of an example, using a nonconformity function, that can be the \gls{mae} (\(\|y_i - \hat{y}_i\|\)) for regression,\cite{papadopoulos_reliable_2011} on a calibration set (green in Figure~\ref{fig:conformal})
\begin{equation}
    \alpha = \frac{\|y_i - \hat{y}_i\|}{\exp\left(\sigma_i \right)},
\end{equation}
and that can be scaled by a measure of uncertainty, like the variance \(\sigma\) between the different trees in a random forest.\cite{papadopoulos_regression_2011, cortes-ciriano_concepts_2019}
One then sorts this list of nonconformity scores and can then choose the \(n\)th percentile (e.g., 60th percentile \(\alpha_\mathrm{CL}\) corresponding to a confidence level of \SI{60}{\percent}) and compute the prediction region for a test example (red in Figure~\ref{fig:conformal})
\begin{equation}
    \hat{y}_i \pm \left(\exp\left(\sigma_i\right) \cdot \alpha_\mathrm{CL} \right).
\end{equation}
The review by Cortés-Ciriano and Bender gives a more detailed overview of the possibilities and limitations of conformal prediction in the chemical sciences, especially for drug discovery,\cite{cortes-ciriano_concepts_2019}  and a tutorial by Shafer and Vovk provides more theoretical background.\cite{shafer_tutorial_2008}
A Python package that implements the conformal prediction framework is \texttt{nonconformist}.\cite{linusson_nonconformist_2019}

\subsection{Comparing Models}\label{sec:comparing_models}
One of the reasons why we focus on developing robust metrics and measures of variance is to be able to compare the predictive performance of different models.
Even though, as it is sometimes done, one could simply compare the metrics, such a comparison is not meaningful given that the predictions are random variables with an error bar around them.
The task of the modeler is to identify statistically significant and relevant differences in model performance.
There are a range of statistical tools that try to identify significant differences.\cite{dietterich_approximate_1998} Some of the fallacies and the most common techniques are discussed in a seminal paper by Dietterich.\cite{dietterich_approximate_1998}

If the difference between the error of two models is small, or not even statistically significant, one usually prefers, following Occam's Razor, the simpler model. One popular rule-of-thumb is the one-standard error rule according to which one chooses the simplest model within one standard error of the best performing one.\cite{raschka_model_2018, tibshirani_elements_2017}

The simplest approach to compare two models is to perform a \(z\)-test which practically means to check if their confidence intervals overlap---but this tends to often show differences even if there are none (due to not independent training and/or test sets in resampling approaches which results in a variance estimate that is too small).

It was found that one of the most reliable estimates is the \(5\times2\)-fold cross-validated \(t\)-test in which the data is split into training and test set five times.
For each fold, the two models that shall be compared are fitted on the training set and evaluated on the test set (and the sets are rotated afterward) which results in two performance difference estimates per fold.
The variance of this procedure can be used to calculate a \(t\)-statistic which was shown to have a low type-1 error---but also low replicability, i.e., different results are obtained when the test is rerun.\cite{bouckaert_choosing_2003}
Using statistical tests for model comparison leads to another problem when one does not only compare two models: Namely, the problem of multiple comparisons for which reasons additional corrections, like the Bonferroni correction, need to be applied.
Also, problems with the interpretability of \(p\)-values are also widely discussed outside the \gls{ml} domain. For this reason, it is not practical to use such statistical tests and estimation statistics might be the method of choice.\cite{halsey_reign_2019, claridge-chang_estimation_2016,claridge-chang_estimation_2016, halsey_fickle_2015}
It is more meaningful to compare effect sizes, e.g., differences between the accuracies of two classifiers, and the corresponding confidence interval than relying on a dichotomous decision based on the \(p\)-value.
A convenient format to do this can be a Gardner-Altman plot for bootstrapped performance estimates. Here, each measurement is plotted together with the means and the bootstrapped confidence interval of the effect size---with is particularly useful if the main focus of a study is to compare algorithms.
A Python package that create such plots is \texttt{DABEST}.\cite{ho_moving_2019}

\subsubsection{Ablation Studies}\label{sec:ablation}
When designing a new model, one often changes multiple parameters at the same time: the network architecture, the optimizer or the hyperparameters.
But to understand what caused an improvement, ablation studies, where one removes one part of the set of changes and monitors the change in model performance, can be used.
In several instances, it was shown that not a more complex model architecture but rather a better hyperparameter optimization is the reason for improved model performance.\cite{lipton_troubling_2018, melis_state_2017, sculley_winners_2018}
Understanding and reporting where the improvement stems from is especially important when the main objective of the work is to report a new model architecture.

\subsection{Randomization Tests: Is the Model Learning Something Meaningful?}\label{sec:randomization_tests}
\begin{figure}
    \centering
    \includegraphics{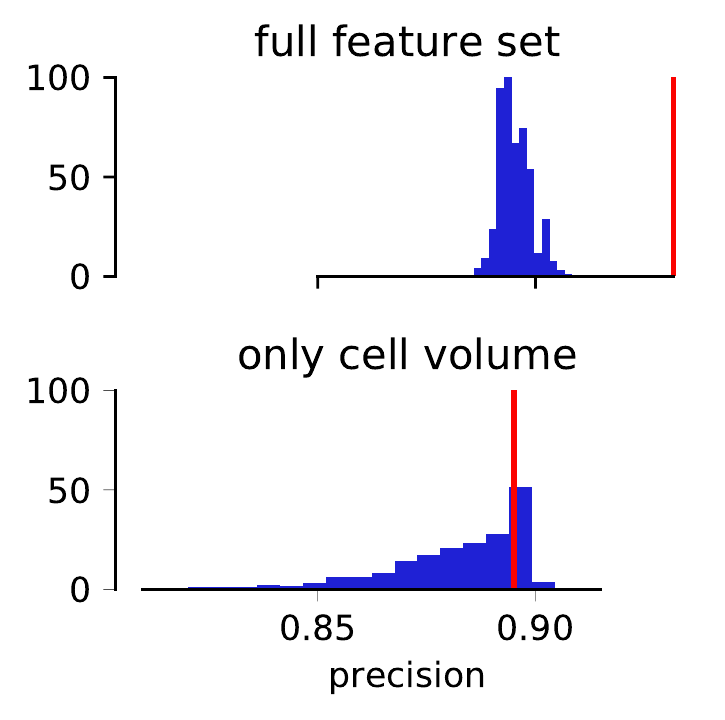}
    \caption{Example of a \(y\)-scrambling analysis to assess the significance of a performance metric. For this example, we built two simple \gls{gbdt} classifiers that attempt to classify the materials from Boyd et al.\protect{\cite{boyd_data_2019}} into structures with high and low \ce{CO2} uptake, respectively.
        We trained one of them using \glspl{rac} and pore property descriptors and the other one using only the cell volume as descriptor.
        We also measure the performance using the \gls{auc} and can observe that the model, trained on the full feature set, can capture a relationship in the real data (red line) and significantly (\(p<0.01\)) outperforms the models with permuted labels (bars). The model trained only on the cell volume does not perform better than random.}\label{fig:permutation}
\end{figure}

With the number of tested variables the probability of chance correlation increases---but ideally, we want a meaningful model.
Randomization tests, where either the labels or the feature vectors are randomized, are powerful ways to ensure that the model learned something for the right or at least reasonable reasons. \(y\)-scrambling,\cite{rucker_y-randomization_2007} where the labels are randomly shuffled is hence known as the \enquote{probably most powerful validation strategy} for \gls{qsar} (cf.\ Figure~\ref{fig:permutation}).\cite{kubinyi_qsar_2008} A web app available at \url{go.epfl.ch/permutationplotter} allows performing basic permutation analysis online and to explore how easy it is to generate \enquote{patterns} using random data.
The importance of randomization tests has recently been demonstrated for a model for C-N cross-coupling reactions.\cite{ahneman_predicting_2018}
Chuang and Keiser showed that \enquote{straw} models which use random fingerprints perform similarly to the original model trained on chemical features.\cite{chuang_comment_2018}
This showcases that randomization tests can be a powerful tool to understand if the model learns causal chemical relationships or not.

%% file: main/6_interpretation.tex
\section{How to Interpret the Results: Avoiding the Clever Hans}\label{sec:interpretatbility}

Clever Hans was a horse that was believed to be able to perform intellectual tasks like arithmetic operations (it was later shown that it did this by observing the questioner).
In \gls{ml}, there is also the risk that the user of a model can be deceived by the model and (unrightfully) believe that a model makes predictions based on physical or chemical rules it (supposedly) learned.\cite{lapuschkin_unmasking_2019}
In the following, we describe methods that can be used to avoid \enquote{black boxes} or to at least peek inside them to debug models, understand problems with the underlying dataset or to extract design rules.
This is especially valuable when high-level, physical and interpretable, features are used.

Unfortunately, the term \enquote{interpretable} is not well-defined.\cite{lipton_mythos_2016}
Sometimes, the term might be used to describe efforts to understand how the model works (e.g., if one could replicate what the model does using pen and paper) and in other instances it might be used to generate \textit{post-hoc} explanations that one could hope to use for inferring general design rules.
Still, one needs to keep in mind that we draw conclusions and interpretations only based on the model's reasoning (and the underlying training data) which can be a crude approximation of nature and without prove of predictive ability of the underlying models, such analyzes remain inutile.\cite{breiman_statistical_2001}
For a more comprehensive overview over the field of interpretable \gls{ml} we recommend the book from Molnar.\cite{molnar2019}

\subsection{Consider Using Explainable Models}
Cynthia Rudin makes a strong point against post-hoc explanations.\cite{rudin_stop_2019}
If they were completely faithful, there would be no need for the original model in the first place. Especially for high-stakes decisions a post-hoc explanation that is right \SI{90}{\percent} of the time is not trustworthy.
To avoid such problems, one can attempt to first use simple models that might be intrinsically interpretable, e.g., in terms of their weights.
Obviously, simple models such as linear regression reach their limitations of expressivity for some problems, especially if the feature sets are not optimal.

\begin{figure}
    \centering
    \includegraphics{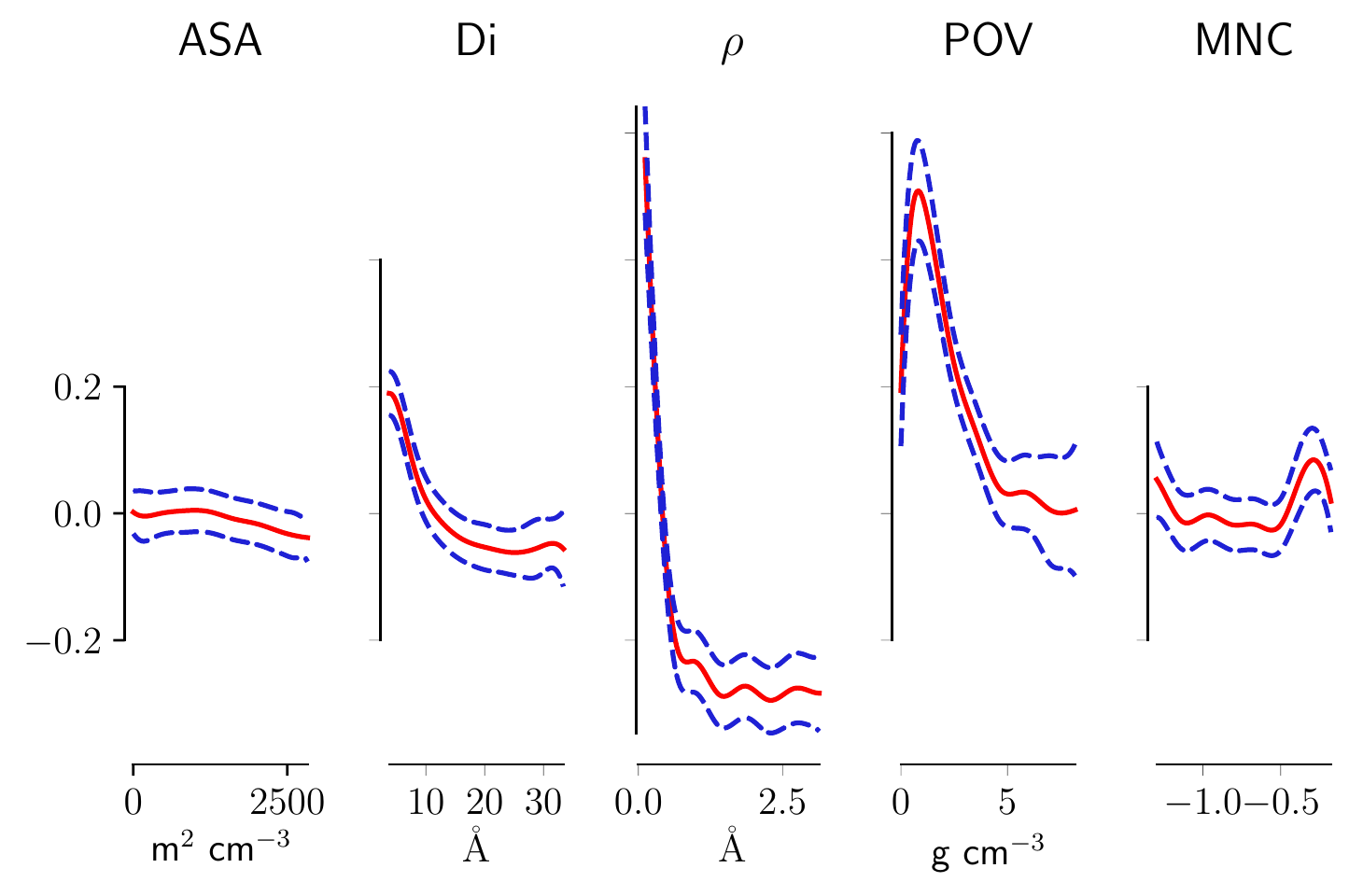}
    \caption{Examples for the splines for features that we used in a \gls{gam} to predict the \ce{N2} uptake for structures in the database of Boyd et al.\protect{\cite{boyd_data_2019}}
        Overall, we can observe that the surface area (ASA) and the minimum negative charge (MNC) have only a small influence on the prediction, whereas an increase in density leads to a stark decrease in the model outcome. }
    \label{fig:gams}
\end{figure}

\Glspl{gam} try to combine the advantages of linear models---for each feature one can analyze the weight (due to the additivity) and get confidence intervals around it---with flexibility to describe non-linear patterns (cf.\ Figure~\ref{fig:gams}).
This can be achieved by using the features via smooth, nonparametric functions, like splines:
\begin{equation}
    g(E_Y(y|x))=\beta_0+f_1(x_{1})+f_2(x_{2})+\cdots+f_p(x_{p}).
\end{equation}
\Glspl{gam} are hence additive models that describe the outcome by adding up smooth relationships between the target and the label.
Linear models can be seen as special case of \glspl{gam}, where the \(f\) are restricted to be linear.
One drawback of such additive models is that interaction effects have to be incorporated by creating a specific interaction feature like \(f(\text{density} \cdot \text{surface area})\) (in case one assumes that the interaction between the density and the surface area is important).
A modification of Caruana et al.\ includes pairwise interactions in the form of \(f(x_1, x_2)\) by default,\cite{caruana_intelligible_2015} and is implemented in the \texttt{interpret} package.\cite{interpretml_team_interpret_2019}

Similar to \gls{dt}---which we do not recommend due to their instability, and the fact that they are only interpretable when they are short---decision rules formulate if-then statements.
The simplest approach to create such rules is to discretize continuous variables and then create cross tables between feature values and model outcomes.
Afterwards, one can attempt to create decision rules based on the frequency of the outcomes, e.g., \enquote{if \(\rho > \SI{2}{\gram\per\centi\meter\cubed}\) then deliverable capacity low and if \(\SI{1}{\gram\per\centi\meter\cubed} < \rho < \SI{2}{\gram\per\centi\meter\cubed}\) then deliverable capacity high}.
Further developments provide safeguards against overfitting and multiple features can be taken into account by deriving rules from small \gls{dt}.
One of the main disadvantages of this method is that it needs discretization of features and targets, which induces steps in the decision surfaces. The \texttt{skater} Python library implements this technique.\cite{oracle_community_skater_2019}
Short \glspl{dt} are also used in the RuleFit algorithm.\cite{friedman_predictive_2008} Here, Friedman and Popescu propose to create a linear model with additional features that have been created by decomposing decision trees. The model is then sparsified using the \gls{lasso}.
The problem using this approach is that, although the features and rules themselves might be interpretable, there might be problems in combining them when there are overlapping rules.
This is the case since the interpretation of weights of linear models assumes that all other weights remain fixed (e.g., there can be problems with co-linear features).

Another form of interpretability can be achieved using \gls{knn} models.
As the model does not learn anything (cf.\ section~\ref{sec:instance}) the explanation for any prediction are the \(k\) closest examples from the training set---which works well if the dimensionality is not too high (cf.\ section~\ref{sec:curse_dimensionality}).

This also illustrates the two different levels of interpretation one might aim for.
Some methods like the coefficients of linear models or the feature importance rankings for tree models (see below) give us global interpretations (integrated over all data points), whereas other techniques like \gls{knn} give us local explanations for each sample and some techniques can give us both (like \gls{shap}, see below).

\subsection{Post-Hoc Techniques to Shine Light Into Black Boxes}
The most popular approach to extract interpretation from \gls{ml} models in the materials informatics domain is to use feature importance---often based on where in a tree model a feature contributed to a split (an early split is more important) or how good this split was, e.g., by measuring how much it reduces the model's variance.
Most of these methods fall under the umbrella of sensitivity analysis,\cite{cortez_using_2013, saltelli_sensitivity_2002} which is also widely known as the study of how uncertainty in the output of models is related to the uncertainties in the inputs by studying how the model reacts to changes in the input.
Unfortunately, there are problems with several of those techniques---like the fact that some of them are biased towards the high-variance features.\cite{strobl_bias_2007, altmann_permutation_2010}

\begin{figure}
    \centering
    \includegraphics[width=\textwidth]{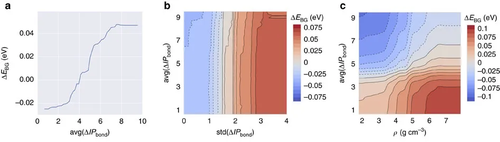}
    \caption{Partial dependence plots of \(\Delta \mathrm{IP}_\mathrm{bond}\).
        The first plot reflects physical intuition that more polar bonds (larger ionization potential difference) have larger band gaps.
        Interactions between two features are shown in b and c.
        For example, we can observe that materials with higher density, \(\rho\), and lower average \(\Delta \mathrm{IP}_\mathrm{bond}\) statistically have a larger band gap.
        Figure reprinted from Isayev et al.\protect{\cite{isayev_universal_2017}}}\label{fig:isayev_partial_dependence}
\end{figure}

There are several model-agnostic alternatives that attempt to avoid this problem.
Isayev et al.\ used partial dependence plots (cf.\ Figure~\ref{fig:isayev_partial_dependence}) to interrogate the influence of the features and their interaction on the model outcome.\cite{isayev_universal_2017}
This can be done by marginalizing over all the other features \(x_c\) which are not plotted:
\begin{equation}
    \hat{f}_{x_s}(x_s) = \int \hat{f}(x_s, x_c) \, \mathrm{d}\mathbb{P}(x_c).
\end{equation}
The integral over all the other features  \(x_c\) is in practice estimated using \gls{mc} integration.
By integration over all but two variables, one can generate heatmaps that show how the target property varies as a function of the features assuming that those features are independent of all the other features.
The latter assumption is the biggest problem with partial dependence plots.

Another powerful method, the permutation technique, shares this problem.
In the permutation technique one tries to estimate the global importance of features by measuring the difference between the error of a model trained with fully intact feature columns and one where the values for the feature of interest are randomly permuted.
To remedy issues due to correlated features\cite{hooker_please_2019} one can permute them together.
The permutation technique was for example used by Moosavi et al.\ to capture the importance of synthesis parameters in the synthesis in the of HKUST-1.\cite{moosavi_capturing_2019}

\begin{figure}
    \centering
    \includegraphics[width=\textwidth]{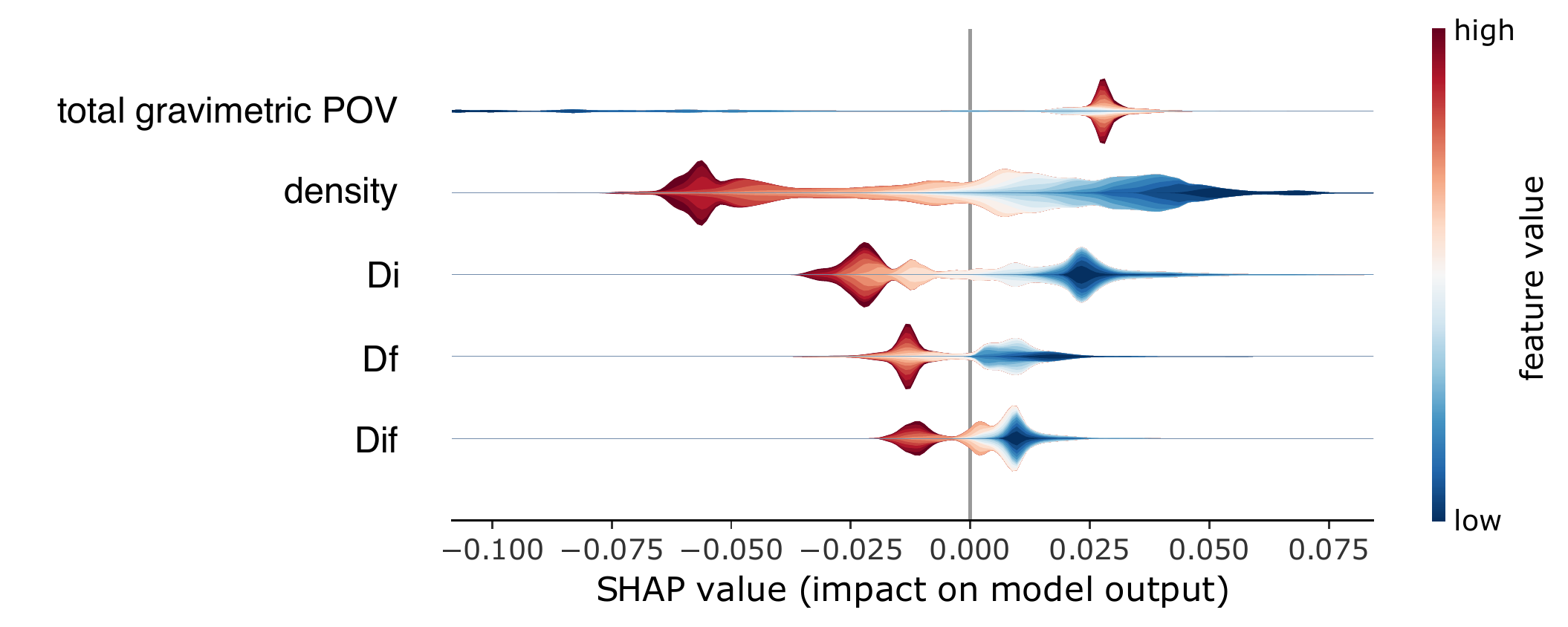}
    \caption{Summary plot of \gls{shap} feature importance for a \gls{gbdt} model, trained using pore properties descriptors (POV: pore occupiable volume, D\(_i\): diameter of the largest included sphere, D\(_f\): diameter of the largest free sphere, \(D_{if}\): diameter of the largest included sphere along the free path)  to predict \ce{N2} uptake from the \ce{CO2}/\ce{N2} mixture data from Boyd et al.\protect{\cite{boyd_data_2019}}
        Note that we chose the \ce{N2} uptake as one expects that the pore geometry is more important than the chemistry, which simplifies the example.
        The violins in this plot show the distributions of the importance, i.e., the spread of the \gls{shap} values (along the abscissa) and how many samples we have for different \gls{shap} values (the thickness of the violin).
        The coloring encodes the value of the features, red meaning high feature values whereas blue represents low feature values (e.g., high vs. low density).
        The \gls{shap} value is shown on the abscissa and reflects how a particular feature (one feature per row) with a value represented by the color impacts the prediction. For example, a high density (red color in the second row) leads to lower predictions for \ce{N2} uptake (indicated by negative \gls{shap} values).}\label{fig:shap_example}
\end{figure}

One technique that attempts to provide consistent interpretations, avoiding most of the aforementioned problems, on both local and global level is the use of Shapley values.
The idea is based on a game-theoretical problem in which one wants to estimate the optimal payout for a player.
The players in the case of \gls{ml} are the features.
Again, this involves marginalization over all the features we are not interested in but considering all possible ways in which the feature can enter the model (similar to all possible teams a player could be in).
But considering all possible combinations of features is computationally unfeasible wherefore Lundberg and Lee developed new algorithms to calculate it efficiently (exact for trees and approximate for kernel methods, see Figure~\ref{fig:shap_example} for an example).\cite{lundberg_unexpected_2016, lundberg_unified_2017, lundberg_consistent_2018}
In contrast to partial dependence plots, which show average effects, the plots of the feature values against the importance will appear dispersed in the case of the Shapley technique, which can give more insight into interaction effects.
This technique started to find use in materials informatics.
For example, Korolev et al.\ used \gls{shap} values to probe their \gls{ml} model for partial charges of \glspl{mof}.
There they, for example, find that the model (a \gls{gbdt}) correctly recovers that the charge should decrease with increasing electronegativity.\cite{korolev_transferable_2019}
But it also highlights that (post-hoc) interpretability methods are not the only puzzle-stone towards interpretability. If the features themselves are not intuitive quantities (like the \gls{rdf}) no post-hoc interpretability technique will make it easier to create design rules---but it still can be useful for debugging of models.

Still, one should keep in mind that it has also been shown that there can be stability problems with \gls{shap}.\cite{alvarez-melis_robustness_2018}

For \glspl{nn} techniques that analyze the gradients are popular.
The magnitude of the partial derivative of the outputs with respect to the input was for example also used by Esfandiari et al.\ to assign importance values to the features they used for their \gls{nn} that predicts the \ce{CO2}/\ce{CH4} separation factor.\cite{esfandiari_using_2017}

Related is work by Umehara et al.\ who used gradient analysis to visualize the predictions of neural networks and showed that this analysis can reveal structure-property relationships for the design of photoanodes.\cite{umehara_analyzing_2019}
This technique, where one calculates the partial derivative in the \(i\)th feature dimension for the \(j\)th sample
\begin{equation}
    G_{ij} = \left| \frac{\partial f (\mathbf{x}) }{\partial x_i} \right|,
\end{equation}
is also known as saliency mapping.
Thanks to libraries like \texttt{tf-explain}\cite{meudec_tf-explain_2019} and \texttt{keras-vis}\cite{kotikalapudi_keras-vis_2019} appealing visualizations of model explanations are often only one function call away, but one should be aware that there are many caveats wherefore some sanity checks (like randomization tests or addition of noise) should be used before relying on such a model interpretation.\cite{adebayo_sanity_2018, alvarez-melis_robustness_2018}

\subsection{Auditing Models: What Are Indirect Influences?}\label{fig:indirect_feature_importance}
In the mainstream \gls{ml} community algorithmic fairness, e.g., to prevent racial bias, is a pressing problem.
One might expect that this is not a problem in scientific datasets.
Jia et al.\ showed that also reaction datasets are anthropogenically biased, e.g.\ by experimenters selecting reactants and reaction conditions that they know to work (Matthew effect mechanism\cite{merton_matthew_1968})---which is similar to the bias towards certain reaction types which Schneider et al.\ found in the U.S. patent database.\cite{schneider_big_2016}
Jia et al.\ trained \gls{ml} models on randomly selected reaction conditions and on larger, human-selected reaction conditions from the chemical literature and found that the models trained on random conditions outperform the models trained on (anthropogenically biased) conditions from the literature for the prediction of crystal formation of amine-templated metal oxides---due to a better sampling of feature space.\cite{jia_anthropogenic_2019}

Some features in our feature set might encode such anthropogenic biases.
Auditing techniques, as for example implemented in the \texttt{BlackBoxAuditing} package,\cite{adler_auditing_2016} try to estimate such indirect influences.
In a high-stake decision case, an example for indirect influence might be a zip-code feature that is a proxy for ethnicity---which we then should drop to avoid that our model is biased due to the ethnicity.

In scientific datasets, such indirect influences might stem from artifacts in the data collection process or non-uniqueness of specific identifiers (which could be interpreted in different ways by different tools).\cite{ramakrishnan_quantum_2014}
The estimation of indirect influences works by perturbing a feature in such a way (typically by random perturbation) that it no longer can be predicted by the other features.
Similar to the perturbation techniques discussed above for (direct) feature importance, one then measures the drop in performance between the original model and the one with the perturbed feature.
And indeed Jia et al.\ found the indirect feature importance for models trained for the reaction conditions in literature conditions to be linearly correlated to those for models trained on randomly selected conditions---except for the features that describe the chemistry of the amines.\cite{jia_anthropogenic_2019}

%% file: main/9_applications_to_mof.tex
\section{Applications of Supervised Machine Learning}\label{sec:applications}

As we mentioned in the introduction, \gls{ml} in the field \glspl{mof}, \glspl{cof}, and related porous materials relies on the availability of tens of thousands experimental structures,\cite{chung_computation-ready_2014, chung_advances_2019} and to a large extent on the large libraries of (hypothetical) structures that have been assembled and scrutinized with computational screenings.\cite{boyd_computational_2017, wilmer_large-scale_2012, simon_materials_2015, ahmed_exceptional_2019, simon_what_2015, boyd_data_2019, rosen_identifying_2019}
But even with the most efficient computational techniques, like force-field-based simulations, the total number of materials has become so large that it is prohibitive to screen all possible materials for any given application.
In addition, brute force screening is not the best way to uncover structure-property relationships.
More importantly, other phenomena, especially electronic properties or fuzzy concepts such as synthesis or reactivity, are so complex that there is no good theory to describe the phenomenon (reaction outcomes) or that the theory is too expensive for a large-scale screening (electronic phenomena).
For these reasons, researchers started to employ (supervised) \gls{ml} for porous materials.

In Table~\ref{tab:overview_table} we give an overview of the techniques which we discussed in the first part and some examples where they have been used in the field of porous materials and will discuss those examples in more detail in the following. It is striking that many of the techniques that we discussed in the first part did not find an application for porous materials. We discuss those possibilities in more detail in the following and the outlook.
\input{main/connection_table}

\subsection{Gas Storage and Separation}
Gas storage is one of the simplest screening studies.
Most screening studies focus on designing a material with the highest deliverable capacity, which is defined as the difference between the amount of gas a material can adsorb at the high, charging, pressure minus the amount of gas that stays in the material at the lowest operational pressure.\cite{simon_optimizing_2014}
Hence, these screening studies typically require two data points on the adsorption isotherms.
Most of the studies for gas storage have focused on methane\cite{makal_methane_2012,mason_evaluating_2013,getman_review_2012,gomez-gualdron_exploring_2014,simon_materials_2015,simon_optimizing_2014} and hydrogen.\cite{suh_hydrogen_2012,getman_review_2012,goldsmith_theoretical_2013}

Gas separations are another important application of porous materials\cite{li_selective_2009,li_metalorganic_2012}.
Given the importance of reducing \ce{CO2} emission,\cite{bui_carbon_2018, smit_introduction_2014} a lot of research has focused on finding materials for carbon capture, both experimentally\cite{dalessandro_carbon_2010,ding_carbon_2019, sumida_carbon_2012, trickett_chemistry_2017} as well as by means computational screening studies.\cite{yazaydin_screening_2009,lin_silico_2012,jain_virtual_2019}
Gas separations require the (mixture) adsorption isotherms of the gases one would like to separate.
In most screening studies, the mixture isotherms are predicted from the pure component isotherms using ideal adsorbed solution theory.
For gas separations, the objective function is less obvious.
Of course, one can argue that for a good separation the selectivity and working capacity are important, but one often has to carry out a more detailed design of an actual separation process to find what are the key performance parameters one would like to screen.

Most screening studies focus on thermodynamic properties.
Yet, if the diffusion coefficients of the gases that need to be adsorbed are too low, excellent thermodynamic properties are of little use.
Therefore, it is also important to screen for transport properties. However, only a few studies have been reported that study the dynamics.\cite{keskin_screening_2007,keskin_efficient_2009,kim_large-scale_2013,mace_automated_2019}
The conventional method to compute transport properties, like diffusion coefficients, is molecular dynamics.
However, depending on the value of the diffusion coefficients these simulations can be time-consuming.\cite{keskin_efficient_2009}
Because of these limitations, free energy-based methods have been developed to estimate the diffusion coefficients from transition state theory (cf.\ ref.~\citep{kim_large-scale_2013,mace_automated_2019}).

A popular starting point is methane storage, a topic which has been studied extensively.\cite{makal_methane_2012,mason_evaluating_2013}
As in most of the screening studies methane is considered a united atom without net charge, and without dipole or quadruple, the interactions with the framework atoms are described by the Van der Waals interactions.\cite{simon_materials_2015}
As these interactions do not vary much from one atom in the framework to another, one can expect that methane storage is dominated by the pore topology rather than the specific chemistry.
Hence, most of the \gls{ml} models are trained using simple geometric properties such as the density, the pore diameter or the surface area.
These characteristics are obviously directly related to physisorption, but sometimes multicolinear, which can lead to problems with some algorithms as we discussed above (cf.\ section~\ref{sec:correlation}).

For gases like \ce{CO2} or \ce{H2O}, the specific chemistry of the material will be more significant.
For these gases, the pore geometry descriptors will not be sufficient and we will need descriptors that can describe phenomena that involve specific chemical interactions.
One also has to keep in mind that conventional high-throughput screenings can have difficulties to properly describe the strong interactions of \ce{CO2} with \glspl{oms}.\cite{ongari_origin_2017}
For example, especially for the low-pressure regime of the adsorption isotherm of \ce{CO2}, the method used to efficiently (i.e., avoiding \gls{dft} calculations for each structure) assign partial charges to the framework atoms can lead to systematic errors in the results.

One also needs to realize that descriptors that are only based on geometric properties have limited use for materials' design.
Even if we find a model that relates pore properties with the gas uptake and then use optimization tools (like particle swarm optimization, genetic algorithms or random searches\cite{ohno_machine_2016}) to maximize the uptake with respect to the pore properties there still remains the burden of proof as a given combination of pore properties might optimize gas adsorption in our model but might not be feasible or synthesizable (cf.\ section~\ref{sec:hypothetical_databases}).

\subsubsection{Starting on Small Datasets}
As in other fields of chemistry, \gls{ml} for porous materials developed from \gls{qspr} on small datasets (tens of data points) to the use of more complex models, such as neural networks, on large datasets with hundred thousands of data points.
Generally, one needs to keep in mind that all boundaries or trends that are observed in \gls{qspr} studies can either be due to underlying physics or limitations of the dataset, which necessarily does not explore some areas of the enormous design space of \glspl{mof}.\cite{wilmer_structureproperty_2012}

As in \gls{cadd}, the first studies also used high-level descriptors.
Kim reported one of the first \gls{qspr} for gas storage in \glspl{mof}.\cite{kim_quantitative_2007}
Inspired by previous works in \gls{cadd}, they calculated descriptors like the polar surface area and the molar refractivity but also used the iso-value of the electrostatic potential to create a model for the \ce{H2} adsorption capacity of ten \glspl{mof}.
Similar to that, Amrouche et al.\ built models based on descriptors of the linker chemistry of \glspl{zif}, like the dipole moment, as well as descriptors of the adsorbing gas molecules to predict the heat of adsorption for 15 \glspl{zif} and eleven gas molecules.\cite{amrouche_prediction_2012}
Also Duerinck et al.\ used descriptors like polarizability and dipole moment, which are familiar from cheminformatics, to build a model for the adsorption of aromatics and heterocyclic molecules on a set of 22 functionalized MIL-47 and found that polarizability and dipole moment are the most important features.\cite{duerinck_pulse_2012}

\paragraph{Pore Geometry Descriptors}
Sezginel et al.\ used a small set of 45 \glspl{mof} and trained multivariate linear models to predict the methane uptake based on geometric properties,\cite{sezginel_multivariable_2015} and also Yilidz and Uzun used a small set of 15 structures to train a \gls{nn} to predict methane uptakes in \glspl{mof} based on geometric properties.\cite{yildiz_prediction_2015}
Wu et al.\ increased the number of structures in their study to 105 and built a model that can predict the \ce{CO2}/\ce{N2} selectivity of \gls{mof} based on the heat of adsorption and the porosity.\cite{wu_revealing_2012}
They used this relationship to create a map of the interplay between the porosity and the heat of adsorption and their impact on the selectivity which showed that simultaneously increasing the heat of adsorption while decreasing the porosity is a route to increase selectivity for this separation.

\subsubsection{Moving to Big Data}
\paragraph{Development of New Descriptors}
Fernandez et al.\ started working with considerably larger sets of structures and also introduced more elaborate techniques like \gls{dt} or \glspl{svm}, which reflect the shift from cheminformatics with (multi)linear models on small datasets to complex nonlinear models trained on large datasets, that also other fields of chemistry experienced.\cite{fernandez_large-scale_2013}

In their first work,\cite{fernandez_large-scale_2013} they used geometric descriptors such as the density or the pore volume to predict the methane uptake but then realized\cite{fernandez_atomic_2013} the need to introduce more chemistry to build predictive models for carbon dioxide adsorption.
They did so by introducing the \gls{ap} weighted \gls{rdf} (\gls{ap}-\gls{rdf}).
For different fields of chemistry different encodings of the \gls{rdf} emerged as powerful descriptors (cf.\ section~\ref{sec:rdf}) and also Fernandez et al.\cite{fernandez_atomic_2013} achieved good predictive performance for gas adsorption using this descriptor and could also show that the principal components of this descriptor show good discrimination of geometrical and gas adsorption properties.
Importantly, they also demonstrated that \gls{ml} techniques can be used for pre-screening purposes to avoid running \gls{gcmc} simulations for low-performing materials.
For this, they trained a \gls{svc} using their \gls{ap}-\gls{rdf} as descriptors and found that this classifier correctly identifies 945 of the top 1,000 \glspl{mof} while only flagging \SI{10}{\percent} for further investigation with \gls{gcmc} simulations.
Recently, also Dureckova et al.\ used this descriptor to screen a database of hypothetical materials with more than 1000 topologies for \ce{CO2}/\ce{N2} selectivity.\cite{dureckova_robust_2019}

\begin{figure}
    \centering
    \includegraphics[width=.75\textwidth]{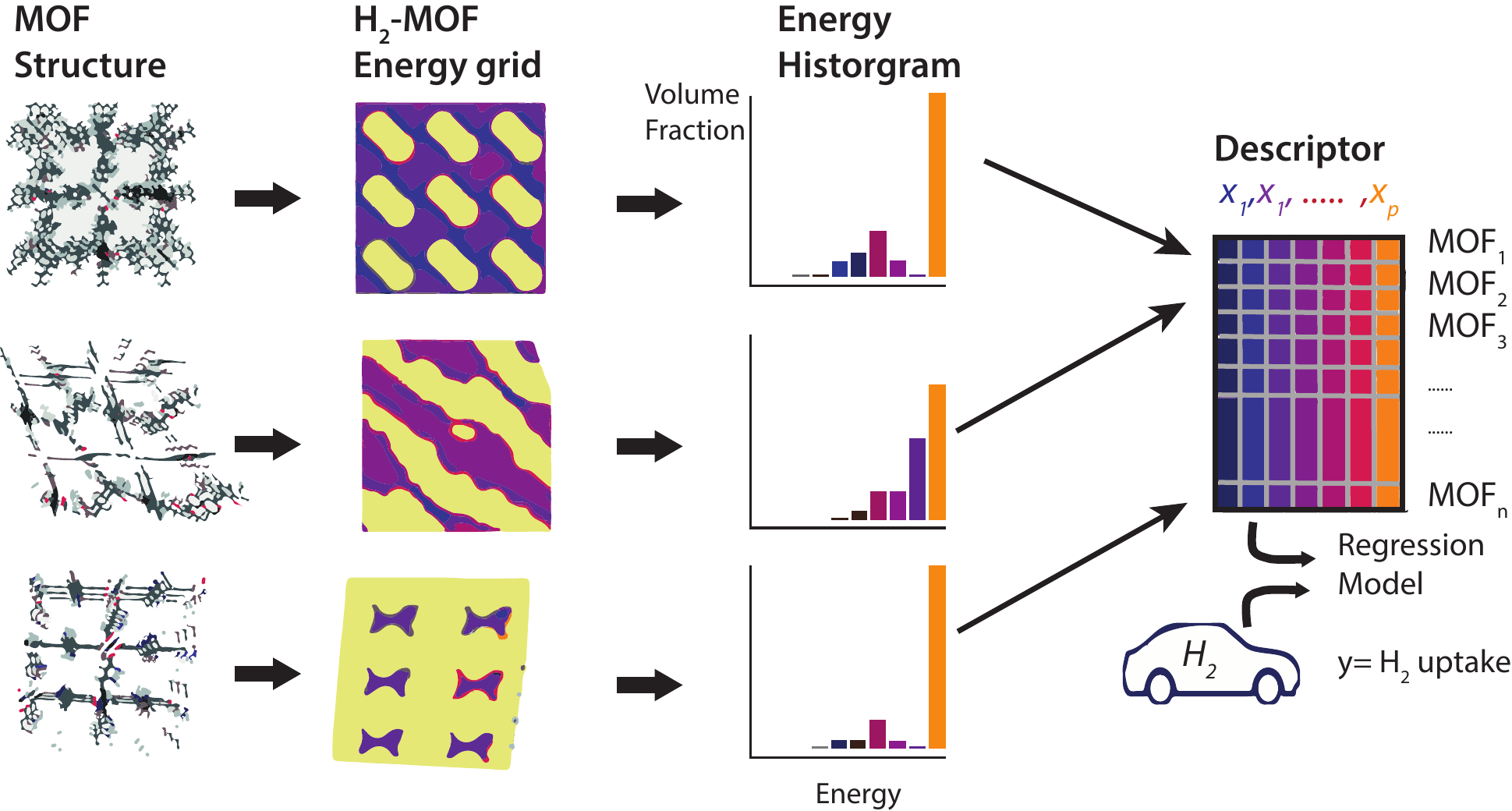}
    \caption{Overall machine learning workflow used by Bucior et al.\protect{\cite{bucior_energy-based_2019}} to predict the \ce{H2} storage capacity of \glspl{mof}.
        For each \gls{mof}, an energy grid within the \gls{mof} unit cell was computed, from which an energy histogram was obtained, which is a feature in their regression model that they used to predict the \ce{H2} uptake.
        Figure adopted from Bucior et al.\protect{\cite{bucior_energy-based_2019}}}\label{fig:bucior_energy}
\end{figure}

\paragraph{Interaction Energy Based Descriptors}
Related to the Voronoi energy introduced by Simon et al.\cite{simon_what_2015} is the energy histogram Bucior et al.\ developed\cite{bucior_energy-based_2019} (see Figure~\ref{fig:bucior_energy}.
In this descriptor, the interaction energy between gas and the framework is binned and used as input for the learning algorithm which the group around Snurr used to learn the \ce{H2} uptake for a large library of hypothetical structures and more than 50,000 experimental structures from the \gls{csd}.
Notably, the authors also investigated the limits of the domain of applicability by training a model only on hypothetical structures---from only one database as well as a random mix of two databases---and evaluating its performance on experimental structures from the \gls{csd}.
Overall, they found better performance for the \enquote{mixed} model that was trained on data from two different hypothetical databases.

Fanourgakis developed a descriptor that uses ideas similar to the ones used for the interaction energy histogram from Bucior et al.
Instead of using the actual probe atom, they decided to use multiple probes with different Lennard-Jones parameters and to compute the average interaction energy for each of them by randomly inserting the probes into the framework, basically computing void fractions for different probe radii.\cite{fanourgakis_robust_2019}
In doing so, Fanourgakis et al.\ observed an improvement in predictive performance in the low methane loading regime compared to conventional descriptors such as void fraction, density, and surface area.

Closely related is the use of the heat adsorption as a descriptor in \gls{ml} models. Similar to the interaction energy captured by the energy histograms, it is a crude estimate of the target.
It was for example used in recent studies on adsorption-based heat pumps, where a working fluid is adsorbed by the adsorbent and the released heat is used to drive the heat pump.
\Glspl{mof} are an interesting alternative for the conventional adsorbents.\cite{de_lange_adsorption-driven_2015}
The most commonly used working fluid is water but for applications below \SI{0}{\celsius} one would like to use an alternative fluid.\cite{del152}
Shi et al.\cite{shi_machine_2020} used \gls{ml} to identify that the density and the heat of adsorption are the most important features from their descriptor set (including geometric properties and the maximal working capacity) for models for identifying the optimal \gls{mof} for a methanol-based adsorption-driven heat pump.
Li et al.\cite{li_screening_2019} used a similar approach, using the Henry coefficient \(K_\mathrm{H}\)  as a surrogate for the target, to build \gls{ml} models that identify promising \glspl{cof} and \glspl{mof} for ethanol-based adsorption.

\paragraph{Geometric Descriptors}
As we already indicated, most of the works on \gls{ml} of the adsorption of non-polar gases in porous materials simply trained their models using geometric descriptors.\cite{aghaji_quantitative_2016, pardakhti_machine_2017, esfandiari_using_2017}

Following the idea that \gls{mof} databases are likely to contain redundant information, Fernandez \textit{et al.} performed \gls{aa} and clustering on geometrical properties to identify the \enquote{truly significant} structures.\cite{fernandez_geometrical_2016}
\Gls{aa} is a matrix decomposition technique that deconstructs the feature matrix, in their case built from geometric properties, into archetypes that do not need to be contained in the data and which can be linearly combined to describe all the data.
They trained classifiers on the \SI{20}{\percent} of structures that are closest to the archetypes and cluster centroids and propose the rules which their \glspl{dt} learned as rules of thumb for enhancing \ce{CO2} and \ce{N2} uptake.

\begin{figure}
    \centering
    \includegraphics[width=.5\textwidth]{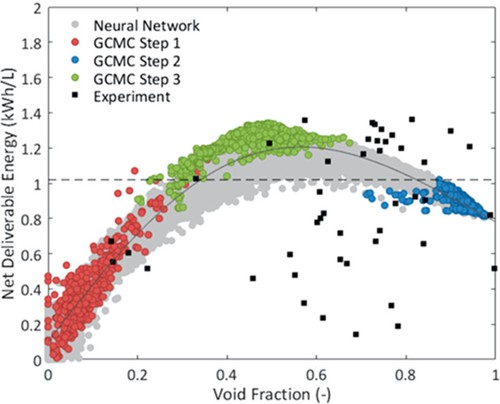}
    \caption{Net deliverable energy as a function of the void fraction for the data predicted using the \gls{nn} and experimental data.
        The solid dark line shows the Langmuir model. Figure reproduced from Thornton et al.\protect{\cite{thornton_materials_2017}}}\label{fig:thornton_mp_iterations}
\end{figure}

Using only geometric descriptors, Thornton et al.\ developed an
iterative prescreening workflow to explore the limits of hydrogen storage in the Nanoporous Materials Genome.
After running \gls{gcmc} simulations on a diverse set of zeolites, they trained a \gls{nn} on that data and used it to predict a set of 1,000 promising candidates, for which they again ran \gls{gcmc} simulations and repeated this cycle two more times to reduce the computational time (cf.\ Figure~\ref{fig:thornton_mp_iterations}).

\paragraph{Using the Building Blocks as Features}
In contrast to all aforementioned studies, Borboudakis et al.\ chose a featurization approach that is not based on geometric properties but that encodes the presence (and absence) of building blocks.
In this way, it is not possible for the model, which they trained with an automated \gls{ml} tool (cf.\ section~\ref{sec:automl}), to perform predictions for structures with building blocks that are not in the training set.\cite{borboudakis_chemically_2017} This approach was recently generalized by Fanourgakis et al.\ who use statistics over atom types (e.g., minimum, maximum and average of triple bonded carbon per unit cell), that would usually be used to set up force field topologies, as descriptors to predict the methane adsorption in \glspl{mof}.\cite{fanourgakis_universal_2020}

\paragraph{Graph-Based Descriptors}
Ohno and Mukae used a different set of descriptors, which have also been used with great success in other parts of chemistry.
They decided to use molecular graphs to describe the building blocks of the structures (cf.\ section~\ref{sec:structure_graphs}) and then used a kernel-based technique (Gaussian process regression, cf.\ section~\ref{sec:bayesian}) to measure similarities between the structures.\cite{ohno_machine_2016}
They used this kernel in a multiple kernel approach together with pore descriptors and then performed a random search to find the combination of linkers and pore properties that maximizes the prediction (methane uptake) of their model. \\
Recently, Korolev et al.\ benchmarked \gls{gcnn} (cf.\ section~\ref{sec:structure_graphs}) on different materials classes and also considered the prediction of the bulk and shear modulus of pure-silica zeolites and the Xe/Kr selectivity of \glspl{mof}.\cite{korolev_graph_2020} For both applications they found worse performance than with the \gls{gbdt} baselines which let the authors to conclude that pore-centered descriptors are more suitable for porous materials than atom centered descriptors. Still, \gls{gcnn} are a promising avenue as the same framework can be applied to many structure classes without tedious feature engineering.

\paragraph{Describing the Pore Shape Using Topological Data Analysis}
A different approach for the description of a similarity between pores has been developed by Lee et al.
Using topological data analysis, they create persistent homology barcodes (see section~\ref{sec:persistent_homology}).
By means of this pore-shape analysis, the authors could find hypothetical zeolites that have similar methane uptake as the top-performing experimental structures.\cite{lee_quantifying_2017, lee_high-throughput_2018}
Lee and co-workers recently also used this descriptor to train machine learning models to predict the methane deliverable capacity of zeolites and \glspl{mof}.\cite{zhang_machine_2019}
To do so, they had to derive fixed-length descriptors based on the original persistent homology barcodes which cannot easily be used in \gls{ml} applications as the number non-zero elements of the barcodes are of varying lengths.
They worked around this limitation by using the distances with respect to landmarks, which are a selection of the most diverse structures, as well as some statistics describing the persistent homology barcode (like the mean survival time, the latest birth time).
A approach related to the distance to barcodes has been chosen by Moosavi, Xu et al.\  who used the distance between barcodes to define a kernel which they then used to train a \gls{krr} model for the  methane deliverable capacities of porous molecular crystals.\cite{moosavi_geometric_2020}

\paragraph{Predicting Full Isotherms}
The works we described so far were built to predict one specific point on a gas adsorption isotherm (i.e., at one specific temperature and pressure).
But in practice, one often wants multiple points on the isotherm, or even the full isotherm, for process development.
In principle, one could imagine training one model per pressure point.
But we also all know that this is a waste of resources as there are laws that connect the pressure and the loading (e.g., Langmuir adsorption).
This motivated researchers to investigate whether one single \gls{ml} model can be used to predict the full isotherm.

Recently, Sun et al.\ reported a multitask deep \gls{nn} (SorbNet) for the prediction of binary adsorption isotherms on zeolites.\cite{sun_deep_2019}
Their idea was to use a model architecture in which the two components have two independent branches in the neural network close to the output and share layers close to the inputs, which are the initial loading, the volume, and the temperature.
They then used this model to optimize process conditions for desorptive drying, which highlights that such models can help avoid the need for iteratively running simulations for the optimization of process conditions (we discuss the connection between materials simulation and process engineering in more detail in the next section).
A limitation of the reported model is that it does not use any descriptors of the sorbate or the porous framework and is therefore limited to a specific combination of sorbates and framework and needs to be retrained for new systems.
A recent work by Desgranges uses the same inputs (\(N, V, T\), or \(N_1, N_2, V, T\), respectively) to predict the partition function, which in principle gives them access to all thermodynamic quantities.\cite{desgranges_ensemble_2020} But similar to the work of Sun et al.\ the model remains limited to the systems (gas and framework) it was trained on.
An interesting avenue might be to combine this approach with the ideas from Anderson et al.\ who encode the sorbates by training with different achemical species (e.g., varying the Lennard-Jones interaction strength, \(\epsilon\)).\cite{anderson_attainable_2019, anderson_adsorption_2019} \\

Most of the works we discussed so far trained their models on data that were generated with \glspl{ff}. But in some cases this is not accurate enough.
Correlated method such as \gls{rpa} might enable simulations to reach chemical accuracy (\SI{1}{\kilo cal\per\mole}).
Unfortunately, those methods are prohibitively expensive for use in \gls{md} simulations. For this reason, Chehaibou et al.\ combined several (\gls{ml}) techniques to predict adsorption energies of \ce{CO2} and \ce{CH4} in zeolites.\cite{chehaibou_computing_2019}
First, they ran \gls{md} simulations with an affordable \gls{dft} functional, then they selected a few distant snapshots on which they performed \gls{rpa} calculations.
They used these calculations to train a \gls{krr} model for which they used a \gls{soap} kernel to describe the similarity between structures. Interestingly, they also use the $\Delta$-ML approach in which they predict the different between the \gls{rpa} and \gls{dft} energy.
This is based on the reasoning that the \gls{dft} result already gives the majority of the contribution to the \gls{rpa} total energy, wherefore it is not necessary to learn this part (cf.\ section~\ref{sec:transfer}). Using thermodynamic pertubation theory, they reweighted the \gls{dft} trajectory using the \gls{rpa} energies predicted using the \gls{krr} model to get ensemble averages on \gls{rpa} level.

\subsubsection{Bridging the Gap Between Process Engineering and Materials Science}
Materials' design is nearly always a multiobjective optimization in which the goal is to find an optimal spot on the Pareto front of multiple performance metrics. One issue with performance metrics is that it is not always clear how they relate to the actual performance on a process level, e.g.\ in a pressure swing adsorption system.
This is also reflected in the 2018 Mission Innovation report that highlights the need to \enquote{understand the relationship between material and process integration to produce optimal capture designs for flexible operation---bridging the gap between process engineering and materials science}.\cite{mission_innovation_accelerating_2017}
\Gls{ml} might help to bridge this gap.\cite{tsay_110th_2019, psichogios_hybrid_1992, oliveira_combining_2004,pai_experimentally_2020}
Motivated by the need to integrate materials science and process engineering, Burns et al.\ performed molecular simulations and detailed simulations for a vacuum swing adsorption process for carbon capture on 1632 \gls{mof}.\cite{burns_prediction_2020}
When attempting to build \gls{ml} models that can predict the process level performance metrics they realized that they can predict the ability of a material to reach the \SI{95}{\percent} \ce{CO2} purity and \SI{90}{\percent} \ce{CO2} recovery targets (95/90-PRT)---but not the parasitic energy, which is the energy needed to recover the sorbent and to compress the \ce{CO2}. Furthermore, using their \gls{rf} models they found the \ce{N2} adsorption properties to be of the highest importance for the prediction of the 95/90-PRT.

\subsubsection{Interpreting the Models}
Over the years \gls{qspr} has evolved from visual inspection of relationships,\cite{wilmer_structureproperty_2012} over the use of more and more complex models to the interpretation of these models, e.g., using some feature importance analysis.
On the one hand, these analyses can give potentially more insights, also for new materials but, on the other hand, they introduce new error sources.
As we discussed in section~\ref{sec:interpretatbility}, we not only have to consider the limitations of the dataset for such analyses but also the limitations of the \gls{ml} model, that might not be able to capture these relationships.

The use of tree-based models,\cite{qiao_high-throughput_2017, liang_combining_2019, aghaji_quantitative_2016, li_large-scale_2019, li_screening_2019, shi_machine_2020, deng_large-scale_2020} and the feature importance that can be extracted from them (e.g., based on how high in the tree a feature was used for a split) have evolved to the most popular techniques to interrogate \gls{ml} models in the \gls{mof} community.\cite{wu_understanding_2019, simon_what_2015, pardakhti_machine_2017, anderson_role_2018}

For example, Gülsoy fitted decision trees for the \ce{CH4} storage capacity of \glspl{mof} using two different feature sets\cite{gulsoy_analysis_2019}. Similar trees were also derived by Fernandez and Barnard as \enquote{rules of thumb} for \ce{CO2} and \ce{N2} uptake in \glspl{mof}\cite{fernandez_geometrical_2016}.

\begin{figure}
    \centering
    \includegraphics[width=\textwidth]{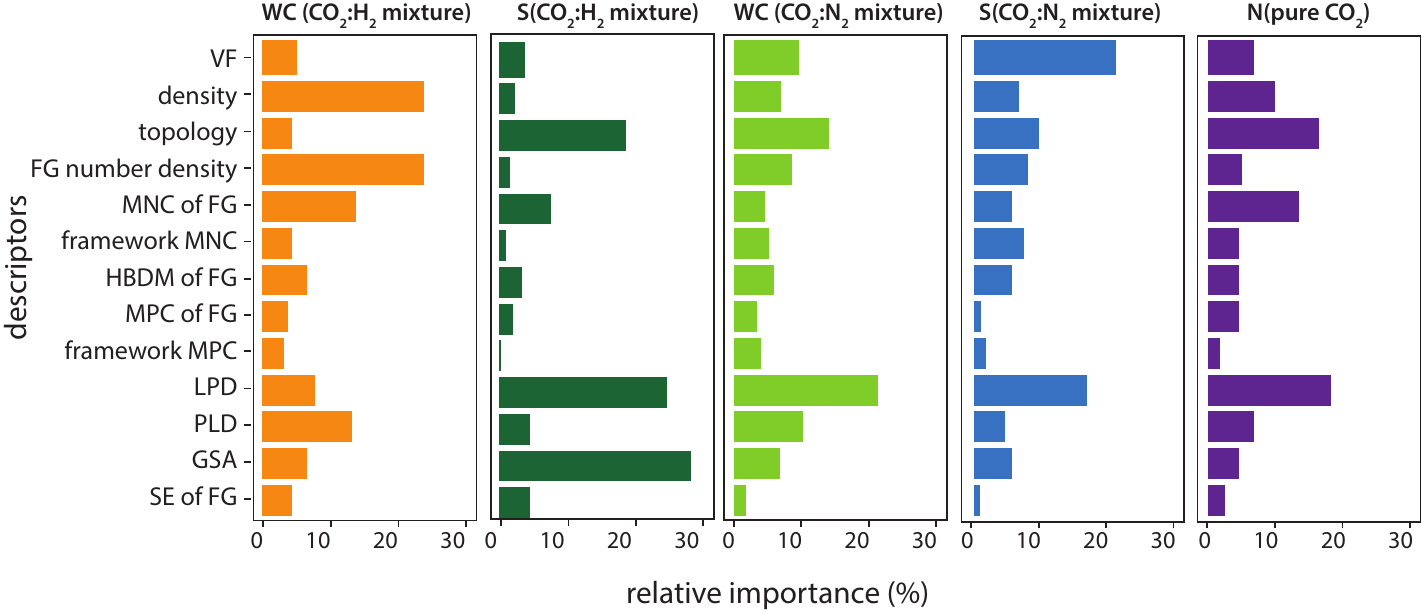}
    \caption{Relative importance of the different material descriptors for carbon capture, as obtained for \gls{gbdt} models trained by Anderson et al.\protect{\cite{anderson_role_2018}}
        In these plots S = selectivity, WC = working capacity, N = adsorption loading. FG = functional group, VF = void fraction, HDBM = highest dipole moment, MPC = most positive charge, MNC = most negative charge, LPD = largest pore diameter, PLD = limiting pore diameter, SE = sum of epsilons, GSA = gravimetric surface area. Figure adopted from Anderson et al.\protect{\cite{anderson_role_2018}}.}\label{fig:anderson_2018_role}
\end{figure}

Anderson et al.\ used feature importance analysis on a library of hypothetical databases for a selection of storage and separation tasks and found that the importance of different features depends on the task.
For example, they found chemistry-related metrics (such as the maximum charges) to be more important for \ce{CO2}/\ce{N2} mixtures than for only the uptake of \ce{CO2}\cite{anderson_role_2018} (see Figure~\ref{fig:anderson_2018_role}).
One advantage of \gls{ml} models is that they can potentially be used for materials' design, i.e., to design a material with an optimal performance from scratch.
Anderson et al.\ attempted to do so by using a genetic algorithm to find feature combinations that maximize the performance indicators.

\subsection{Stability}
But also the \gls{mof} with the best gas adsorption properties is not of much use if it is not stable.
One needs to distinguish between chemical stability and mechanical stability.\cite{mouchaham_stability_2018}

The issue of chemical stability is one of the most asked questioned after a \gls{mof} presentation.
Indeed, MOF-5, one of the first published \glspl{mof}, is not stable in water and therefore there is a strong perception that therefore all \glspl{mof} have a water issue.
However, one has to realize that \gls{mof} are, like polymers, a class of materials.
Some can be boiled in acids for months without losing their crystallinity while others readily dissolve in water.\cite{wang_applications_2016}
For most practical applications it is important, however, to know whether a structure is stable in water.
For this reason, there have been efforts to develop models that are able to predict the stability of porous materials based on readily available descriptors.
This is a typical example of a less well-defined property as can be seen by the different proxies that are used to mimic the notion of stability.
Most of these proxies are based on the idea that for a chemically unstable \gls{mof} it is favorable to replace a linker by water.
To the best of our knowledge, no \gls{ml} studies have been reported that investigate the chemical stability.
Yet this is a complex topic in which \gls{ml} might give us some interesting insights.

Sufficient mechanical stability is also of considerable practical importance.
In most practical applications \glspl{mof} need to be processed, and during this processing there will be pressure and shear forces applied on the crystal.
If this causes the pores to deform, the properties of the material may change significantly.
Therefore, sufficient mechanical stability is an important practical requirement.
Yet, it is not a property that is often studied.\cite{tan_chemical_2010,tan_mechanical_2011, moosavi_improving_2018}

Evans and Coudert took on this challenge by training a \gls{gbdt} to predict the bulk and shear moduli based on geometrical properties for 121 training points calculated using \gls{dft}.\cite{evans_predicting_2017}
Moghadam et al.\ followed up this work by training a \gls{nn} on bulk moduli of more than 3000 \glspl{mof} that they obtained from \gls{ff}-based simulations.\cite{moghadam_structure-mechanical_2019}
Their model uses geometric descriptors and also information about the topology, which their \gls{eda} showed to be of utter importance. Recently, the group around Coudert extended their analysis of the mechanical properties of zeolites using \gls{ff}-derived mechanical properties for all structures from Deem's database of hypothetical zeolites\cite{pop131} for a subset of which they also computed the mechanical properties using \gls{dft}.
Motivated by the lackluster performance of the \gls{ff} to describe the mechanical properties, they trained a \gls{gbdt} (using the same approach which they also used in their first work) on the data derived with \gls{dft}. And they found that, on average, their model can predict the  Poisson's ratio better than the \gls{ff}.

For a related family of porous materials, organic cages, mechanical stability is even a bigger problem as they lack 3D bonding.
Turcani et al.\ built models to predict the stability of the cages based on the precursors to focus more elaborate investigations on materials that are likely mechanically stable.\cite{turcani_machine_2019}

Such a tool would certainly also benefit screenings of \glspl{mof}, but the lack of good training data makes it difficult to create such a model and also explains the scarcity of the studies in this field.
An important part of a solution for this problem is the adoption of standardized computing protocols---such that different databases can be combined into one training set---and sharing of the data in a \gls{fair} compliant way.\cite{coudert_materials_2019}

\subsection{Reactivity and Chemical Properties}
One of the emerging topics in \glspl{mof} is catalysis.\cite{lee_metalorganic_2009,huang_multifunctional_2017, jiao_metalorganic_2018, kang_metalorganic_2019}
\Glspl{mof} are interesting for catalysis as the presence of \gls{oms} or the specifics of the linker can be combined with concepts of shape selectivity known from zeolite catalysis.\cite{smit_towards_2008}

For reactivity on surfaces,\cite{studt_identification_2008} but also in zeolites,\cite{brogaard_methanolalkene_2014, wang_reactivity_2014, wang_transition_state_2015} scaling relations (that often incorporate the heat of adsorption of the reactants) have been proven to be a powerful tool to predict and rationalize chemical reactivity.
Rosen et al.\ recently introduced such relationships, for example, based on the H-affinity of open metal sites, for methane activation in \glspl{mof}.\cite{rosen_structureactivity_2019}
As Andersen et al.\ recently pointed out, more elaborate \glspl{ml} techniques like compressed sensing (cf.\ section~\ref{sec:lasso}) might help us to go beyond scaling relationships and discover hidden patterns in big data.
This approach is motivated by the realization that some phenomena might not be describable by a simple equation and that data-driven techniques might be able to approximate those complex relationships.\cite{andersen_beyond_2019}

\subsection{Electronic Properties}
Other emerging applications of \glspl{mof} are photocatalysis,\cite{zhang_metalorganic_2014} luminescence,\cite{cui_luminescent_2012,rocha_luminescent_2011} and sensing.\cite{kreno_metalorganic_2012,hu_luminescent_2014}
For these properties it is important to know the electronic (band) structure.
However, \gls{ml} studies on the electronic properties of \glspl{mof} are scarce due to the lack of training data in open databases, and the fact that this data is expensive to create using \gls{dft} due to the large unit cells of many \glspl{mof}.
This motivated He et al.\ to attempt to use transfer learning.\cite{he_metallic_2018}
They trained four different classifiers on inorganic structures from the \gls{oqmd} in which the band gaps have been calculated for about 52,300 materials using \gls{dft}, and then retrained the model to classify nearly 3,000 materials from the \gls{core}-\gls{mof} database as either metallic or non-metallic using their \gls{ml} model.

A key descriptor for the chemistry of materials, that is also needed as input for electronic structure calculations, is the oxidation state of a material.
Jablonka et al.\ retrieved the oxidation states assigned in the chemical names of \glspl{mof} in the \gls{csd} and trained an ensemble of classifiers to assign the oxidation state,\cite{jablonka_using_2020} using features that, amongst other, describe the geometry of local coordination environments.\cite{r.zimmermann_local_2020} Using the ensemble they not only made the model more robust (cf.\ section~\ref{sec:ensemble_models}) but also obtained an uncertainty measure. In this way, they could not only assign oxidation states with high predictive performance but also find some errors in the underlying training data.

\subsection{ML for Molecular Simulations}
In other parts of chemical science, \glspl{hdnpp} received a lot of attention as they promise to create potentials in ab initio quality that can be used to run simulations at a cost of \gls{ff} based simulation with the additional advantage of the ability to describe reactions (bond breaking and formation).
Also, popular molecular simulation codes such as \gls{lammps} have been extended to perform simulations with such potentials.
However, such models are usually trained on \gls{dft} reference data which can make it a demanding task to create a training a set given the large unit cells of \glspl{mof}.
\begin{figure}
    \centering
    \includegraphics[width=.4\textwidth]{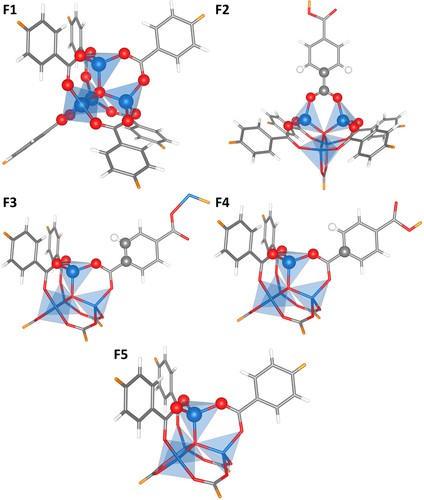}
    \caption{Molecular basis fragments used by Eckhoff and Behler as starting points for the generation of reference structures for the training of the \gls{hdnpp} for MOF-5.
        Based on the five fragments, more than 4,500 other fragments were generated by scaling of the coordinates and small random displacements.
        All atoms that have complete bulk-like environments within a cutoff radius of \SI{12}{\angstrom} are shown as balls, capping hydrogen atoms, to saturate broken bonds, are shown in orange.
        Figure reprinted from Eckhoff and Behler.\protect{\cite{eckhoff_molecular_2019}}}\label{fig:fragments_behler}
\end{figure}
Eckhoff and Behler attempted to avoid this problem by constructing a potential based on more than 4,500 small molecular fragments (the base fragments are shown in Figure~\ref{fig:fragments_behler}) that were constructed by cutting out fragments from the crystal structure of MOF-5.
The \gls{hdnpp} which they trained in this way was able to correctly describe the negative thermal expansion and the phonon density of states.\cite{eckhoff_molecular_2019}

Besides a potential that describes the interatomic interactions, the assignment of partial charges is needed to calculate the Coulomb contribution to the energy in molecular simulations.
The most reliable methods to assign those charges rely on \gls{dft} derived electrostatic potentials and in this way can easily become the bottleneck for molecular simulations.
As an alternative, Xu and Zhong proposed to use connectivity-based atom types, for which it is assumed that atoms with the same connectivity have the same charge.\cite{xu_general_2010}
Korolev and co-workers attempted to solve the main limitation of the connectivity-based atom types, namely that all relevant atom types need to be included in the training set, using a \gls{ml} approach.\cite{korolev_transferable_2019}
To do so, they trained a \gls{gbdt} on 440,000 partial charge assignments using local descriptors such as the electronegativity of the atom or local order parameters, which are based on a Voronoi tessellation of the neighborhood of a given site.

\subsection{Synthesis}
Synthesis is at the heart of chemistry.
Still, it is unfeasible to use computational approaches to predict reactivity or to suggest ideal reaction conditions---also because for example crystallization is a complex interfacial phenomenon that is influenced by structure-directing agents or modifiers.\cite{moliner_machine_2019}
For this reason, chemical reactivity is one of the most promising fields for \gls{ml}.

Nevertheless, there are only a few reports that try to use artificial intelligence techniques in the synthesis of \glspl{mof}.
This is likely due to the same reasons as for reactivity and electronic properties, for which there are also no large open databases of properties and for which the training data is expensive to generate.

Some of the early works in the field set out to optimize the synthesis of zeolites.
Corma et al.\ attempted to make high-throughput synthesis (e.g., using robotic systems) more efficient, i.e., improve on classical \gls{doe} techniques like full factorial design (generating all possible combinations of experimental parameters, cf.\ section~\ref{sec:doe})\cite{akporiaye_combinatorial_1998, choi_combinatorial_1999} by reducing the number of low-promising experiments.\cite{corma_new_2006}
First, they attempted to use simple statistical analysis to estimate the importance of different experimental parameters and then moved to actual predictive modeling.
After training a \gls{nn} on synthesis descriptors to predict and optimize crystallinity,\cite{moliner_application_2005, corma_new_2006} they combined a \gls{ga} with a \gls{nn} to guide the next experiments suggested by the \gls{ga} with the knowledge extracted by the \gls{nn}\cite{corma_optimisation_2005} (using the \gls{nn} to predict the fitness).\cite{corma_optimisation_2005}
A related approach was introduced to the field of \gls{mof} synthesis by Moosavi et al.\ where the synthesis parameters were optimized using a \gls{ga}.
To make this more efficient, the authors introduced the importance of variables derived from a \gls{rf} model, that was also trained on the failed experiments, as weights for the distance metric for the selection of a diverse set of experimental parameters.
In this way, they could synthesize the HKUST-1 with the highest \gls{bet} surface area reported so far.\cite{moosavi_capturing_2019}

In a similar vein, Xie et al.\cite{xie_machine_2019} analyzed failed and partly successful experiments and used a \gls{gbdt} to determine the importance of experimental variables that determine the crystallization of \glspl{monc}, which are compounds that can self-assemble and form porous crystals in some cases.\cite{dalgarno_metallo-supramolecular_2008}

Given the large body of experimental procedures for the synthesis of porous materials many works attempted to mine or extract this collective knowledge to create structured datasets that can be used to train \gls{ml} models for reaction condition prediction.
\begin{figure}
    \centering
    \includegraphics[width=\textwidth]{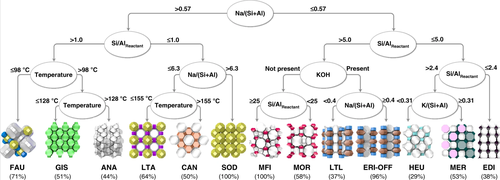}
    \caption{First four layers of a decision tree for zeolite synthesis that Muraoka et al. extracted from a GBDT model fitted on literature data. The percentages give the fractions of the dominant phases for the complete tree with 12 layers. According to this tree, the most important factor for predicting the synthesis result is the Na/(Si + Al) ratio. Figure reprinted from Muraoka et al.\protect{\cite{muraoka_linking_2019}}}\label{fig:zeolite_decisions_tree}
\end{figure}
A recent study of Muraoka et al.\ was enabled by a literature review on the synthesis of zeolites.
Using this data, they trained \gls{ml} models to predict the phase based on parameters describing the synthetic conditions, producing decision trees, as shown in Figure~\ref{fig:zeolite_decisions_tree}, that reflect chemically reasonable knowledge extraction from the literature data.
For example, the authors compare the early split based on the Si/Al ratio with Löwenstein’s rule that forbids Al-O-Al bonds.
By optimizing the structural fingerprint by re-weighting the similarity between zeolites to be similar in the synthesis and structure space, they could build a similarity network in which they could uncover an overlooked similarity between zeolites that also manifested itself in the synthesis conditions.\cite{muraoka_linking_2019}

Jensen et al.\ developed algorithms to retrieve the synthesis conditions from 70,000 zeolite papers and used this to build a model that can predict the framework density of germanium zeolites based on the synthetic conditions.\cite{jensen_machine_2019}
Also, Schwalbe-Koda mined the literature about polymorphic transformations between zeolites to enable their work in which they showed that graph isomorphism can be used as a metric for these transformations.\cite{schwalbe-koda_graph_2019}

For \glspl{mof}, Park et al.\cite{park_text_2018}, as well as Tayfuroglu et al.\cite{tayfuroglu_silico_2019}, parsed the literature to retrieve surface areas and pore volumes for a large collection of \gls{mof}.
But so far, the data generated from these studies have not yet been used to build predictive models for \gls{mof} properties and synthesis.

\begin{figure}
    \centering
    \includegraphics[width=0.3\textwidth]{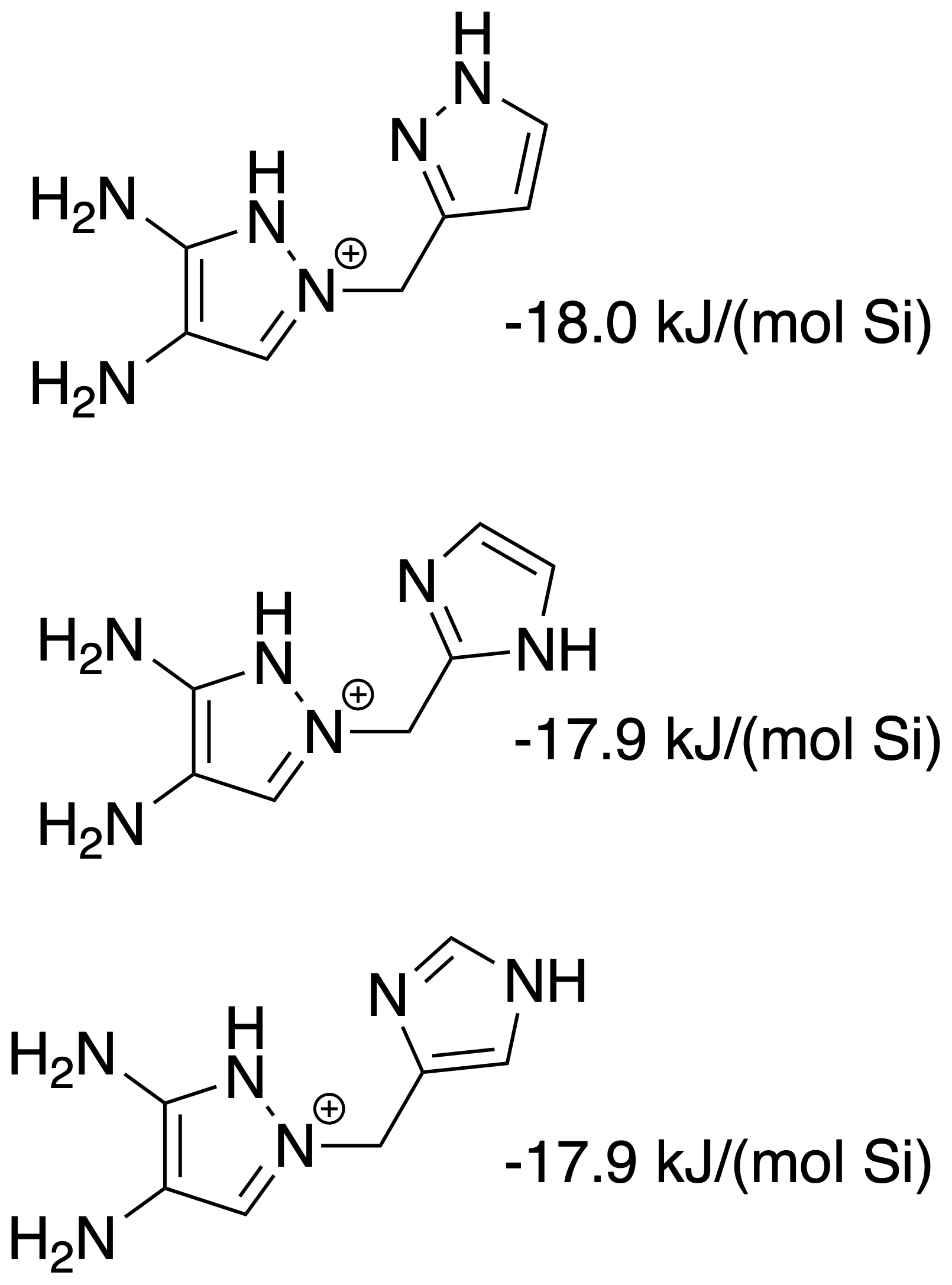}
    \caption{Deem and co-workers used a \gls{ml} approach to identify chemically synthesizable \gls{osda} for zeolite beta, which is one of the top-six zeolites of commercial interest.
        The figure shows the top-three \glspl{osda} that Deem and co-workers discovered.\protect{\cite{daeyaert_machine-learning_2019}}
        The scores in the figure are the binding energy in kJ/(mol Si). Figure adopted from ref.\protect{\cite{daeyaert_machine-learning_2019}}}\label{fig:daeyaert_machine-learning_2019}
\end{figure}

Another approach was taken by Deem and co-workers who addressed the design of \glspl{osda}.\cite{daeyaert_machine-learning_2019}
Zeolites are all isomorphic structures and \glspl{osda} are used during the synthesis to favor the formation of the desired isomorph.
Finding the right \gls{osda} to synthesize a particular zeolite is seen as one of the bottlenecks.
To support this effort, Deem and co-workers developed a materials' design program to generate synthetically accessible \gls{osda}.\cite{pop131}
To expedite this process, Deem and co-workers developed a \gls{ml} approach, in which they calculated the stabilization energy of different \glspl{osda} inside of zeolite beta and then trained a \gls{nn} using molecular descriptors derived from ideas of electron diffraction.\cite{schuur_coding_1996}
In this way, they could speed up the search for novel \gls{osda} by a factor of 350 and suggest 469 new and promising \gls{osda} (see Figure~\ref{fig:daeyaert_machine-learning_2019}).
Daeyaert and Deem\cite{daeyaert_design_2019} further extended this work to find an \gls{osda} for some of the hypothetical zeolites that were found to perform optimally in a screening study for the separation of \ce{CO2} and \ce{CH4}.\cite{kim_large-scale_2013}

Even if one manages to create some material it is not always trivial what the material is. To address this, Wang et al.\ build models, including \glspl{cnn} similar to the one we described in section~\ref{sec:xrd_cnn} to identify the material based on its experimental \gls{xrpd} pattern. To do so, they predicted diffraction patterns for structures deposited in the \gls{csd} and used data augmentation techniques (cf.\ section~\ref{sec:augmentation}) like the addition of noise and then tested their model using experimental diffraction patterns.\cite{wang_rapid_2019}

\subsubsection{Synthesizability}
One question that always arises in the context of hypothetical materials is the question of synthesizability. In the context of zeolites, this question received a lot of attention.
Early works proposed that low framework energies are the distinctive criterion\cite{bushuev_feasibility_2010, foster_structural_2003, akporiaye_relative_1989}---akin to the recent attempt of Anderson and Gómez-Gualdrón to assess the synthetic feasibility of \glspl{mof}.\cite{anderson_large-scale_2020}
But this quickly got overturned with the discovery of high-energy zeolites and replaced by a \enquote{flexibility window},\cite{sartbaeva_flexibility_2006} which was eventually also found to not be reliable and replaced by criteria that focus on local interatomic distances.\cite{li_criteria_2013}
A library of such criteria was used in a screening study of Perez et al.\ to reduce the pool of candidate materials from over 300,000 to below 100. As a conclusion of their study, they suggest using the overlap between the distribution of descriptors of experimental materials and those generated in silico as a metric to evaluate how feasible the materials are which an algorithm produces.\cite{salcedo_perez_high-throughput_2019}
Such an approach, which is related to approaches suggested for benchmarking of generative techniques for small molecules,\cite{brown_guacamol_2019, gao_synthesizability_2020} might also be useful for evaluation of the generative models that we discuss in the following.

\subsection{Generative Models}\label{sec:gan}
The ultimate goal of materials' design is to build a model that, given desired (application) properties, can produce candidate structures using generative techniques like \glspl{gan}.
Though this flavor of \gls{ml} is formally not supervised learning, on which we focused in this review, we give a short overview of recent progress in this promising application of \gls{ml} to porous materials.
One model architecture that is often used in this context are \glspl{gan} where a first \gls{nn} acts as generator and tries to \enquote{deceive} a discriminator \gls{nn} that tries to distinguish real data (structures) from the \enquote{fake} ones that the generator generated.
For molecules, this approach received wide attention,\cite{elton_deep_2019, sanchez-lengeling_inverse_2018} but the works on nanoporous solids proved to be more difficult due to the periodicity and the non-unique representation of the unit cell.
Kim and co-workers started building \glspl{gan} that can generate energy grids of zeolites\cite{lee_predicting_2019} and recently extended their model to predict the structure of all-silica zeolites.\cite{kim_inverse_2020}
\begin{figure}
    \centering
    \includegraphics[width=\textwidth]{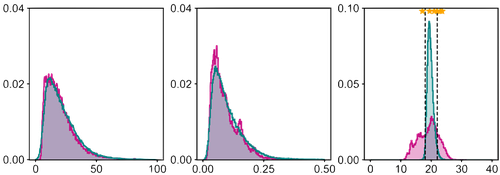}
    \caption{Distribution of Henry coefficient, void fraction and heat of adsorption for generated structures with a user-defined target range of \SIrange{18}{22}{\kilo\joule\per\mole} for the heat of adsorption. Reproduced from ref.~\cite{kim_inverse_2020}.}
    \label{fig:zeogan}
\end{figure}
To do so, they used a separate channel (as it is used for the RGB channels in color images) for oxygen and silicon atom positions which they encoded by placing Gaussian at the atom positions. By adjusting the loss function to target structures with a specific heat of adsorption, they could observe a drastic shift in the shape of the distribution of this property but not in the one for the void fraction or the Henry coefficient.

%% file: main/connection_table.tex
\begin{longtable}{p{12em}p{3em}p{20em}}
	\caption{Overview of learning methods that we discussed in Section~\ref{sec:learning_algorithms} and examples of their use in the field of porous materials.
		For some methods there has been no application reported in the field of porous materials, and we instead provide ideas of possible applications.}\label{tab:overview_table}                                                                                                                                                                                                                                                                                     \\
	\toprule
	\textbf{method}                                  & \textbf{section}          & \textbf{application to porous materials}                                                                                                                                                                                                                                                                                                                                         \\
	\midrule
	\endhead
	\midrule
	\multicolumn{3}{r}{{Continued on next page}}                                                                                                                                                                                                                                                                                                                                                                                                                    \\
	\midrule
	\endfoot
	\bottomrule
	\endlastfoot
	\multicolumn{3}{l}{\emph{representation learning}}                                                                                                                                                                                                                                                                                                                                                                                                              \\
	\midrule
	\gls{hdnpp}                                      & \ref{sec:ann}             & trained on fragments for MOF-5 by Behler and co-workers\cite{eckhoff_molecular_2019}                                                                                                                                                                                                                                                                                             \\
	message-passing \gls{nn}                         & \ref{sec:messagepassing}  & not used for porous materials so far                                                                                                                                                                                                                                                                                                                                             \\
	convolutional or recurrent \gls{nn}              & \ref{sec:cnn}             & Wang et al. used CNN to classify \glspl{mof} based on their \gls{xrpd} pattern\cite{wang_rapid_2019}                                                                                                                                                                                                                                                                             \\
	crystal-graph based models                       & \ref{sec:graph_model}     & Korolev et al.\ use them to predict bulk and shear moduli of pure silica zeolites and Xe/Kr selectivity of \glspl{mof}\cite{korolev_graph_2020}                                                                                                                                                                                                                                  \\
	generative models                                & \ref{sec:gen_intro}       & ZeoGAN by Kim and co-workers\cite{kim_inverse_2020} (cf.\ section~\ref{sec:gan})                                                                                                                                                                                                                                                                                                 \\
	\midrule
	\multicolumn{3}{l}{\emph{classical statistical learning}}                                                                                                                                                                                                                                                                                                                                                                                                       \\
	\midrule
	linear models                                    & \ref{sec:linear_model}    & predicting gas uptakes based on tabular data of simple geometric descriptors\cite{fernandez_large-scale_2013}                                                                                                                                                                                                                                                                    \\
	kernel methods                                   & \ref{sec:kernel_methods}  & predicting gas uptakes based on graphs and geometric properties,\cite{ohno_machine_2016} might be also interesting in the \gls{soap}-\gls{gap} framework, as work by Ceriotti and co-workers as well as Chehaibou et al.\ showed\cite{helfrecht_new_2019,chehaibou_computing_2019}                                                                                               \\
	ensemble models                                  & \ref{sec:ensemble_models} & often used in form of \gls{rf} or \gls{gbdt} to predict gas uptakes based on tabular data of simple geometric descriptors, ensemble used to estimate uncertainty when predicting oxidation states\cite{jablonka_using_2020}                                                                                                                                                      \\
	Bayesian methods                                 & \ref{sec:bayesian}        & have been used e.g., in the form of \gls{gpr}\cite{ohno_machine_2016} or Bayesian \gls{nn}\cite{thornton_towards_2015, thornton_materials_2017} but not all features, like the uncertainty measure, have been fully exploited so far. This might be useful for active learning, e.g.\ for \gls{md} simulations in the Bayesian formulation of the \gls{soap}-\gls{gap} framework \\
	\gls{tda}                                        & \ref{sec:tda}             & Moosavi, Xu et al. built \gls{krr} models for gas uptake in porous organic cages,\cite{moosavi_geometric_2020} or Zhang et al. for gas uptake in \gls{mof}\cite{zhang_machine_2019}, Lee et al.\ for similarity analysis\cite{lee_quantifying_2017}                                                                                                                              \\
	\midrule
	\multicolumn{3}{l}{\emph{other \gls{ml} techniques}}                                                                                                                                                                                                                                                                                                                                                                                                            \\
	\midrule
	automated machine learning                       & \ref{sec:automl}          & Tsamardinos et al.\cite{tsamardinos_automated_2020} use the Just Add Data tool to predict the \ce{CH4} and \ce{CO2} capacity of \glspl{mof}, Borboudakis et al.\ use the same tool to predict \ce{CO2} and \ce{H2} uptakes\cite{borboudakis_chemically_2017}                                                                                                                     \\
	data augmentation                                & \ref{sec:datasource}      & Wang et al.\ used it for the detection of \glspl{mof} based on their diffraction patterns\cite{wang_rapid_2019}                                                                                                                                                                                                                                                                  \\
	transfer learning                                & \ref{sec:transfer}        & He et al.\ used it for the prediction of band gaps\cite{he_metallic_2018}                                                                                                                                                                                                                                                                                                        \\
	active learning                                  & \ref{sec:active_learning} & could be used for \gls{md} simulations using \gls{ml} forcefields\cite{jinnouchi_phase_2019}, or to guide the selection of next experiments or computations                                                                                                                                                                                                                      \\
	capturing the provenance of \gls{ml} experiments & \ref{sec:reproducibility} & Jablonka et al.\ used \texttt{comet.ml} to track the experiments they ran for building models that can predict the oxidation state of metal centers in \glspl{mof}\cite{jablonka_using_2020}                                                                                                                                                                                     \\
	$\Delta$-ML                                      & \ref{sec:transfer}        & Chehaibou et al.\ used a $\Delta$-ML approach to predict \gls{rpa} adsorption energies in zeolites\cite{chehaibou_computing_2019}                                                                                                                                                                                                                                                \\
\end{longtable}

%% file: main/7_outlook.tex
\section{Outlook and Concluding Remarks}
One of the aims of this review is to provide a comprehensive overview of the state of the art of \gls{ml} in the field of materials science.
In our review, we not only discuss the technical details, but we also try to point out the potential caveats that are more specific for material science.
As part of the outlook, we discuss some techniques that are, as of yet, little, if at all, used for porous materials.
Yet, these methods can address some of the issues that we have discussed in the previous sections.

\subsection{Automatizing the Machine Learning Workflow}\label{sec:automl}
Given that the complete process from structure to prediction, which we discussed in this review, is quite laborious, there is a significant barrier for scientists with little computational background to enter the field.
To lower the entrance barrier, a lot of effort is spent to automatize the \gls{ml} process.\cite{zoller_survey_2019}
In the \gls{ml} community tools like H2O's \texttt{autoML},\cite{h2o.ai_automl_2019} \texttt{TPOPT}\cite{squillero_automating_2016} or Google's \texttt{AutoML} are widely known and receive mounting attention.\cite{zoph_learning_2017}
In the materials science community especially the \texttt{chemml}\cite{vishwakarma_towards_2019, haghighatlari_chemml_2019} and the \texttt{automatminer} package\cite{dunn_automatminer_2019} are worth mentioning.
Latter uses \texttt{matminer} to calculate descriptors that are relevant for materials science, performs the feature selection (using \texttt{TPOPT}) as well as training and cross-validation.\cite{dunn_automatminer_2019}
Such tools will lower the barrier for domain experts even more and also help practitioners of \gls{ml} to expedite tedious tasks.

\subsection{Reproducibility in Machine Learning}\label{sec:reproducibility}
Reproducibility, and being able to build on top of previous results, is one of the hallmarks of science.
And it is also one of the main technical debts of \gls{ml} systems, where technical debt describes costs due to (code) rework that are caused by choosing an easy solution now instead of a proper one that might take longer to be developed.\cite{sculley_hidden_2015}
If one cannot even replicate published experiments one can ask if we are making any progress as a community.
This question was posed by a recent study that found that they could only reproduce 7 from 18 recommender algorithms. Moreover, six of the algorithms which were reproducible could be outperformed by simple heuristics.\cite{dacrema_are_2019}

It is also the authors' personal experience that reproducing computational data from the literature can be a painful process.
Even, if the literature is an article from the same group, reproducing the results from only a few years earlier can be a difficult search for the information that was not reported in the original article.
Often, the reason for being unable to reproduce the data is that many programs use default settings.
These default settings can be hidden in the input files---or in the code itself---and since they are never changed during the reported studies, these settings get overlooked and do not get reported.
However, if in a new release or for any other reasons the defaults get changed, the results become nearly impossible to reproduce.
Of course, if we had realized the importance of these unknown unknowns, we, and any other author, would have reported the values in the original article.
The only way to avoid these issues is to rigorously report all input and output files as well as workflows for all computations.\cite{coudert_reproducible_2017}
In \gls{ml} the same holds---for example different implementations of performance measures (e.g., in off-the-shelf \gls{ml} libraries) can lead to different, biased, estimates that hinder comparability and reproducibility.\cite{forman_apples--apples_2010}

In computational materials science there are ongoing efforts, like the AiiDA infrastructure\cite{pizzi_aiida_2016} or the Fireworks workflow management system,\cite{jain_fireworks_2015} to make computational workflows more reproducible and to lower the barrier of applying the \gls{fair} principles of data sharing.\cite{wilkinson_fair_2016}
For example, Ongari et al.\cite{ongari_building_2019} developed a workflow to optimize and screen experimental \glspl{cof} structures for their potential for carbon capture.\cite{ongari_building_2019}
\begin{figure}
    \centering
    \includegraphics[width=\textwidth]{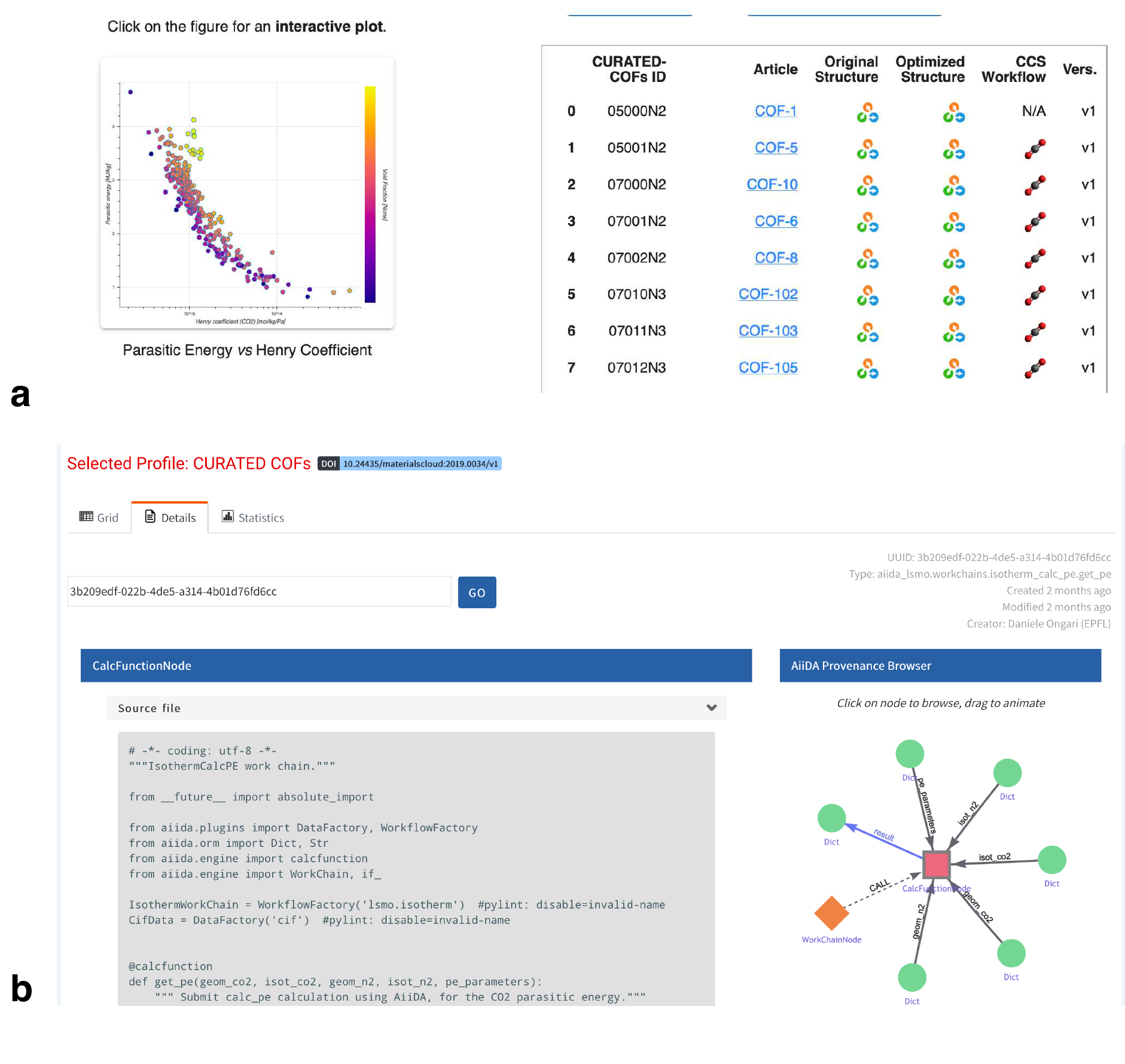}
    \caption{Screenshot of the Materials Cloud (\url{https://archive.materialscloud.org/2019.0034/v2}) pages for the screening of \glspl{cof} for carbon capture.
        In the \enquote{Discover} section (a) there is an interactive plot and table that links to the original references and structures as well as to plots of the relaxation and the process optimization which are all linked to the \enquote{Explore} section (b), where one can find the source code for the workflows and interactive provenance graphs.}
    \label{fig:cofCO2}
\end{figure}
Figure~\ref{fig:cofCO2} shows a snapshot from the Materials Cloud website where, by clicking on a data point, one not only obtains all data that have been computed for this particular material, but also the complete provenance.
This provenance includes an optimization step of the experimental structure, the computation of the charges on the framework, the \gls{gcmc} simulations to compute the isotherms and heats of adsorption, and finally the program that computes the objective function used to rank the materials for carbon capture.
The idea here is that anybody in the world can reproduce the data by simply downloading the AiiDA scripts and running the programs on a local computer.
Or, by adding more structures, extending to work to other materials, or reproducing the complete study using a different force field, by simply replacing the force field input file. Given that the data contains rich metadata, and all parameters of the calculations, it is easy to identify with which other databases it could be combined to create a training set for a \gls{ml} algorithm.

But these workflow management tools, and even version control system like \texttt{git}, are not easily applicable to \gls{ml} problems, where one usually wants to share and curate data separately from the code, but still retain the link between data, hyperparameters, code, and metrics.
Tools like \texttt{comet},\cite{cometml} \texttt{neptune},\cite{neptune} \texttt{provenance},\cite{mabey_provenance_2019}, \texttt{Renku},\cite{swiss_data_science_center_renku_2020} \texttt{mlflow},\cite{databricks_mlflow_2019} \texttt{ModelDB},\cite{vartak_m_2016} and \texttt{dvc}\cite{petrov_dvc_2019} try to make \gls{ml} more reproducible by providing parts of this solution, like data version control or automatic tracking of hyperparameter and metrics together with data hashes.

We consider both reproducibility and sharing of data as essential for the progress in this field.
Therefore, to promote the adaptation of good practices, we encourage using tools such as the data-science cookiecutter\cite{drivendata_home_2019} that automatically sets up a \gls{ml} development environment that promotes good development practices.

Journals in the chemical domain might also encourage good practices by providing \enquote{reproducibility checklists}, similar to the major \gls{ml} conferences like NeurIPS.\cite{beygelzimer_neurips_2019}

Publishing the full provenance of the model development process, as it can be done for example with tools like \texttt{comet}, can to some extent also remedy the problem that negative results (e.g., plausible architectures that do not work) are usually not reported.

\subsubsection{Comparability and Reporting Standards}
One factor that makes it difficult to build on top of previous work is the lack of standardization.
In the \gls{mof} community, many researchers use hypothetical databases to build their models.
But unfortunately, they typically use different databases, or different train/test splits of the same database.
This makes it difficult to compare different works as the chemistry in some databases might be less diverse and easier to learn than for example in the \gls{core}-\gls{mof} database, which contains experimental structures.
Also, in comparing the protocols with which the various labels (\(y\)) for different databases are created, one often finds worrying differences, e.g., in the details of the truncation of the interaction potential\cite{jablonka_applicability_2019} or the choice of the method for assigning partial charges.
This can make it necessary to recompute some of the data, as the discrepancy between the two approaches will dictate the Bayes basis error.\cite{ng_machine_2018, huo_unified_2017}
Unfortunately, there are no widely accepted benchmark sets in the porous materials community---even though the \gls{ml} efforts on (small) molecules greatly benefited from such benchmark sets (see e.g.\ \url{http://quantum-machine.org/datasets/} or MoleculeNet\cite{wu_moleculenet_2018}) which allow for a fair comparison between studies.\cite{ramakrishnan_quantum_2014}
We are currently working on assembling such sets for \gls{ml} studies on porous materials.

In addition to the lack of benchmark sets, there is also a lack of common reporting standards.
Not all works provide full access to data, features, code, trained models,
and choice of hyperparameters---even though this would be needed to ensure replicability.
The \texttt{crystals.ai} project is an effort to create a repository for such data.\cite{materials_virtual_lab_crystals.ai_2019}
Again, reproducibility checklists, like the one for NeurIPS, might be beneficial for our community to ensure that researchers stick to some common reporting standard.

\subsection{Transfer Learning and Multifidelity Optimization}\label{sec:transfer}
A problem of \gls{ml} for materials science, and in particular \glspl{mof} with their large unit cells, is that the datasets of the ground truth (the experimental results) are scarce and only available for a few materials.
Often, experimental data are replaced by estimates from computations, and these computational data necessarily introduce errors due to approximations in the theories.\cite{hutchinson_overcoming_2017}
Similarly, it is much easier to create large datasets using \gls{dft} than using expensive, but more accurate, wavefunction methods.
But even \gls{dft} can still be prohibitively expensive for large libraries of materials with large unit cells.
This is why multifidelity optimization (which combines low and high-fidelity data, like semi-empirical and \gls{dft}-level data) and transfer learning are promising avenues for materials science.

Transfer learning has found widespread use in the \enquote{mainstream} \gls{ml} community, e.g., for image recognition, where models are trained on abundant data and then partially (re)-trained on the less abundant (and more expensive) data.
Hutchinson et al.\ used transfer learning techniques to predict experimental band gaps and activation energies using \gls{dft} labels as the main data source and showed that transfer learning generally seems to be able to improve predictive performance.\cite{hutchinson_overcoming_2017}
Related to this is a recent physics-based neural network from the Pande group in which a cheap electron density, for example from \gls{hf}, is used to predict the energetics and electron density on the \enquote{gold standard} level of theory (\gls{ccsdt}).\cite{sinitskiy_physical_2019}
The authors relate the expensive electron density \(\rho\) to the cheap one using a Taylor expansion and use a \gls{cnn} to learn the \(\Delta \rho\) and \(\Delta E\).
Since both Taylor expansions for \(\Delta E\) and \(\Delta \rho\) share terms like \(\left(\frac{\delta^2 E}{\delta \rho (\br) \delta \rho(\br')}\right)\) they can use the same first layers and then branch into two separate output channels for \(\Delta \rho\) and \(\Delta E\), respectively.
The \gls{nn} was first trained using less expensive \gls{dft} data, and then transfer learning was used to refine the weights using the more expensive and less abundant CCSD(T) densities.
This is similar to the approach which was used to bring the ANI-1 potential to CCSD(T) accuracy on many benchmark sets.\cite{s_smith_approaching_2019}

But for transfer learning to find more widespread use in the materials science domain it would be necessary to share the trained models, and the training as well as evaluation data, in an interoperable way.

The fact that inaccurate, but inexpensive, simulation data is widely available motivated the development of the \(\Delta\)-ML technique, where the objective of the \gls{ml} model is to predict the difference between the result of a cheap calculation and one obtained at a higher level of theory.\cite{ramakrishnan_big_2015}
This approach was subsequently formalized and extended in multiple dimensions using the sparse grid combination technique, which combines models trained on different subspaces (e.g.\ combination of basis set size and correlation level) such that only a few samples are needed on the highest, target, level of accuracy.\cite{zaspel_boosting_2019}

A different multifidelity learning approach, known as co-kriging, can combine low- and high-fidelity training data to predict properties at the highest fidelity level---without using the low-fidelity data as features or baseline.
This technique was used by Pilania et al.\ to predict band gaps of elpasolites on hybrid functional level of theory using a training set of properties on both \gls{gga} and hybrid functional level.\cite{pilania_multi-fidelity_2017}

All these methods are promising avenues for \gls{ml} for porous materials.

\subsection{Multitask Prediction}
In the search for new materials, we usually do not only to want to optimize one property but multiple.
Also, we usually not only have training data for only one target but also for related targets, e.g., for Xe, Kr and \ce{CH4} adsorption.
Multitask models are built around this insight and that models, particularly \glspl{nn}, might learn similar high-level representations to predict related properties (e.g., one might expect the gas uptake for noble gases and \ce{CH4} follow the same basic relationship).
Hence, training a model to predict several properties at the same time might improve its generalization performance due to the implicit information captured between the different targets.
In the chemical sciences, Zubatyuk et al.\ used multimodal training to create an information-rich representation using a message-passing \gls{nn}.\cite{zubatyuk_accurate_2019}
This representation could then be used to efficiently (with less training data) learn new properties.
Similar benefits of multitask learning were also observed in models trying to predict properties relevant for drug discovery.\cite{kearnes_modeling_2016, ramsundar_massively_2015}

%% file: main/8_summary_conclusion.tex
\subsection{The Future of Big-Data Science in Porous Materials}\label{sec:concluding_remarks}

It is tempting to conclude that \glspl{mof} and related porous material are synthesized for \gls{ml}.
\Glspl{mof} are among the most studied materials in chemistry and the number of \glspl{mof} that are being synthesized is still growing.
In addition, the number of possible applications of these materials is also increasing.
We are already in a situation that if a group has synthesized a novel \gls{mof} it is in practice impossible to test this novel material for all possible applications.
One can then clearly envision the role of \gls{ml}.
If we can capture the different computational screening studies using \gls{ml}, we should be able to indicate the potential performance of a novel material for a range of different applications.
Clearly, a lot of work needs to be done to reach this aim; with this review we intended to show that the foundations for such an approach are being built.

The other important domain where we expect significant progress is in the field of \gls{mof} synthesis.
The global trend in science is to share more data, and technology makes it easier to share large amounts of data. But the common practice to only publish successful synthesis routes is throwing away lots of valuable information.
For example, an essential step in \gls{mof} synthesis is finding the right conditions for the material to crystallize. At present, this is mainly trial and error.
Moosavi et al.\cite{moosavi_capturing_2019} have shown how to learn from the failed and partially successful experiments. Interestingly, they used as example HKUST-1, which is one of the most synthesized \glspl{mof}, but they had to reproduce the failed experiments to be able to analyze the data using \gls{ml} techniques.
One can only dream about the potential of such studies if all synthetic \gls{mof} groups would share their failed and partially successful experiments.
This would open the possibility to use \gls{ml} to find correlations between linker/metal nodes and crystallization conditions, and would allow us to make predictions of the optimal synthesis conditions for novel \glspl{mof}.
Also here, \gls{ml} methods have the potential to change the way we do chemistry, but the challenges are enormous in solving the practical issues in creating an infrastructure and change of mind set that all synthesis attempts are shared in such a way that the data are publicly accessible.

Hence, a key factor in the success of \gls{ml} in the field of \glspl{mof} will be the extent in which the community is willing and able to share data.
If all data on these hundreds of thousands porous materials are shared, it will open up possibilities that go beyond the conventional ways of doing science.
We hope that the examples of \gls{ml} applied to \glspl{mof} we discussed in this review, illustrate how \gls{ml} can change the way we do and think science.

%% file: main/biographies.tex
\section*{Author Information}
\subsection*{Corresponding author}
Berend Smit; e-mail \href{mailto:berend.smit@epfl.ch}{berend.smit@epfl.ch}.

\subsection*{Biographies}

\paragraph{Kevin Maik Jablonka} received his undergraduate degree in chemistry from the Technical University of Munich and then joined EPFL for his graduate studies, supported by the Alfred Werner fund, during which he also obtained a degree in data science.
Currently, he is a PhD student in Berend Smit's group, investigating the use of data-driven methods for the design and discovery of new materials for energy-related applications.

\paragraph{Daniele Ongari} received his diploma in chemical engineering from Politecnico di Milano. In 2019, he completed his PhD  under the supervision of Prof.\ Berend Smit. His research focuses on the investigation of microporous materials, in particular \glspl{mof} and \glspl{cof}, using computational methods to assess their performances for molecular adsorption and catalysis.

\paragraph{Seyed Mohamad Moosavi} was born in Boroujerd, Iran. He received his undergraduate degree in mechanical engineering from Sharif University of Technology in Tehran, Iran. He has recently defended his PhD in chemistry and chemical engineering at EPFL under the supervision of Prof.\ Berend Smit. He visited Prof.\ Kulik's group at MIT on an SNSF fellowship award. His research interests focus on computational and data-driven design and engineering of novel materials for energy-related applications.

\paragraph{Berend Smit}
received a MSc in Chemical Engineering in 1987 and a MSc in Physics both from the Technical University in Delft (the Netherlands). In 1990, he received a cum laude PhD in Chemistry from Utrecht University (the Netherlands). He was a (senior) Research Physicists at Shell Research before he joined the University of Amsterdam (the Netherlands) as Professor of Computational Chemistry.
In 2004, he was elected Director of the European Center of Atomic and Molecular Computations (CECAM) Lyon France. Since 2007 he is Professor of Chemical Engineering and Chemistry at U.C. Berkeley and Faculty Chemist at Materials Sciences Division, Lawrence Berkeley National Laboratory.
Since July 2014 he is full professor at EPFL.\@

Berend Smit's research focuses on the application and development of novel molecular simulation techniques, with emphasis on energy-related applications.
Together with Daan Frenkel he wrote the textbook Understanding Molecular Simulations and together with Jeff Reimer, Curt Oldenburg, and Ian Bourg the textbook Introduction to Carbon Capture and Sequestration.

\section*{Acknowledgments}
The research in this article was supported by the European Research Council (ERC) under the European Union’s Horizon 2020 research and innovation programme (grant agreement 666983, MaGic) and by the NCCR-MARVEL, funded by the Swiss National Science Foundation.